\documentclass[pra,twocolumn,amsmath,amssymb,superscriptaddress]{revtex4-1}
\usepackage{hyperref}
\pdfoutput=1
\usepackage{graphicx}
\usepackage{epsfig}
\usepackage{epstopdf}

\usepackage{amsmath}
\usepackage{bm}
\usepackage{amssymb}
\usepackage{textcomp}
\usepackage{stmaryrd} 
\usepackage{ulem}

\usepackage{import}
\usepackage{color}
\usepackage{verbatim}

\usepackage{tabularx}
\usepackage{nccmath}

\usepackage{float}
\newcommand{\tr}[1]{{\textrm{#1}}}

\def\XXint#1#2#3{{\setbox0=\hbox{$#1{#2#3}{\int}$}
     \vcenter{\hbox{$#2#3$}}\kern-.5\wd0}}

\begin{document}

\title{Unified description of the optical phonon modes in  $N$-layer MoTe$_2$}

\author{Guillaume Froehlicher}
\affiliation{Institut de Physique et Chimie des Mat\'eriaux de Strasbourg and NIE, UMR 7504, Universit\'e de Strasbourg and CNRS, 23 rue du L\oe{}ss, BP43, 67034 Strasbourg Cedex 2, France}

\author{Etienne Lorchat}
\affiliation{Institut de Physique et Chimie des Mat\'eriaux de Strasbourg and NIE, UMR 7504, Universit\'e de Strasbourg and CNRS, 23 rue du L\oe{}ss, BP43, 67034 Strasbourg Cedex 2, France}

\author{Fran\c{c}ois Fernique}
\affiliation{Institut de Physique et Chimie des Mat\'eriaux de Strasbourg and NIE, UMR 7504, Universit\'e de Strasbourg and CNRS, 23 rue du L\oe{}ss, BP43, 67034 Strasbourg Cedex 2, France}

\author{Chaitanya Joshi}
\affiliation{Physics and Materials Science Research Unit, University of Luxembourg, 162a avenue de la Fa\"iencerie, L-1511 Luxembourg, Luxembourg}
\affiliation{Department of Physics, Indian Institute of Technology Bombay, Powai, Mumbai-400076, India}

\author{Alejandro Molina-S\'{a}nchez}
\affiliation{Physics and Materials Science Research Unit, University of Luxembourg, 162a avenue de la Fa\"iencerie, L-1511 Luxembourg, Luxembourg}

\author{Ludger Wirtz}
\affiliation{Physics and Materials Science Research Unit, University of Luxembourg, 162a avenue de la Fa\"iencerie, L-1511 Luxembourg, Luxembourg}

\author{St\'ephane Berciaud}
\email{stephane.berciaud@ipcms.unistra.fr}
\affiliation{Institut de Physique et Chimie des Mat\'eriaux de Strasbourg and NIE, UMR 7504, Universit\'e de Strasbourg and CNRS, 23 rue du L\oe{}ss, BP43, 67034 Strasbourg Cedex 2, France}

\begin{abstract}
$N$-layer transition metal dichalcogenides provide a unique platform to investigate the evolution of the physical properties between the bulk (three dimensional) and monolayer (quasi two-dimensional) limits. Here, using high-resolution micro-Raman spectroscopy, we report a unified experimental description of the $\bm \Gamma$-point optical phonons in $N$-layer $2H$-molybdenum ditelluride (MoTe$_2$). We observe a series of $N$-dependent low-frequency interlayer shear and breathing modes (below $40~\rm cm^{-1}$, denoted LSM and LBM) and well-defined Davydov splittings of the  mid-frequency modes (in the range $100-200~\rm cm^{-1}$, denoted iX and oX), which solely involve displacements of the chalcogen atoms. In contrast, the high-frequency modes (in the range $200-300~\rm cm^{-1}$, denoted iMX and oMX), arising from displacements of both the metal and chalcogen atoms, exhibit considerably reduced splittings. The manifold of phonon modes associated with the in-plane and out-of-plane displacements are quantitatively described by a force constant model, including interactions up to the second nearest neighbor and surface effects as fitting parameters. The splittings for the iX and oX modes observed in $N$-layer crystals are directly correlated to the corresponding bulk Davydov splittings between the $E_{2u}/E_{1g}$ and $B_{1u}/A_{1g}$ modes, respectively, and provide a measurement of the frequencies of the bulk silent $E_{2u}$ and $B_{1u}$ optical phonon modes. Our analysis could readily be generalized to other layered crystals.


\end{abstract}

%\pacs{63.22.Np,78.67.-n,~78.30.-j,63.20.kd}
%02.20.-a
%63.20.Kr Phonon-electron and phonon-phonon interactions
%63.22.Np 	Layered systems
%73.22.Lp (colllective excitations), 63.20.kd, 78.30.Na, 78.67.-n
%73.22.Pr 	Electronic structure of graphene
%73.63.-b Electronic transport in nanoscale materials and structures
%78.67.Wj 	Optical properties of graphene
%71.70.Di 	Landau levels
%73.22.Lp Collective excitations
% electronic transport graphene 72.80.Vp
%phonons in graphene, 63.22.Rc
%,~63.20.kd,~63.20.kg,~78.30.Na
%73.63.-b 	Electronic transport in nanoscale materials and structures
%78.67.-n 	Optical properties of low-dimensional, mesoscopic, and nanoscale materials and structures (for magnetic properties of nanostructures, see 75.75.-c; for electronic transport in nanoscale structures, see 73.63.-b; for mechanical properties of nanoscale systems, see 62.25.-g)
%61.72.Dd 	Experimental determination of defects by diffraction and scattering
%82.45.-h 	Electrochemistry and electrophoresis
%82.47.-a 	Applied electrochemistry (see also 88.30.G- Fuel cell systems, and 88.30.P- Types of fuel cells in renewable energy resources and applications)
%82.45.Mp 	Thin layers, films, monolayers, membranes (for anodic films, see 82.45.Cc; for surface double layers, see 73.30.+y in electronic structure of surfaces)

\maketitle

{\textit{\textbf{Introduction~}}}
In the wake of graphene, a vast family of layered materials is attracting tremendous attention~\cite{Geim2013}. Now available in the form of $N$-layer crystals, the latter exhibit peculiar physical properties that complement the assets of graphene and offer exciting perspectives to design van der Waals heterostructures~\cite{Geim2013}. Semiconducting transition metal dichalcogenides (MX$_2$, with M = Mo, W and X = S, Se, Te) are among the most actively investigated layered crystals~\cite{Wilson1969}. Indeed, although bulk MX$_2$ exhibit indirect bandgaps, monolayer MX$_2$ are direct bandgap semiconductors~\cite{Mak2010,Splendiani2010} with remarkable spin, valley~\cite{Xu2014} and optoelectronic properties~\cite{Peng2015}. More Generally, $N$-layer MX$_2$ crystals provide an ideal platform to uncover the impact of symmetry breaking and interlayer interactions on the  electronic, optical and vibrational properties, from the bulk (three-dimensional) to  the monolayer (quasi two-dimensional) limit.

\begin{figure*}[!tbh]
\begin{center}
\includegraphics[width=1\linewidth]{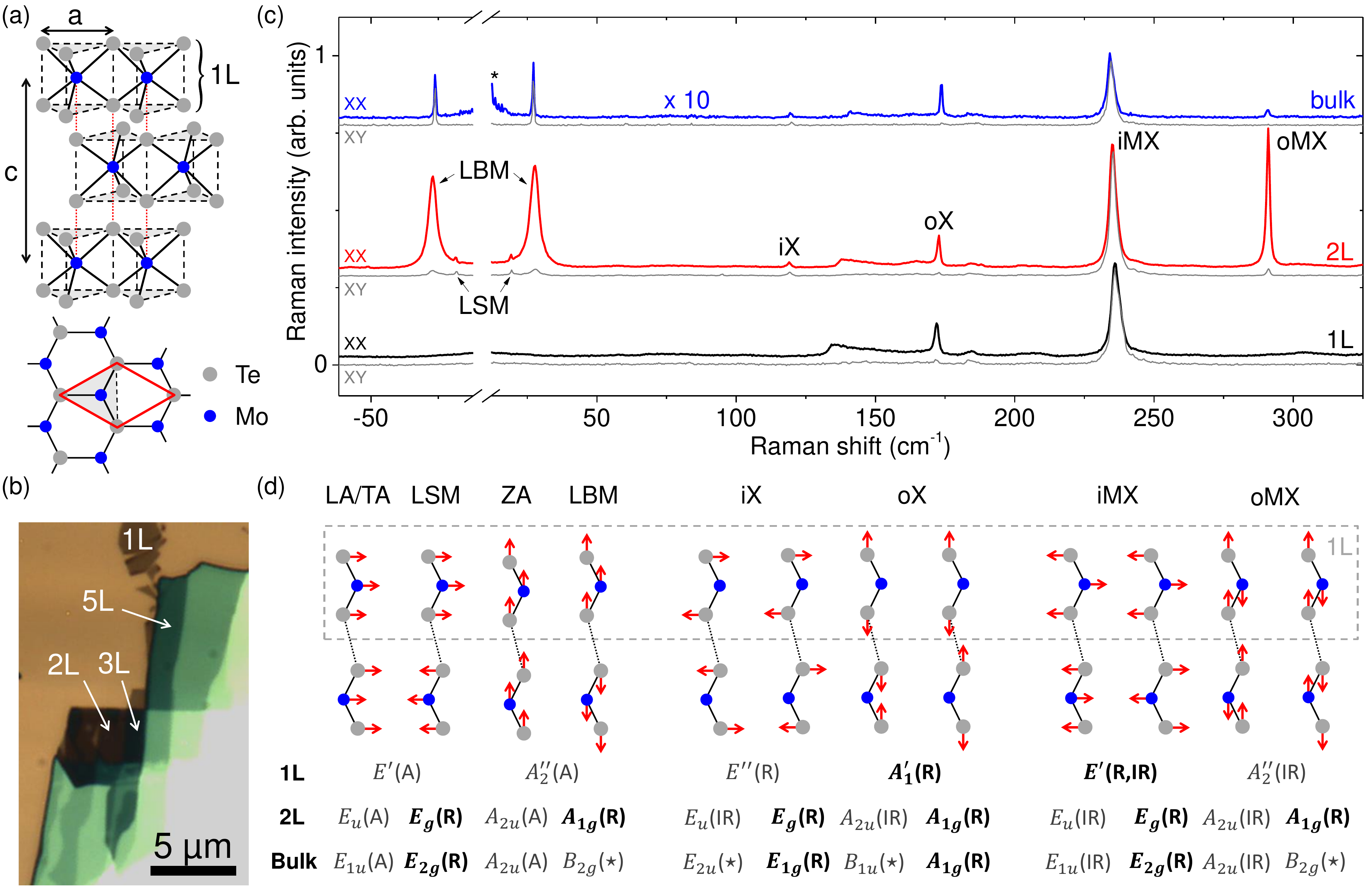}
\caption{(a) Side and top view of the crystal structure of $2H$-MoTe$_2$. The red rhombus represents the unit cell. (b) Optical image of a $N$-layer MoTe$_2$ crystal deposited onto a Si/SiO$_2$ substrate. (c) Raman spectra of monolayer, bilayer and bulk MoTe$_2$ in the parallel (XX, thick colored solid lines) and perpendicular (XY, thin grey lines) polarization configuration. The spectra are vertically offset for clarity and the asterisk highlights residual contributions from the exciting laser beam.  (c) Atomic displacements and irreducible representations associated with the $\bm \Gamma$ point phonon modes in monolayer, bilayer and bulk $\rm{MoTe}_2$. The Raman (R) and/or infrared (IR) activity are indicated, and stars denote silent modes. The zero frequency acoustic (A) modes (LA, TA, ZA) and their irreducible representations are also shown for clarity.}
\label{Fig1}
\end{center}
\end{figure*}

In particular, in $N$-layer MX$_2$, interlayer interactions result in a splitting of all the \textit{monolayer} phonon modes~\cite{Wieting1971,Molina2011,Luo2013,Luo2013b,Terrones2014,Ribeiro2014,Yamamoto2014,Scheuschner2015,Zhang2015b,Zhao2013} (see Table~\ref{TabRep}). The latter effect is known as the Davydov splitting~\cite{Davydov1971} and is closely related to the force constants that govern the vibrational properties of MX$_2$~\cite{Luo2013}. The Davydov splitting has been previously studied in polyaromatic molecules~\cite{Khelladi1975}, thin films~\cite{Aroca1987}, and bulk layered crystals, including MX$_2$~\cite{Wieting1971,Wieting1980,Ghosh1983,Molina2011,Luo2013}. Recently, low-frequency Raman spectroscopy has been employed in $N$-layer MoS$_2$ ~\cite{Plechinger2012,Zeng2012,Zhao2013,Zhang2013,Boukhicha2013} and WSe$_2$~\cite{Zhang2013} to uncover the \textit{fan diagrams} associated with the frequencies of the \textit{interlayer} shear (LSM) and breathing (LBM) modes, which arise from the Davydov splitting of the zero frequency acoustic modes (see Fig.~\ref{Fig1}(d)). Related splittings of the higher frequency modes involving \textit{intralayer} atomic displacements have been reported  in $N\leq 5$-layer MoSe$_2$~\cite{Tonndorf2013,Chen2015} and  WS$_2$~\cite{Staiger2015}. Thus far, such splittings have solely been reported for the out-of-plane Raman-active phonon with $A_{1g}$ symmetry in the bulk, whereas other high-frequency modes, such as the in-plane phonon with $E_{2g}$ symmetry in the bulk, exhibit anomalous $N$-dependent frequency shifts~\cite{Lee2010,Li2012,Molina2011,Luo2013,Luo2013b,Tonndorf2013,Berkdemir2013,Yamamoto2014} but no splitting. However, a unified description of the Davydov splitting in $N$-layer MX$_2$ is still lacking.

In this letter, we quantitatively investigate the Davydov splitting of \textit{all} the $\bm \Gamma$-point optical phonon modes, over the range $4-300~\rm cm^{-1}$ in $N$-layer $2H$ molybdenum ditelluride (MoTe$_2$) crystals. As in other MX$_2$, $N$-layer MoTe$_2$ displays a set of $\lfloor{N/2}\rfloor$ (herein $\lfloor\,\rfloor$ ($\lceil\,\rceil$) denote the floor (ceil) functions that map $N/2$ to the largest previous (smallest following) integer) well-separated  LSM and LBM  below $40~\rm cm^{-1}$. In the range $100-200~\rm cm^{-1}$, the in-plane and out-of-plane modes involving solely displacements of the chalcogen (Te) atoms (herein denoted iX and oX) exhibit pronounced splittings into $\lfloor{N/2}\rfloor$ and $\lceil{N/2}\rceil$ Raman-active modes, respectively. The Davydov splitting between the bulk in-plane $E_{1g}$ Raman-active and $E_{2u}$ silent modes (out-of-plane $A_{1g}$ Raman-active and $B_{1u}$ silent modes) is $2.7~\rm cm^{-1}$ ($4.7~\rm cm^{-1}$).  In the range $200-300~\rm cm^{-1}$, the
modes involve motion of both the metal (Mo) and chalcogen (Te) atoms. The in-plane iMX mode exhibits no observable splitting.
The out-of-plane oMX mode exhibits a small splitting of approximately $1\rm ~cm^{-1}$, which is actually not a Davydov splitting but a splitting due to surface effects.

All the observed trends are fitted using a finite linear chain model that includes inter- and intralayer force constants up to the second nearest neighbor, as well as surface effects at the edges of the chain~\cite{Luo2013}. This model naturally explains why the interlayer force constants, which directly determine the manifold of rigid layer modes, also give rise to sizable Davydov splittings for the iX and oX modes but to reduced splittings for the iMX and oMX modes. Conversely, surface effects are responsible for the apparent downshift of the iMX mode as $N$ augments and for the slight splitting of the oMX mode. 

\setlength{\tabcolsep}{0.3cm}
\renewcommand{\arraystretch}{0.8}

\begin{table*} [ht!]
\begin{center}
\small
\begin{tabular}{ccccccc}
\hline
\hline
\\[-5pt]
\textbf{Number of} &  \textbf{LSM} & \textbf{LBM} & \textbf{iX} & \textbf{oX} & \textbf{iMX} & \textbf{oMX}  \\
\textbf{layers} &  $\leqslant 30~\textrm{cm}^{-1}$ & $\leqslant 40~\textrm{cm}^{-1}$ & $\sim 120~\textrm{cm}^{-1}$ & $\sim 170~\textrm{cm}^{-1}$ & $\sim 235~\textrm{cm}^{-1}$ & $\sim 290~\textrm{cm}^{-1}$ \\
\\[-5pt]
\hline
\hline
\\[-5pt]
1  &  $-$  & $-$ & $E''$ & $\boldsymbol{A'_1}$ & $\boldsymbol{E'}$ & $A''_2$   \\[5pt]

\hline
\\[5pt]

2  & $\boldsymbol{E_{g}}$ & $\boldsymbol{A_{1g}}$ & $\boldsymbol{E_{g}}$ & $\boldsymbol{A_{1g}}$ & $\boldsymbol{E_{g}}$ & $\boldsymbol{A_{1g}}$ \\[5pt]
$-$ & $-$ & $-$ & $E_{u}$ & $A_{2u}$ & $E_{u}$ & $A_{2u}$   \\[5pt]

\hline
\\[5pt]

odd N &  $\frac{N-1}{2}\boldsymbol{E'}$  & $\frac{N-1}{2}\boldsymbol{A'_1}$ & $\frac{N-1}{2}\boldsymbol{E'}$ & $\frac{N+1}{2}\boldsymbol{A'_1}$ & $\frac{N+1}{2}\boldsymbol{E'}$ & $\frac{N-1}{2}\boldsymbol{A'_1}$   \\[5pt]
$-$ &  $\frac{N-1}{2}E''$  & $\frac{N-1}{2}A''_2$ & $\frac{N+1}{2}E''$ & $\frac{N-1}{2}A''_2$ & $\frac{N-1}{2}E''$ & $\frac{N+1}{2}A''_2$    \\[5pt]

\hline
\\[5pt]

even N  & $\frac{N}{2}\boldsymbol{E_{ g}}$ & $\frac{N}{2}\boldsymbol{A_{ 1g}}$ & $\frac{N}{2}\boldsymbol{E_{ g}}$ & $\frac{N}{2}\boldsymbol{A_{ 1g}}$ & $\frac{N}{2}\boldsymbol{E_{ g}}$ & $\frac{N}{2}\boldsymbol{A_{1g}}$ \\[5pt]

$-$ & $\left(\frac{N}{2}-1\right)E_{u}$ & $\left(\frac{N}{2}-1\right)A_{2u}$ & $\frac{N}{2}E_{u}$ & $\frac{N}{2}A_{2u}$ & $\frac{N}{2}E_{u}$ & $\frac{N}{2}A_{2u}$\\ [5pt]

\hline
\\[5pt]

bulk & $\boldsymbol{E_{2g}}$ & $B_{2g}~\star$ & ${E_{1g}}$ & $\boldsymbol{A_{1g}}$ & $\boldsymbol{E_{2g}}$ & $B_{2g}~\star$   \\[5pt]
$-$ & $-$ & $-$ & $E_{2u}~\star$ & $B_{1u}~\star$ & $E_{1u}$ & $A_{2u}$  \\[5pt]

\hline
\hline
\end{tabular}
\end{center}  
\caption{Irreducible representations of the optical phonon modes at $\Gamma$ for single-, bi-, $N$- layer and bulk MoTe$_2$. Bold characters denote Raman-active modes in a backscattering  geometry. Note that modes with $E_{1g}$ and $E''$ symmetry are Raman-active but not observable in a backscattering geometry~\cite{Loudon1964} and that modes with $E'$ are Raman- and infrared-active. Stars ($\star$) denote silent modes. All the other modes are infrared active.}
\label{TabRep}
\end{table*}

\begin{figure*}[!tbh]
\begin{center}
\includegraphics[width=1\linewidth]{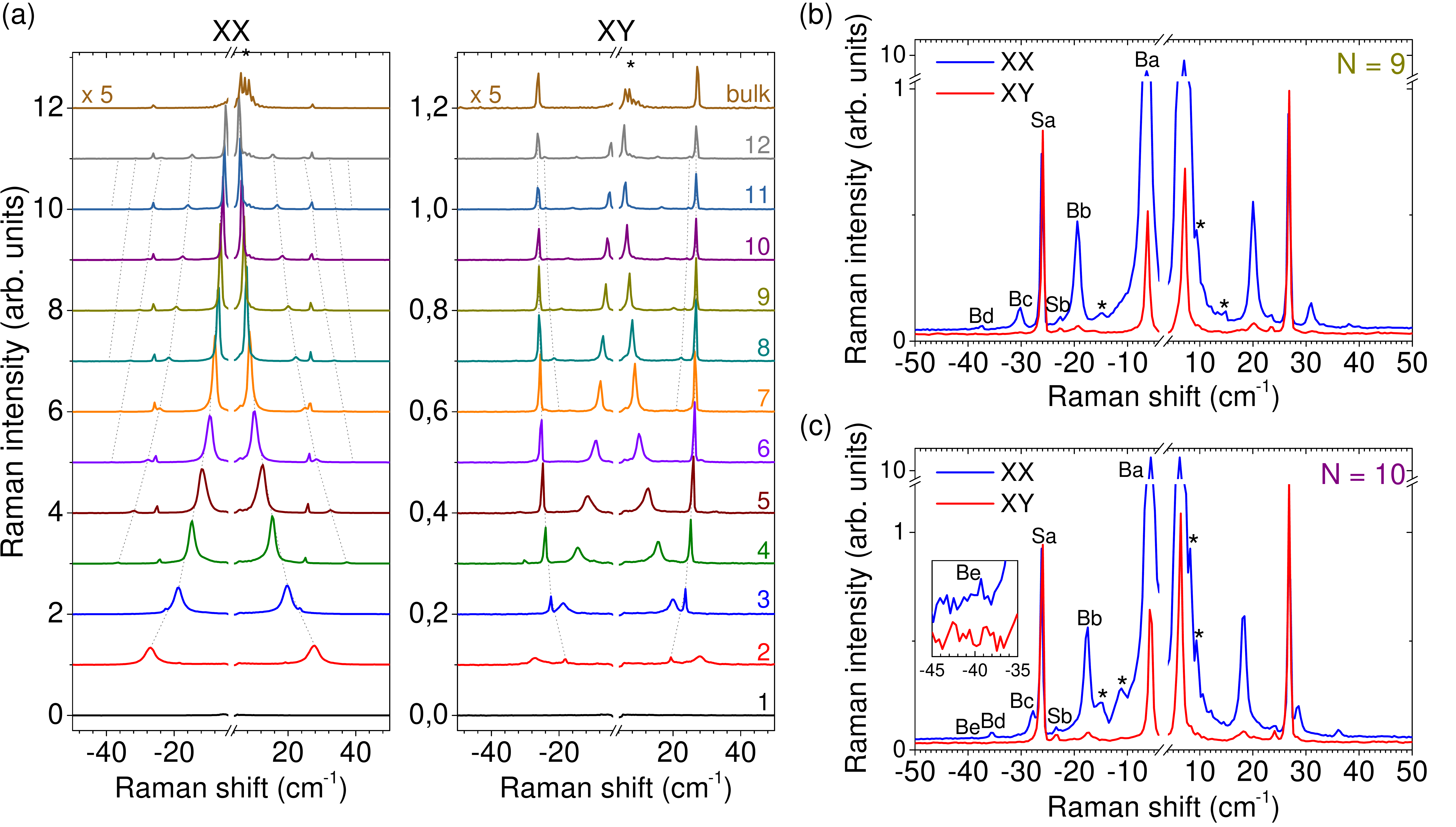}
\caption{(a) Polarization-resolved low-frequency Raman spectra of $N=1$ to $N=12$ layer MoTe$_2$ and of bulk $2H$-MoTe$_2$, in the parallel (XX) and transverse (XY) configuration recorded at $E_{\rm L}=2.33~\rm eV$. The interlayer breathing modes (LBM) largely dominate the spectra in the XX configuration and their intensity is reduced by more than one order of magnitude in the XY configuration. The shear modes (LSM) are not sensitive to the polarization configuration. The dashed lines follow the frequencies of each LSM and LBM (see also Fig.~\ref{Fig5}). (c-d) Polarization-resolved low-frequency micro-Raman spectra of (b) $N=9$ and (c) $N=10$ layer MoTe$_2$ in the parallel (XX) and transverse (XY) configuration. The four (five) expected Raman-active LBM (denoted Ba to Bd (Ba to Be) for $N=9$ $(N=10)$ and two LSM (denoted Sa and Sb) are observed. The asterisks highlight residual contributions from the exciting laser beam. }
\label{Fig2}
\end{center}
\end{figure*}

\begin{figure*}[!tb]
\begin{center}
\includegraphics[width=1\linewidth]{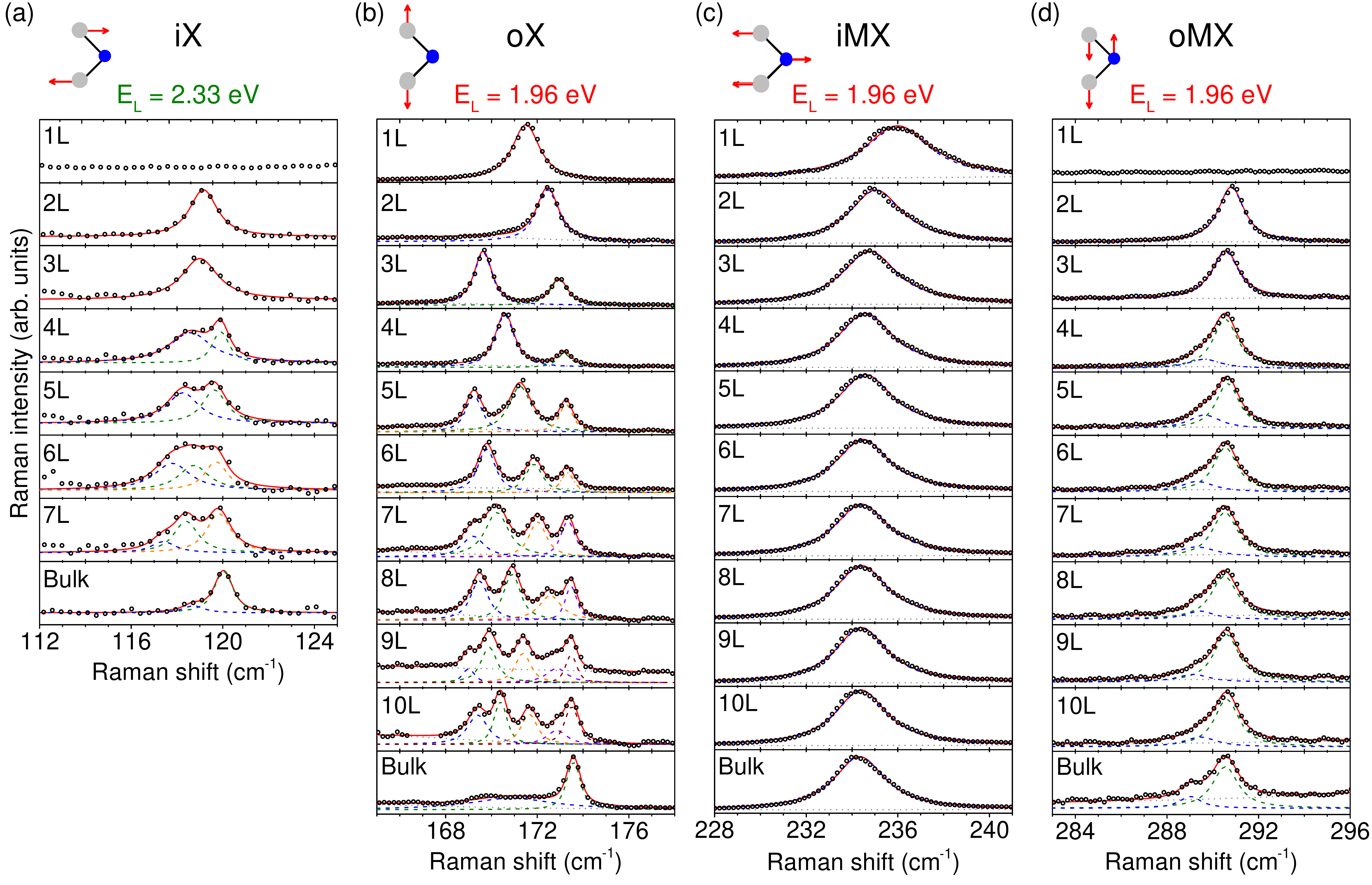}
\caption{Normalized micro-Raman spectra of the  (a) iX, (b) oX, (c) iMX, and (d) oMX mode-features in $N$-layer MoTe$_2$. The laser phonon energy $E_{\rm L}$ and the elementary intralayer displacements are indicated. The measured Raman features (symbols) are fit to Voigt profiles (solid lines). For the modes that show a Davydov splitting, each subpeak is represented with a colored dashed line. A featureless background (grey dashed line) has been considered when necessary.}
\label{Fig3}
\end{center}
\end{figure*}

In its stable form, bulk MoTe$_2$ displays a trigonal prismatic ($2H$ polytype) structure with AbA/BaB stacking (see Fig.~\ref{Fig1}) and shares many properties with the widely studied MoS$_2$, MoSe$_2$, WS$_2$ and WSe$_2$ crystals~\cite{Wilson1969,Ribeiro2014}. MoTe$_2$ has recently attracted particular interest due to its lower transport bandgap~\cite{Fathipour2014,Lezama2014,Lin2014,Pradhan2014} and near-infrared emission~\cite{Ruppert2014,Lezama2015}. Indeed, at room temperature, monolayer MoTe$_2$ exhibits a direct optical bandgap at 1.1 eV, whereas the bulk material has an indirect optical bandgap slightly below 1.0~eV~\cite{Wilson1969,Ruppert2014,Lezama2015}. Bulk MoTe$_2$ exhibits 3 acoustic and 15 optical phonon modes at $\bm \Gamma$ as is the case for MoS$_2$ and MoSe$_2$~\cite{Wieting1971,Molina2011,Luo2013,Ribeiro2014,Yamamoto2014,Scheuschner2015,Zhang2015b,Zhao2013}. As illustrated in Fig~\ref{Fig1}, the manifold of optical phonons breaks down into 5 (doubly degenerate) in-plane modes and 5 out-of-plane modes. According to group theory~\cite{Loudon1964}, the integrated intensity of the in-plane Raman-active modes is expected to be independent on the angle $\theta$ between the polarization of the incoming and scattered photons, whereas the integrated intensity of the out-of-plane modes should exhibit a $\cos^2\theta$ dependence.

\textit{\textbf{Results~}}
Figure~\ref{Fig1} shows the micro-Raman spectra of monolayer, bilayer and bulk MoTe$_2$ recorded at $E_{\rm L}=2.33~\rm eV$ for $\theta=0$ (parallel configuration, XX) and $\theta=\pi/2$ (cross-polarized configuration, XY). We first address the low-frequency range below $40~\rm cm^{-1}$. In bulk MoTe$_2$, we observe only a single feature at $26~\rm cm^{-1}$ that shows similar intensities in the XX and XY configurations. The latter is assigned to the in-plane interlayer shear mode (LSM) with $E_{2g}$ symmetry~\cite{Wieting1980}. For $N=2$, we observe a prominent feature at $28~\rm cm^{-1}$ that shows strong extinction in the XY polarization and a fainter feature at $18~\rm cm^{-1}$ whose intensity is similar in the XX and XY configurations. The former is thus assigned to the out-of-plane layer breathing mode (LBM) with $A_{1g}$ symmetry, whereas the latter is assigned to the LSM with $E_{g}$ symmetry. The LBM has $B_{2g}$ symmetry in the bulk and is silent. As expected, we do not observe any interlayer mode for $N=1$.

In the mid- $(100-200~\rm cm^{-1})$ and high-frequency $(200-300~\rm cm^{-1})$ ranges, the Raman spectra of bilayer MoTe$_2$ displays four one-phonon features, which have previously been identified as originating from the following \textit{intralayer} displacements: (i) the in-plane, out-of-phase vibration of the Te planes, with $E_{1g}$ symmetry in the bulk (iX mode, near $120~\rm {cm^{-1}}$), (ii) the out-of-plane, out-of-phase vibration of the Te planes, with $A_{1g}$ symmetry in the bulk (oX mode, near 170~cm$^{-1}$), (iii) the in-plane vibration of the Mo and Te planes against each other, with $E_{2g}$ symmetry in the bulk (iMX mode, near 230~cm$^{-1}$), and (iv) the out-of-plane vibration of the Mo and Te planes against each other, with $B_{2g}$ symmetry in the bulk, (oMX mode, near 290~cm$^{-1}$)~\cite{Yamamoto2014,Guo2015,Wieting1980}. The bulk iX and oMX modes are predicted to be Raman inactive in a backscattering geometry and silent, respectively. However, both modes appear as faint features in thick MoTe$_2$ flakes (considered as a bulk reference). This surprising observation, also reported recently on other MX$_2$ might be a consequence of the finite penetration depth of our laser due to the strong optical absorption of MoTe$_2$~\cite{Wilson1969,Ruppert2014} or may arise from a breakdown of the Raman selection rules due to resonance effects~\cite{Luo2013b,Lee2015,Scheuschner2015}. As predicted by group theory~\cite{Ribeiro2014,Scheuschner2015,Yamamoto2014}, the iX and oMX modes are not observed in monolayer MoTe$_2$ (see Figs.~\ref{Fig1} and \ref{Fig3}(a)). We verified that the oX and oMX features nearly vanish in the XY configuration, whereas the integrated intensities of the iX and iMX features do not change (see Fig.~\ref{Fig1}).

Figure~\ref{Fig2} shows the evolution of the low-frequency modes (LSM and LBM) from $N=2$ to $N=12$-layer MoTe$_2$. As previously reported in multilayer graphene~\cite{Tan2012,Lui2014} and recently in other MX$_2$,~\cite{Plechinger2012,Zeng2012,Zhao2013,Boukhicha2013,Zhang2013,Chen2015}, a set of $N$-dependent low-frequency Stokes and anti-Stokes peaks appears for $N \geq 2$. The number of detected peaks increases with $N$ and the peaks can be separated into branches that seemingly \textit{stiffen} or \textit{soften} with increasing $N$ (see the dashed lines in Fig~\ref{Fig2}(a)). Interestingly, compared to a reference recorded in the XX configuration, the integrated intensity of peaks belonging to a branch that softens with increasing $N$ drops by more than one order of magnitude in the XY configuration, whereas the integrated intensity of the peaks that belong to a branch that stiffens with increasing $N$ is marginally affected. Therefore, the branches that soften (stiffen) with increasing $N$ are assigned to the LBM (LSM). We are able to resolve two branches of LSM and, remarkably, five branches of LBM, \textit{i.e.}, the complete manifold of Raman-active LBM up to $N=11$~(see Table~\ref{TabRep} and Fig.~\ref{Fig5}).

The Raman spectra of the mid- (iX, oX) and high-frequency (iMX and oMX) modes in $N$-layer MoTe$_2$ are shown in Fig.~\ref{Fig3}. For $N\geq3$-layer MoTe$_2$, we observe a prominent splitting of the oX-mode feature (see Fig.~\ref{Fig3}(b)), whereas in the bulk limit, one recovers a single symmetric feature (assigned to the Raman-active $A_{1g}$ mode). Interestingly, as shown in Fig.~\ref{Fig3}(a), the iX-mode  feature also splits, but only for $N\geq4$. We can resolve up to three subfeatures for $N=6$ and 7, but the Raman signal in $N\geq 8$-layer MoTe$_2$ becomes too small to perform a quantitative analysis. At $E_{\rm L}=1.96~\rm eV$, the oMX-mode feature also exhibits a modest splitting, on the order of $1~\rm cm^{-1}$, for $N\geq4$. Two subpeaks can be distinguished. However, the evolution of their frequencies does not follow a specific trend as a function of $N$ (see Fig.~\ref{Fig3}(d) and Fig.~\ref{Fig5}(f)). In contrast, the iMX-mode feature exhibits a faint shoulder on its high-energy side (see Fig.~\ref{Fig3}(c)), but no appreciable splitting can be resolved. However, the iMX feature downshifts as $N$ increases, consistently with previous reports on $N$-layer MX$_2$~\cite{Lee2010,Li2012,Molina2011,Luo2013,Luo2013b,Berkdemir2013,Tonndorf2013,Zhao2013,Yamamoto2014,Ruppert2014}.

The splitting of a bulk phonon mode in a $N$-layer $2H$-MX$_2$ can be understood based on a group theory analysis~\cite{Luo2013,Luo2013b,Terrones2014,Ribeiro2014,Yamamoto2014,Scheuschner2015,Staiger2015,Zhang2015b,Zhao2013}. Bulk MoTe$_2$ belongs to the $D_{6h}$ non-symmorphic space group and crystals with odd or even $N$ belong to the $D_{3h}$ and $D_{3d}$ symmorphic space groups, respectively. In bulk $2H$-MX$_2$, the (silent) LBM and (Raman active) LSM correspond to an out-of phase displacement of adjacent layer pairs, whereas the iX, iMX, oX, and oMX atomic displacements  give rise to two modes, in which the relative motion of equivalent atoms   belonging to two adjacent layers is either in-phase or out-of-phase (see Fig.~\ref{Fig1}(d)).  In a $N$-layer system, interlayer interactions and symmetry breaking result in a manifold of $N-1$ LSM and LBM, and  $N$ iX, oX, iMX and oMX modes. Table~\ref{TabRep} summarizes the irreducible representations and the activity of these optical phonon modes for mono-, bi-, $N$-layer, and bulk MoTe$_2$. 

In agreement with group theory predictions, our Raman measurements reveal that (i) secondary LSM and LBM features appear from $N\geq4$ (ii) the oX feature splits for $N\geq3$ and  we observe $\lceil N/2\rceil$ Raman active subfeatures in $N$-layer MoTe$_2$, (iii) the splitting of the iX feature into $\lfloor N/2\rfloor$ Raman active subfeatures is observed from $N\geq4$, (iv) a splitting of the oMX feature appears for $N\geq4$. The frequencies of all the measured Raman features, extracted from Voigt fits (see Figs.~\ref{Fig2}-\ref{Fig3}) are reported in Fig.~\ref{Fig5}.

\begin{figure*}[!t]
\begin{center}
\includegraphics[width=1\linewidth]{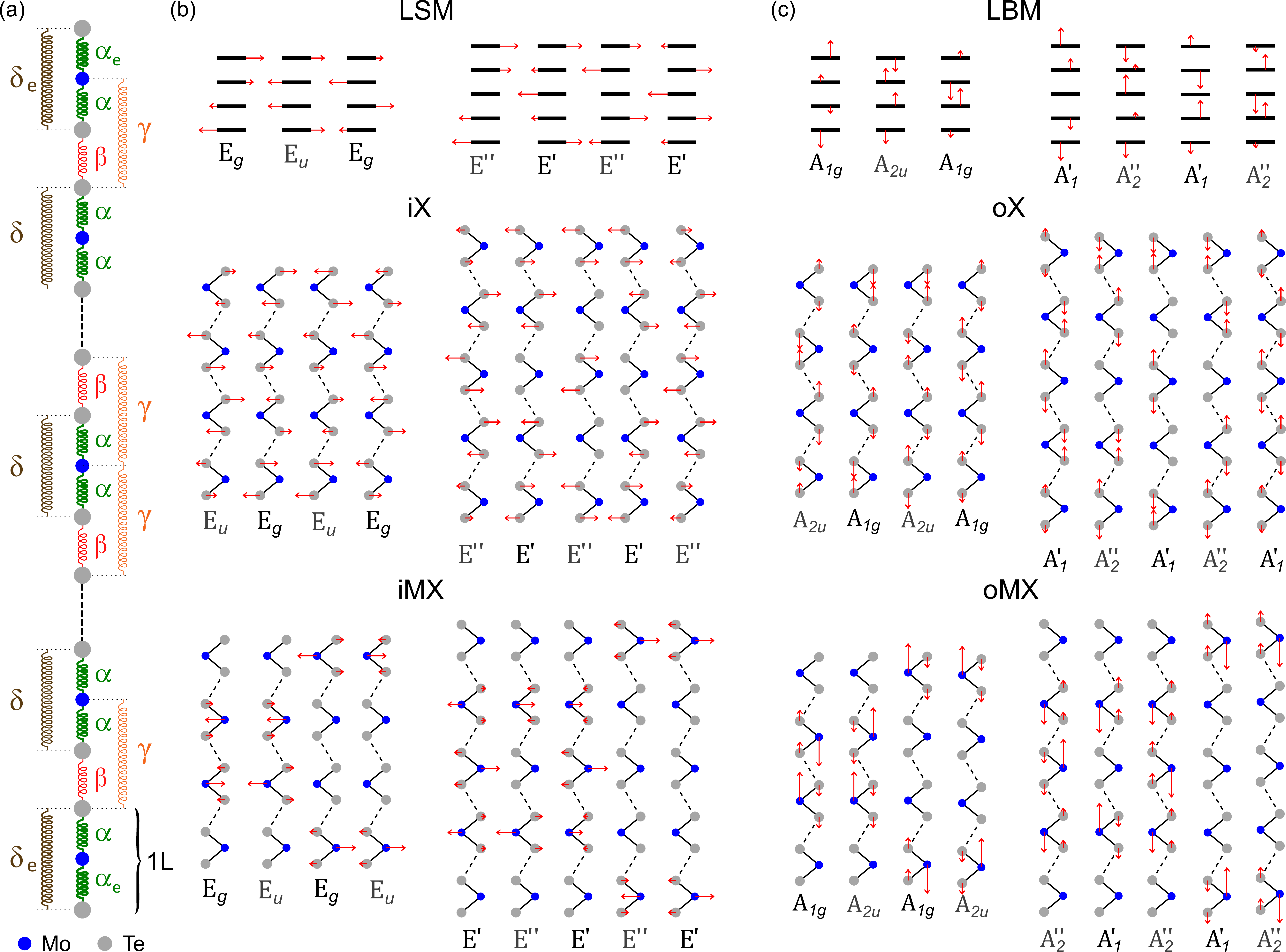}
\caption{(a) Schematic of the finite linear chain model discussed in the text.  Calculated normal displacements associated with all the in-plane (b) and out-of-plane (c)  optical phonon modes in $N=4$ and $N=5$ layer MoTe$_2$. The size of the arrows is proportional to the amplitude $u_{i,j}^k$ of the normal displacement obtained from the solution of Eq.~\eqref{eq1}. At a given $N$, the mode frequencies increase from left to right. The irreducible representation of each normal mode is indicated (see also Supporting Information).}
\label{Fig4}
\end{center}
\end{figure*}

\textit{\textbf{Force constant model~}}
We now introduce a semiempirical model to rationalize the observed splittings. $N$-layer MoTe$_2$ is modeled as a finite linear chain composed of $N$ Mo atoms and $2N$ Te atoms. Within one MoTe$_2$ unit nearest neighbor Mo and Te atoms and the pair of second nearest neighbor Te atoms are connected by springs with force constants per unit area $\alpha$ and $\delta$ respectively~\cite{Luo2013} (see Fig.~\ref{Fig4}(a)). Interlayer interactions are then described by two force constants per unit area, $\beta$ and $\gamma$, between nearest neighbor Te atoms belonging to adjacent layers and between second nearest neighbor Mo and Te atoms, respectively. In addition, finite size effects are known to lead to a slight reduction of the metal-chalcogen bond length on the outer layers (``surface effects'')~\cite{Luo2013}. As a result, to improve our model, effective values $\alpha_{\rm e}>\alpha$ and $\delta_{\rm e}>\delta$ are phenomenologically considered for the first and $N^{\rm th}$ layer. We note that our choice of using the same value of $\alpha_{\rm e}$ at both ends of the chain is consistent with the fact that no significant substrate-induced frequency shifts have been observed on the Raman response of MoTe$_2$ and other MX$_2$~\cite{Lee2010,Yamamoto2014,Luo2013,Luo2013b,Zhao2013}. The normal modes $\mathcal{U}^k =\left(\begin{mmatrix} u_{1,1}^{k}, & u_{2,1}^{k}, & u_{3,1}^{k}, & \hdots, & u_{i,j}^{k}, & \hdots,  & u_{1,N}^{k}, & u_{2,N}^{k}, & u_{3,N}^{k} \end{mmatrix}\right)$, with $u_{i,j}^{k}$ the normal displacement of the $i^{\rm th}$ atom ($i=1,3$ for Te and $i=2$ for Mo) in the $j^{\rm th}$ MoTe$_2$ layer ($j\in \llbracket 1,N \rrbracket$) associated with the normal mode $k\in \llbracket 1,3N \rrbracket$ (see Fig. \ref{Fig4}(a)), and the eigenfrequencies $\omega_k$ of a $N$-layer system can be obtained from Newton's equations of motion, which lead to the secular equation 

\begin{equation}
\omega_k^2 \:\mathcal{U}^k=  \mathcal{D}\:\mathcal{U}^k,
\label{eq1}
\end{equation}
involving the $3N\times 3N$ dynamical matrix $\mathcal{D}$ (see Supporting Information).

For in-plane (out-of-plane) displacements, Eq.~\eqref{eq1} predicts three manifolds of $N$ normal modes that correspond to (i) the low frequency LSM (LBM)  (including the zero frequency acoustic mode)  (ii) the mid-frequency iX (oX) modes and (iii) the high-frequency iMX (oMX) modes. Figure \ref{Fig4}(b) represents the calculated iX and oX normal modes (using the parameters in Table~\ref{TabFIT}) and their irreducible representations for $N=4$ and $N=5$. For the low- (LSM, LBM) and mid-frequency (iX, oX) modes, the eigenfrequencies increase (decrease) as the layers exhibit more out-of phase (in-phase) relative motion, up to the limit of the highest- (lowest-) frequency mode, which corresponds to an out-of phase (in-phase) oscillation for all layers. This trend is as expected from classical theories of coupled oscillators. In particular, we can readily conclude that the dominant LBM feature corresponds, for even $N$ to the out-of-phase oscillation of two blocks composed of $N/2$ layers that vibrate in-phase; for odd $N$ to the out-of-phase oscillations of two blocks composed of $(N-1)/2$ layers that vibrate in-phase, whereas the central layer stays at rest. In contrast, the dominant LSM feature corresponds to an out-of-phase displacement of adjacent layers as in bulk crystals~\cite{Michel2012,Zhao2013} (see also Supporting Information).

As shown in Fig.~\ref{Fig5}, the fan diagrams associated with the in-plane LSM, iX- and iMX-mode frequencies on the one hand, and out-of-plane LBM, oX- and oMX-mode frequencies on the other hand are very well described by the force-constant model. The force constants (per unit area) used as fitting parameters and the corresponding bulk frequencies (see Supporting Information for their analytical expressions) are  reported in Table~\ref{TabFIT}. We find that all the force constants correspond to a restoring force, except for the in-plane second nearest neighbor force constant between Te pairs. The values obtained here are qualitatively similar to the values that Luo \textit{et al.} reported for MoS$_2$ by fitting the bulk frequencies (obtained from calculations based on the local density approximation (LDA)) to a force constant model~\cite{Luo2013}. 

\begin{figure*}[!tb]
\begin{center}
\includegraphics[width=0.65\linewidth]{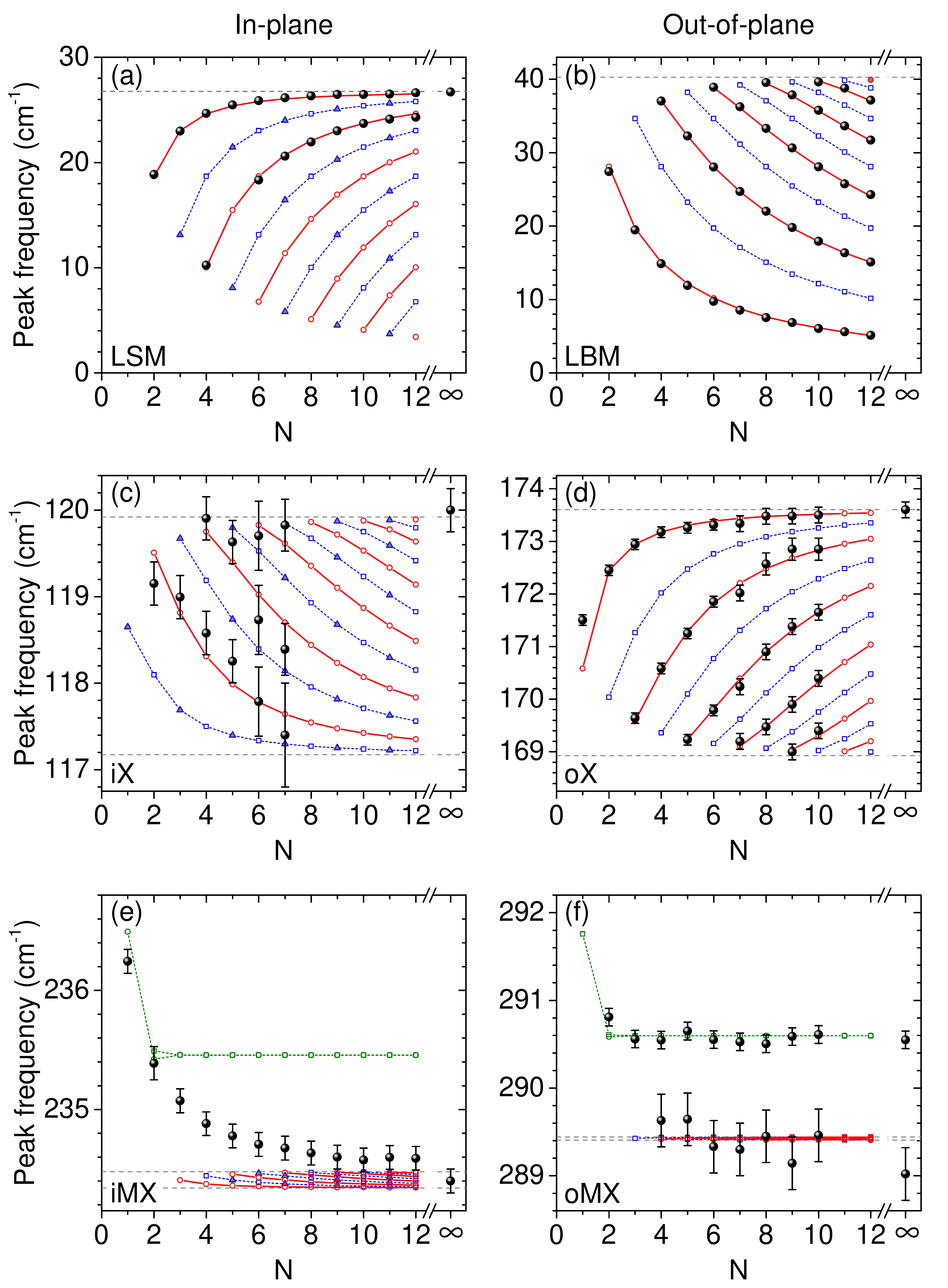}
\caption{Frequencies of the (a) LSM (b) LBM (c) iX, (d) oX, (e) iMX, (f) oMX modes extracted from fits of the spectra displayed in Figs~\ref{Fig2}-\ref{Fig3} (black circles) as a function of the number of layers $N$. The red open cirles in (a)-(f) are the frequencies of the Raman-active modes calculated by solving Eq.\eqref{eq1} with the fitting parameters in Table~\ref{TabFIT}. The grey-filled triangles in (a), (c), (f) (resp. the open squares in (a)-(f)) are the frequencies of the $E''$ modes that are not Raman active in a backscattering geometry (resp. of the infrared-active modes) also predicted by Eq.~\eqref{eq1} using the same fitting parameters. The solid and dashed lines connect the calculated frequencies and are guides to the eye.  The upper and lower horizontal dotted lines in (a)-(b), (c)-(d), and (e)-(f) correspond to the bulk frequencies  $\omega_{\rm low}^{\pm}$, $\omega_{\rm mid}^{\pm}$, and $\omega_{\rm high}^{\pm}$, respectively (see also Table~\ref{TabFIT}). The green open circles (squares) in (e) and (f) correspond to the Raman-active (infrared-active) surface modes (see also Fig.~\ref{Fig4}(b)).}
\label{Fig5}
\end{center}
\end{figure*}

\setlength{\tabcolsep}{0.4cm}
\renewcommand{\arraystretch}{2}
\begin{table} [!htb]
\begin{center}
\begin{tabular}{ccc}

\hline
\hline

~ & in-plane & out-of-plane   \\

\hline

$\alpha (10^{19}\;\rm N/m^3)$ & 105 & 159   \\

$\alpha_e (10^{19}\;\rm N/m^3)$ & 107 & 163   \\

$\beta (10^{19}\;\rm N/m^3)$ & 2.28 & 5.61   \\

$\gamma (10^{19}\;\rm N/m^3)$ & 0.585 & 1.11   \\

$\delta (10^{19}\;\rm N/m^3)$ & -4.53 & 19.8   \\

$\delta_e (10^{19}\;\rm N/m^3)$ & -4.22 & 20.4   \\

\hline

$\omega_{\rm low}^+ (\rm cm^{-1})$ & 26.8~($E_{2g}$) & 40.3~($B_{2g}$)   \\

$\omega_{\rm mid}^- (\rm cm^{-1})$ & 117.2 $(E_{2u})$ & 168.9 $(B_{1u})$   \\
$\omega_{\rm mid}^+ (\rm cm^{-1})$ & 119.9 $(E_{1g})$  & 173.6 $(A_{1g})$   \\

\hline
\hline

\end{tabular}
\end{center}  

\caption{Force constants per unit area and corresponding bulk frequencies of the low-frequency (LSM, LBM) and mid-frequency (iX and oX) modes extracted from the fit of our experimental data to the finite linear chain model (see Eq.~\eqref{eq1} and dashed lines in Fig.~\ref{Fig5}). The irreducible representations of the bulk phonon modes are indicated.}
\label{TabFIT}
\end{table}

\textit{\textbf{Discussion~}}
First, we note that the low-frequency branches of LSM and LBM can also be separately modeled using a linear chain of $N$-oscillators. Indeed, in first approximation, van der Waals interactions between adjacent layers are sufficient to accurately describe the series of interlayer modes without further consideration of the in-plane crystal structure and intralayer force constants~\cite{Tan2012,Michel2012,Zhao2013,Zhang2013,Boukhicha2013}. The LSM and LBM frequencies are then very well approximated by $\omega_{{\rm low,}k}\left(N\right)=\frac{\omega_{\rm low}^+}{\sqrt{2}}\sqrt{1-\cos{\left(\frac{\left(k-1\right)\pi}{N}\right)}}$, with $k\in\llbracket 2, N\rrbracket $ ($k=1$ corresponds to the acoustic mode at $\omega_{\rm low}^-=0$). The observed Raman-active modes correspond to branches  with (i) $k=N,\;N-2$ for the LSM Sa and Sb and (ii) $k=2,\;4,\;6,\;8,\;10$ for the LBM Ba to Be (see Fig.~\ref{Fig2}~(b),(c)). Using the complete model, the bulk frequencies $\omega_{\rm low}^+\approx2\sqrt{\frac{\beta+2\gamma}{\mu}}$, with $\mu$ the MoTe$_{2}$ mass per unit area, then allow to determine effective interlayer force constants per unit area of $\beta+2\gamma=3.5\times10^{19}\rm~ N/m^3$ and $\beta+2\gamma=7.8\times10^{19}\rm~  N/m^3$ for the LSM and LBM, respectively (see Table~\ref{TabFIT}). These values are close to those derived from the LBM and LSM in MoS$_2$~\cite{Zhao2013,Zhang2013,Boukhicha2013} and WSe$_2$~\cite{Zhao2013}. We note that surface effects only affect intralayer force constants and have therefore a negligible influence on the rigid layer modes.

Second, a force constant model restricted to the first nearest neighbor interactions (\textit{i.e.,} $\gamma=\delta=0$) suffices to fit the frequencies of the low- and mid-frequency modes, but would then fail to predict the frequencies of the oMX and iMX modes. Indeed, Table~\ref{TabFIT} reveals that the second nearest neighbor interlayer ($\gamma$) and intralayer ($\delta$) force constants are of the same order of magnitude and larger than the nearest neighbor interlayer force constant $\beta$, respectively. Using the complete model, the bulk Davydov splittings of the iX and oX modes can conveniently be expressed as $\omega_{\rm mid}^+-\omega_{\rm mid}^-\approx\sqrt{\frac{\alpha}{\mu_{\rm X}}}\frac{\beta}{\alpha}\left(1-\frac{\gamma+2\delta}{2\alpha}\right) $, with $\mu_{\rm X}$ the chalcogen (Te) mass per unit area, leading to the values of $2.7~\rm cm^{-1}$ and $4.7~\rm cm^{-1}$ for the iX and oX modes, respectively (see Supporting Information).

According to our calculations and to group theory~\cite{Ribeiro2014,Staiger2015,Scheuschner2015} (see Fig.~\ref{Fig1},\ref{Fig4} and Table~\ref{TabRep}), the mid-frequency Raman-active modes that are observable in our backscattering experiments correspond (i) to the second-lowest, fourth-lowest,\dots frequency mode for the iX phonons and (ii) to the highest, third-highest\dots frequency mode for the oX phonons. These distinct symmetry properties result in a set of \textit{softening} and \textit{stiffening} branches in the experimentally measured fan diagrams in Fig.~\ref{Fig5}(c) and \ref{Fig5}(d), respectively. A remarkable validation of this symmetry analysis is that the highest frequency iX mode that can be observed has the highest frequency ($E_g$ symmetry) for even $N$ and the second highest frequency ($E'$ symmetry) for odd $N$. As a result, the iX-mode frequency  for $N=3$ is lower than for $N=2$, and the two observed iX-mode subfeatures for $N=4$ are slightly upshifted relative to their counterparts recorded for $N=5$ (see Figs.~\ref{Fig3} and \ref{Fig5}). We also note that our model predicts a somewhat lower frequency for the oX ($A'_{1}$) mode in monolayer MoTe$_2$ (see Fig.~\ref{Fig5}(e)). This discrepancy is due to a stronger bond length contraction in the limit of a monolayer~\cite{Luo2013}, as compared to the bilayer or to the outer layers in $N\geq2$-layer samples.

Third, we focus on the high-frequency iMX and oMX modes. Our model predicts  very small \textit{anomalous} Davydov splittings (below $0.2~\rm cm^{-1}$) for the iMX and oMX modes (consistent with previous studies of the iMX mode in bulk MX$_2$~\cite{Wieting1980,Ghosh1983,Molina2011,Luo2013}) and a critical influence of surface effects. Indeed we find that among the $N$ iMX and oMX normal displacements, two modes predominantly involve atomic displacement of the first and/or $N^{\rm th}$ layer and can be regarded as surface modes. Such surface modes are not predicted for the iX and oX displacements (see Fig.~\ref{Fig4} and the Supporting Information). As a result, the surface terms $\alpha_e$ and $\delta_e$, will  only bring a small correction to the fan diagrams associated with the mid-frequency modes, but will shape the manifold of high-frequency phonon modes. For $N\geq3$, our model predicts a fan diagram for $N-2$ quasi-degenerate iMX and oMX modes, where atomic displacement chiefly occurs in the inner layers and two surface modes (see Fig.~\ref{Fig4}(b) and Fig.~\ref{Fig5}(c),(f). The strength of the surface effects will set the frequencies of the surface modes relative to that of the other modes. 

A tentative explanation for the pronounced downshift observed for the iMX mode is then that the surface mode dominates for small $N$ and that the modes arising from the inner layers provide most of the Raman intensity as $N$ increases. This scenario is consistent with calculations based on a bond polarizability model~\cite{Umari2001}, which was also recently used for layered BN and BN nanotubes~\cite{Wirtz2005} and which is explained in the Supporting Information. The model predicts that the surface mode will dominate in the iMX peak for $N<5$. Let us also note that the iMX-mode feature in $N$-layer MoTe$_2$ is appreciably broader (with a full width at half maximum (FWHM) decreasing from 3.5~cm$^{-1}$ for $N=1$ down to 2.6~cm$^{-1}$ in the bulk) than the iX, oX and oMX peaks, whose FWHM are approximately 1~cm$^{-1}$. This broadening prevents us from unveiling any splitting of the iMX-mode feature in $N$-layer MoTe$_2$. This splitting, according to our force-constant fit is very small anyway (see Fig.~\ref{Fig5} (e)) and also our {\it ab-initio} calculations for phonons of bulk MoTe$_2$ yield a minimal anomalous Davydov splitting of 0.1 cm$^{-1}$ (see Supporting Information).

For the oMX mode, we also find that the surface modes have a slightly higher frequency than the modes that are localized on the inner layer. Supported by the results from our bond-polarizability model (see Supporting Information), we assign the dominant high-frequency oMX subfeature to the surface mode, while the faint lower-frequency shoulder is assigned to the inner modes. We note that the low-frequency shoulder is not observed at all at $E_{\rm L}=2.33~\rm eV$  (see Supporting Information) and may thus arise from a resonance effect~\cite{Luo2013b,Scheuschner2015,Lee2015}.  We also point out that {\it ab-initio} calculations (see Supporting Information) yield a sizable Davydov splitting of 5.8 cm$^{-1}$ between the bulk $A_{2u}$ and $B_{2g}$ frequencies (in the absence of the Lydanne-Sachs-Teller interaction). In order to reproduce this splitting, an additional force-constant between Mo atoms of neighboring layers would need to be introduced (which would not modify the splitting between the oX modes $B_{1u}$ and $A_{1g}$ modes because those modes do not involve motion of the Mo atoms). Since the resulting fan-diagram is not visible in our experimental spectra of the oMX mode -- due to almost vanishing intensity -- we did not include this additional force constant here.

Finally, we comment on the possible influence of resonance effects on our measurements. Resonance effects should not impact the phonon frequencies but may strongly alter the integrated intensity of one given Raman feature and the repartition of the spectral weight within a given subfeature. For the iX mode, our bond polarizability model predicts spectral weights that are in-line with group theory predictions and our experimental results (see Supporting Information). However, the same model predicts that the highest frequency oX mode should have a much larger oscillator strength than its lower frequency counterparts. This prediction is clearly in contradiction with our observations at $E_L=1.96~\rm eV$ (see Fig~\ref{Fig3}b), where the oX-mode subfeatures have comparable integrated intensities, but is consistent with the absence of splittings seen in numerous studies on MoS$_2$ using visible photon energies~\cite{Lee2010,Li2012,Luo2013,Lee2015,Zhang2015b}. Noteworthy, a pronounced splitting of the oX feature was also observed  at $E_L=2.33~\rm eV$ (see Supporting Information). However, in this case the high-frequency peak largely dominates the oX-mode feature for $N\geq6$. These intriguing observations, together with the observation of the iX and oMX bulk modes, and a particularly intense oX-mode feature in monoloayer MoTe$_2$ at $E_L=1.96~\rm eV$ (see Supporting Information and Ref.~\cite{Ruppert2014}) provide a strong impetus for a quantitative analysis of resonant, symmetry-dependent exciton-phonon coupling~\cite{Chakraborty2013,Luo2013b,Carvalho2015,Scheuschner2015,Lee2015} in $N$-layer MX$_2$.
 
\textit{\textbf{Conclusion~}}
Using micro-Raman spectroscopy, we have reported a unified description of the optical phonon modes in a $N$-layer $2H$-transition metal dichalcogenide crystal (here, MoTe$_2$), between the  bulk (three-dimensional) and monolayer (quasi-two-dimensional) limits. The manifolds of low-frequency interlayer shear and breathing modes, and of the mid-frequency modes involving out-of-phase intralayer motion of the chalcogen atoms are well understood using classical theories of coupled oscillators. In contrast, the behavior of the high-frequency modes that involve out-of phase motion of the metal and chalcogen planes is largely influenced by surface effects. We have introduced a global fitting procedure based on a linear chain model to derive the force constants up to the second nearest neighbor and to assess the strength of the surface effects. This model allows us to deduce the frequency of optical phonons that are silent in bulk crystals, namely the low-frequency interlayer breathing mode (LBM) with $B_{2g}$ symmetry, and the mid-frequency in-plane (iX) and out-of-plane (oX) modes with $E_{2u}$ and $B_{1u}$ symmetry, respectively (see Table~\ref{TabFIT}). Finally, our work may motivate related studies of Davydov splitting, force constants and surface effects in other layered crystals, such as black phosphorus~\cite{Favron2015}, bismuth selenide and bismuth telluride~\cite{Zhao2014}.

\textit{\textbf{Methods~}}
$N$-layer MoTe$_2$ crystals were prepared by mechanical exfoliation of commercially available bulk crystals (2D semiconductors) onto Si wafers covered with a 90-nm or 285-nm-thick SiO$_2$ epilayer. The number of layers was first estimated from optical contrast and atomic force microscopy measurements, and further characterized using a home-built micro-Raman setup. Micro-Raman scattering studies were carried out in ambient conditions, in a backscattering geometry using using a monochromator equipped with a 2400 grooves/mm holographic grating, coupled to a two-dimensional liquid nitrogen cooled charge-coupled device (CCD) array.  Two laser photon energies ($E_{\rm L}=2.33~\rm eV$ and $E_{\rm L}=1.96~\rm eV$ were employed. Spectral resolutions of $0.6~\rm cm^{-1}$ and $0.4~\rm cm^{-1}$ were obtained at $E_{\rm L}=2.33~\rm eV$, $E_{\rm L}=1.96~\rm eV$, respectively. A laser intensity below $50~ \rm kW/cm^2$ was used in order to avoid photoinduced damage of our samples. In order to attain the low-frequency range (for measurements at $E_{\rm L}=2.33~\rm eV$), a combination of one narrow bandpass filter and two narrow notch filters (Optigrate) was used. After optimization, Raman features at frequencies as low as 4.5~cm$^{-1}$ could be measured. Polarized Raman studies are performed using a Glan-Thomson analyzer placed before the entrance of our spectrometer. An achromatic half-wave plate was placed after the analyzer and adequately rotated such that the Raman backscattered beam enters the spectrometer with a fixed polarization. Thereafter, we will be considering parallel (XX) and perpendicular (XY) polarizations of the incoming and Raman-scattered photons. The observation of residual features from the out of plane modes (LBM, oX, oMX) in the XY configuration~(see Figs.~\ref{Fig1} and.~\ref{Fig2} is due to depolarization effects induced by our high numerical aperture objective (100~x, $\rm NA=0.9$). Finally, the measured Raman features are fit to Voigt profiles, taking into account our spectral resolution. 

\textit{\textbf{Acknowledgments~}}
We are grateful to C. Faugeras and K. Nogajewski for discussions. We acknowledge financial support from the Agence Nationale de la Recherche (under grant QuanDoGra 12 JS10-001-01), from the CNRS and from Universit\'e de Strasbourg, as well as support by the National Research Fund, Luxembourg (projects INTER/ANR/13/20/NANOTMD and C14/MS/7731521/FAST-2DMAT).

%\bibliography{TMD_Davydov_Splitting_resub}

\begin{thebibliography}{51}%
\makeatletter
\providecommand \@ifxundefined [1]{%
 \@ifx{#1\undefined}
}%
\providecommand \@ifnum [1]{%
 \ifnum #1\expandafter \@firstoftwo
 \else \expandafter \@secondoftwo
 \fi
}%
\providecommand \@ifx [1]{%
 \ifx #1\expandafter \@firstoftwo
 \else \expandafter \@secondoftwo
 \fi
}%
\providecommand \natexlab [1]{#1}%
\providecommand \enquote  [1]{``#1''}%
\providecommand \bibnamefont  [1]{#1}%
\providecommand \bibfnamefont [1]{#1}%
\providecommand \citenamefont [1]{#1}%
\providecommand \href@noop [0]{\@secondoftwo}%
\providecommand \href [0]{\begingroup \@sanitize@url \@href}%
\providecommand \@href[1]{\@@startlink{#1}\@@href}%
\providecommand \@@href[1]{\endgroup#1\@@endlink}%
\providecommand \@sanitize@url [0]{\catcode `\\12\catcode `\$12\catcode
  `\&12\catcode `\#12\catcode `\^12\catcode `\_12\catcode `\%12\relax}%
\providecommand \@@startlink[1]{}%
\providecommand \@@endlink[0]{}%
\providecommand \url  [0]{\begingroup\@sanitize@url \@url }%
\providecommand \@url [1]{\endgroup\@href {#1}{\urlprefix }}%
\providecommand \urlprefix  [0]{URL }%
\providecommand \Eprint [0]{\href }%
\providecommand \doibase [0]{http://dx.doi.org/}%
\providecommand \selectlanguage [0]{\@gobble}%
\providecommand \bibinfo  [0]{\@secondoftwo}%
\providecommand \bibfield  [0]{\@secondoftwo}%
\providecommand \translation [1]{[#1]}%
\providecommand \BibitemOpen [0]{}%
\providecommand \bibitemStop [0]{}%
\providecommand \bibitemNoStop [0]{.\EOS\space}%
\providecommand \EOS [0]{\spacefactor3000\relax}%
\providecommand \BibitemShut  [1]{\csname bibitem#1\endcsname}%
\let\auto@bib@innerbib\@empty
%</preamble>
\bibitem [{\citenamefont {Geim}\ and\ \citenamefont
  {Grigorieva}(2013)}]{Geim2013}%
  \BibitemOpen
  \bibfield  {author} {\bibinfo {author} {\bibfnamefont {A.~K.}\ \bibnamefont
  {Geim}}\ and\ \bibinfo {author} {\bibfnamefont {I.~V.}\ \bibnamefont
  {Grigorieva}},\ }\href {\doibase 10.1038/nature12385} {\bibfield  {journal}
  {\bibinfo  {journal} {Nature}\ }\textbf {\bibinfo {volume} {499}},\ \bibinfo
  {pages} {419} (\bibinfo {year} {2013})}\BibitemShut {NoStop}%
\bibitem [{\citenamefont {Wilson}\ and\ \citenamefont
  {Yoffe}(1969)}]{Wilson1969}%
  \BibitemOpen
  \bibfield  {author} {\bibinfo {author} {\bibfnamefont {J.}~\bibnamefont
  {Wilson}}\ and\ \bibinfo {author} {\bibfnamefont {A.}~\bibnamefont {Yoffe}},\
  }\href {\doibase 10.1080/00018736900101307} {\bibfield  {journal} {\bibinfo
  {journal} {Advances in Physics}\ }\textbf {\bibinfo {volume} {18}},\ \bibinfo
  {pages} {193} (\bibinfo {year} {1969})}\BibitemShut {NoStop}%
\bibitem [{\citenamefont {Mak}\ \emph {et~al.}(2010)\citenamefont {Mak},
  \citenamefont {Lee}, \citenamefont {Hone}, \citenamefont {Shan},\ and\
  \citenamefont {Heinz}}]{Mak2010}%
  \BibitemOpen
  \bibfield  {author} {\bibinfo {author} {\bibfnamefont {K.~F.}\ \bibnamefont
  {Mak}}, \bibinfo {author} {\bibfnamefont {C.}~\bibnamefont {Lee}}, \bibinfo
  {author} {\bibfnamefont {J.}~\bibnamefont {Hone}}, \bibinfo {author}
  {\bibfnamefont {J.}~\bibnamefont {Shan}}, \ and\ \bibinfo {author}
  {\bibfnamefont {T.~F.}\ \bibnamefont {Heinz}},\ }\href {\doibase
  10.1103/PhysRevLett.105.136805} {\bibfield  {journal} {\bibinfo  {journal}
  {Physical Review Letters}\ }\textbf {\bibinfo {volume} {105}},\ \bibinfo
  {pages} {136805} (\bibinfo {year} {2010})}\BibitemShut {NoStop}%
\bibitem [{\citenamefont {Splendiani}\ \emph {et~al.}(2010)\citenamefont
  {Splendiani}, \citenamefont {Sun}, \citenamefont {Zhang}, \citenamefont {Li},
  \citenamefont {Kim}, \citenamefont {Chim}, \citenamefont {Galli},\ and\
  \citenamefont {Wang}}]{Splendiani2010}%
  \BibitemOpen
  \bibfield  {author} {\bibinfo {author} {\bibfnamefont {A.}~\bibnamefont
  {Splendiani}}, \bibinfo {author} {\bibfnamefont {L.}~\bibnamefont {Sun}},
  \bibinfo {author} {\bibfnamefont {Y.}~\bibnamefont {Zhang}}, \bibinfo
  {author} {\bibfnamefont {T.}~\bibnamefont {Li}}, \bibinfo {author}
  {\bibfnamefont {J.}~\bibnamefont {Kim}}, \bibinfo {author} {\bibfnamefont
  {C.-Y.}\ \bibnamefont {Chim}}, \bibinfo {author} {\bibfnamefont
  {G.}~\bibnamefont {Galli}}, \ and\ \bibinfo {author} {\bibfnamefont
  {F.}~\bibnamefont {Wang}},\ }\href {\doibase 10.1021/nl903868w} {\bibfield
  {journal} {\bibinfo  {journal} {Nano Letters}\ }\textbf {\bibinfo {volume}
  {10}},\ \bibinfo {pages} {1271} (\bibinfo {year} {2010})}\BibitemShut
  {NoStop}%
\bibitem [{\citenamefont {Xu}\ \emph {et~al.}(2014)\citenamefont {Xu},
  \citenamefont {Yao}, \citenamefont {Xiao},\ and\ \citenamefont
  {Heinz}}]{Xu2014}%
  \BibitemOpen
  \bibfield  {author} {\bibinfo {author} {\bibfnamefont {X.}~\bibnamefont
  {Xu}}, \bibinfo {author} {\bibfnamefont {W.}~\bibnamefont {Yao}}, \bibinfo
  {author} {\bibfnamefont {D.}~\bibnamefont {Xiao}}, \ and\ \bibinfo {author}
  {\bibfnamefont {T.~F.}\ \bibnamefont {Heinz}},\ }\href {\doibase
  10.1038/nphys2942} {\bibfield  {journal} {\bibinfo  {journal} {Nature
  Physics}\ }\textbf {\bibinfo {volume} {10}},\ \bibinfo {pages} {343}
  (\bibinfo {year} {2014})}\BibitemShut {NoStop}%
\bibitem [{\citenamefont {Peng}\ \emph {et~al.}(2015)\citenamefont {Peng},
  \citenamefont {Ang},\ and\ \citenamefont {Loh}}]{Peng2015}%
  \BibitemOpen
  \bibfield  {author} {\bibinfo {author} {\bibfnamefont {B.}~\bibnamefont
  {Peng}}, \bibinfo {author} {\bibfnamefont {P.~K.}\ \bibnamefont {Ang}}, \
  and\ \bibinfo {author} {\bibfnamefont {K.~P.}\ \bibnamefont {Loh}},\ }\href
  {\doibase http://dx.doi.org/10.1016/j.nantod.2015.01.007} {\bibfield
  {journal} {\bibinfo  {journal} {Nano Today}\ }\textbf {\bibinfo {volume}
  {10}},\ \bibinfo {pages} {128 } (\bibinfo {year} {2015})}\BibitemShut
  {NoStop}%
\bibitem [{\citenamefont {Wieting}\ and\ \citenamefont
  {Verble}(1971)}]{Wieting1971}%
  \BibitemOpen
  \bibfield  {author} {\bibinfo {author} {\bibfnamefont {T.~J.}\ \bibnamefont
  {Wieting}}\ and\ \bibinfo {author} {\bibfnamefont {J.~L.}\ \bibnamefont
  {Verble}},\ }\href {\doibase 10.1103/PhysRevB.3.4286} {\bibfield  {journal}
  {\bibinfo  {journal} {Phys. Rev. B}\ }\textbf {\bibinfo {volume} {3}},\
  \bibinfo {pages} {4286} (\bibinfo {year} {1971})}\BibitemShut {NoStop}%
\bibitem [{\citenamefont {Molina-S\'{a}nchez}\ and\ \citenamefont
  {Wirtz}(2011)}]{Molina2011}%
  \BibitemOpen
  \bibfield  {author} {\bibinfo {author} {\bibfnamefont {A.}~\bibnamefont
  {Molina-S\'{a}nchez}}\ and\ \bibinfo {author} {\bibfnamefont
  {L.}~\bibnamefont {Wirtz}},\ }\href {\doibase 10.1103/PhysRevB.84.155413}
  {\bibfield  {journal} {\bibinfo  {journal} {Physical Review B}\ }\textbf
  {\bibinfo {volume} {84}},\ \bibinfo {pages} {155413} (\bibinfo {year}
  {2011})}\BibitemShut {NoStop}%
\bibitem [{\citenamefont {Luo}\ \emph {et~al.}(2013{\natexlab{a}})\citenamefont
  {Luo}, \citenamefont {Zhao}, \citenamefont {Zhang}, \citenamefont {Xiong},\
  and\ \citenamefont {Quek}}]{Luo2013}%
  \BibitemOpen
  \bibfield  {author} {\bibinfo {author} {\bibfnamefont {X.}~\bibnamefont
  {Luo}}, \bibinfo {author} {\bibfnamefont {Y.}~\bibnamefont {Zhao}}, \bibinfo
  {author} {\bibfnamefont {J.}~\bibnamefont {Zhang}}, \bibinfo {author}
  {\bibfnamefont {Q.}~\bibnamefont {Xiong}}, \ and\ \bibinfo {author}
  {\bibfnamefont {S.~Y.}\ \bibnamefont {Quek}},\ }\href {\doibase
  10.1103/PhysRevB.88.075320} {\bibfield  {journal} {\bibinfo  {journal} {Phys.
  Rev. B}\ }\textbf {\bibinfo {volume} {88}},\ \bibinfo {pages} {075320}
  (\bibinfo {year} {2013}{\natexlab{a}})}\BibitemShut {NoStop}%
\bibitem [{\citenamefont {Luo}\ \emph {et~al.}(2013{\natexlab{b}})\citenamefont
  {Luo}, \citenamefont {Zhao}, \citenamefont {Zhang}, \citenamefont {Toh},
  \citenamefont {Kloc}, \citenamefont {Xiong},\ and\ \citenamefont
  {Quek}}]{Luo2013b}%
  \BibitemOpen
  \bibfield  {author} {\bibinfo {author} {\bibfnamefont {X.}~\bibnamefont
  {Luo}}, \bibinfo {author} {\bibfnamefont {Y.}~\bibnamefont {Zhao}}, \bibinfo
  {author} {\bibfnamefont {J.}~\bibnamefont {Zhang}}, \bibinfo {author}
  {\bibfnamefont {M.}~\bibnamefont {Toh}}, \bibinfo {author} {\bibfnamefont
  {C.}~\bibnamefont {Kloc}}, \bibinfo {author} {\bibfnamefont {Q.}~\bibnamefont
  {Xiong}}, \ and\ \bibinfo {author} {\bibfnamefont {S.~Y.}\ \bibnamefont
  {Quek}},\ }\href {\doibase 10.1103/PhysRevB.88.195313} {\bibfield  {journal}
  {\bibinfo  {journal} {Phys. Rev. B}\ }\textbf {\bibinfo {volume} {88}},\
  \bibinfo {pages} {195313} (\bibinfo {year} {2013}{\natexlab{b}})}\BibitemShut
  {NoStop}%
\bibitem [{\citenamefont {Terrones}\ \emph {et~al.}(2014)\citenamefont
  {Terrones}, \citenamefont {Corro}, \citenamefont {Feng}, \citenamefont
  {Poumirol}, \citenamefont {Rhodes}, \citenamefont {Smirnov}, \citenamefont
  {Pradhan}, \citenamefont {Lin}, \citenamefont {Nguyen}, \citenamefont
  {Elías}, \citenamefont {Mallouk}, \citenamefont {Balicas}, \citenamefont
  {Pimenta},\ and\ \citenamefont {Terrones}}]{Terrones2014}%
  \BibitemOpen
  \bibfield  {author} {\bibinfo {author} {\bibfnamefont {H.}~\bibnamefont
  {Terrones}}, \bibinfo {author} {\bibfnamefont {E.~D.}\ \bibnamefont {Corro}},
  \bibinfo {author} {\bibfnamefont {S.}~\bibnamefont {Feng}}, \bibinfo {author}
  {\bibfnamefont {J.~M.}\ \bibnamefont {Poumirol}}, \bibinfo {author}
  {\bibfnamefont {D.}~\bibnamefont {Rhodes}}, \bibinfo {author} {\bibfnamefont
  {D.}~\bibnamefont {Smirnov}}, \bibinfo {author} {\bibfnamefont {N.~R.}\
  \bibnamefont {Pradhan}}, \bibinfo {author} {\bibfnamefont {Z.}~\bibnamefont
  {Lin}}, \bibinfo {author} {\bibfnamefont {M.~a.~T.}\ \bibnamefont {Nguyen}},
  \bibinfo {author} {\bibfnamefont {A.~L.}\ \bibnamefont {Elías}}, \bibinfo
  {author} {\bibfnamefont {T.~E.}\ \bibnamefont {Mallouk}}, \bibinfo {author}
  {\bibfnamefont {L.}~\bibnamefont {Balicas}}, \bibinfo {author} {\bibfnamefont
  {M.~A.}\ \bibnamefont {Pimenta}}, \ and\ \bibinfo {author} {\bibfnamefont
  {M.}~\bibnamefont {Terrones}},\ }\href {\doibase 10.1038/srep04215}
  {\bibfield  {journal} {\bibinfo  {journal} {Scientific Reports}\ }\textbf
  {\bibinfo {volume} {4}} (\bibinfo {year} {2014}),\
  10.1038/srep04215}\BibitemShut {NoStop}%
\bibitem [{\citenamefont {Ribeiro-Soares}\ \emph {et~al.}(2014)\citenamefont
  {Ribeiro-Soares}, \citenamefont {Almeida}, \citenamefont {Barros},
  \citenamefont {Araujo}, \citenamefont {Dresselhaus}, \citenamefont
  {Cançado},\ and\ \citenamefont {Jorio}}]{Ribeiro2014}%
  \BibitemOpen
  \bibfield  {author} {\bibinfo {author} {\bibfnamefont {J.}~\bibnamefont
  {Ribeiro-Soares}}, \bibinfo {author} {\bibfnamefont {R.~M.}\ \bibnamefont
  {Almeida}}, \bibinfo {author} {\bibfnamefont {E.~B.}\ \bibnamefont {Barros}},
  \bibinfo {author} {\bibfnamefont {P.~T.}\ \bibnamefont {Araujo}}, \bibinfo
  {author} {\bibfnamefont {M.~S.}\ \bibnamefont {Dresselhaus}}, \bibinfo
  {author} {\bibfnamefont {L.~G.}\ \bibnamefont {Cançado}}, \ and\ \bibinfo
  {author} {\bibfnamefont {A.}~\bibnamefont {Jorio}},\ }\href {\doibase
  10.1103/PhysRevB.90.115438} {\bibfield  {journal} {\bibinfo  {journal}
  {Physical Review B}\ }\textbf {\bibinfo {volume} {90}},\ \bibinfo {pages}
  {115438} (\bibinfo {year} {2014})}\BibitemShut {NoStop}%
\bibitem [{\citenamefont {Yamamoto}\ \emph {et~al.}(2014)\citenamefont
  {Yamamoto}, \citenamefont {Wang}, \citenamefont {Ni}, \citenamefont {Lin},
  \citenamefont {Li}, \citenamefont {Aikawa}, \citenamefont {Jian},
  \citenamefont {Ueno}, \citenamefont {Wakabayashi},\ and\ \citenamefont
  {Tsukagoshi}}]{Yamamoto2014}%
  \BibitemOpen
  \bibfield  {author} {\bibinfo {author} {\bibfnamefont {M.}~\bibnamefont
  {Yamamoto}}, \bibinfo {author} {\bibfnamefont {S.~T.}\ \bibnamefont {Wang}},
  \bibinfo {author} {\bibfnamefont {M.}~\bibnamefont {Ni}}, \bibinfo {author}
  {\bibfnamefont {Y.-F.}\ \bibnamefont {Lin}}, \bibinfo {author} {\bibfnamefont
  {S.-L.}\ \bibnamefont {Li}}, \bibinfo {author} {\bibfnamefont
  {S.}~\bibnamefont {Aikawa}}, \bibinfo {author} {\bibfnamefont {W.-B.}\
  \bibnamefont {Jian}}, \bibinfo {author} {\bibfnamefont {K.}~\bibnamefont
  {Ueno}}, \bibinfo {author} {\bibfnamefont {K.}~\bibnamefont {Wakabayashi}}, \
  and\ \bibinfo {author} {\bibfnamefont {K.}~\bibnamefont {Tsukagoshi}},\
  }\href {\doibase 10.1021/nn5007607} {\bibfield  {journal} {\bibinfo
  {journal} {ACS Nano}\ }\textbf {\bibinfo {volume} {8}},\ \bibinfo {pages}
  {3895} (\bibinfo {year} {2014})}\BibitemShut {NoStop}%
\bibitem [{\citenamefont {Scheuschner}\ \emph {et~al.}(2015)\citenamefont
  {Scheuschner}, \citenamefont {Gillen}, \citenamefont {Staiger},\ and\
  \citenamefont {Maultzsch}}]{Scheuschner2015}%
  \BibitemOpen
  \bibfield  {author} {\bibinfo {author} {\bibfnamefont {N.}~\bibnamefont
  {Scheuschner}}, \bibinfo {author} {\bibfnamefont {R.}~\bibnamefont {Gillen}},
  \bibinfo {author} {\bibfnamefont {M.}~\bibnamefont {Staiger}}, \ and\
  \bibinfo {author} {\bibfnamefont {J.}~\bibnamefont {Maultzsch}},\ }\href
  {\doibase 10.1103/PhysRevB.91.235409} {\bibfield  {journal} {\bibinfo
  {journal} {Phys. Rev. B}\ }\textbf {\bibinfo {volume} {91}},\ \bibinfo
  {pages} {235409} (\bibinfo {year} {2015})}\BibitemShut {NoStop}%
\bibitem [{\citenamefont {Zhang}\ \emph {et~al.}(2015)\citenamefont {Zhang},
  \citenamefont {Qiao}, \citenamefont {Shi}, \citenamefont {Wu}, \citenamefont
  {Jiang},\ and\ \citenamefont {Tan}}]{Zhang2015b}%
  \BibitemOpen
  \bibfield  {author} {\bibinfo {author} {\bibfnamefont {X.}~\bibnamefont
  {Zhang}}, \bibinfo {author} {\bibfnamefont {X.-F.}\ \bibnamefont {Qiao}},
  \bibinfo {author} {\bibfnamefont {W.}~\bibnamefont {Shi}}, \bibinfo {author}
  {\bibfnamefont {J.-B.}\ \bibnamefont {Wu}}, \bibinfo {author} {\bibfnamefont
  {D.-S.}\ \bibnamefont {Jiang}}, \ and\ \bibinfo {author} {\bibfnamefont
  {P.-H.}\ \bibnamefont {Tan}},\ }\href {\doibase 10.1039/C4CS00282B}
  {\bibfield  {journal} {\bibinfo  {journal} {Chem. Soc. Rev.}\ }\textbf
  {\bibinfo {volume} {44}},\ \bibinfo {pages} {2757} (\bibinfo {year}
  {2015})}\BibitemShut {NoStop}%
\bibitem [{\citenamefont {Zhao}\ \emph {et~al.}(2013)\citenamefont {Zhao},
  \citenamefont {Luo}, \citenamefont {Li}, \citenamefont {Zhang}, \citenamefont
  {Araujo}, \citenamefont {Gan}, \citenamefont {Wu}, \citenamefont {Zhang},
  \citenamefont {Quek}, \citenamefont {Dresselhaus},\ and\ \citenamefont
  {Xiong}}]{Zhao2013}%
  \BibitemOpen
  \bibfield  {author} {\bibinfo {author} {\bibfnamefont {Y.}~\bibnamefont
  {Zhao}}, \bibinfo {author} {\bibfnamefont {X.}~\bibnamefont {Luo}}, \bibinfo
  {author} {\bibfnamefont {H.}~\bibnamefont {Li}}, \bibinfo {author}
  {\bibfnamefont {J.}~\bibnamefont {Zhang}}, \bibinfo {author} {\bibfnamefont
  {P.~T.}\ \bibnamefont {Araujo}}, \bibinfo {author} {\bibfnamefont {C.~K.}\
  \bibnamefont {Gan}}, \bibinfo {author} {\bibfnamefont {J.}~\bibnamefont
  {Wu}}, \bibinfo {author} {\bibfnamefont {H.}~\bibnamefont {Zhang}}, \bibinfo
  {author} {\bibfnamefont {S.~Y.}\ \bibnamefont {Quek}}, \bibinfo {author}
  {\bibfnamefont {M.~S.}\ \bibnamefont {Dresselhaus}}, \ and\ \bibinfo {author}
  {\bibfnamefont {Q.}~\bibnamefont {Xiong}},\ }\href {\doibase
  10.1021/nl304169w} {\bibfield  {journal} {\bibinfo  {journal} {Nano Letters}\
  }\textbf {\bibinfo {volume} {13}},\ \bibinfo {pages} {1007} (\bibinfo {year}
  {2013})}\BibitemShut {NoStop}%
\bibitem [{\citenamefont {Davydov}(1971)}]{Davydov1971}%
  \BibitemOpen
  \bibfield  {author} {\bibinfo {author} {\bibfnamefont {A.~S.}\ \bibnamefont
  {Davydov}},\ }in\ \href
  {http://link.springer.com/chapter/10.1007/978-1-4899-5169-4_2} {\emph
  {\bibinfo {booktitle} {Theory of {Molecular} {Excitons}}}}\ (\bibinfo
  {publisher} {Springer US},\ \bibinfo {year} {1971})\BibitemShut {NoStop}%
\bibitem [{\citenamefont {Khelladi}(1975)}]{Khelladi1975}%
  \BibitemOpen
  \bibfield  {author} {\bibinfo {author} {\bibfnamefont {F.~Z.}\ \bibnamefont
  {Khelladi}},\ }\href {\doibase 10.1016/0009-2614(75)85547-3} {\bibfield
  {journal} {\bibinfo  {journal} {Chemical Physics Letters}\ }\textbf {\bibinfo
  {volume} {34}},\ \bibinfo {pages} {490} (\bibinfo {year} {1975})}\BibitemShut
  {NoStop}%
\bibitem [{\citenamefont {Aroca}\ \emph {et~al.}(1987)\citenamefont {Aroca},
  \citenamefont {Jennings}, \citenamefont {Loutfy},\ and\ \citenamefont
  {Hor}}]{Aroca1987}%
  \BibitemOpen
  \bibfield  {author} {\bibinfo {author} {\bibfnamefont {R.}~\bibnamefont
  {Aroca}}, \bibinfo {author} {\bibfnamefont {C.}~\bibnamefont {Jennings}},
  \bibinfo {author} {\bibfnamefont {R.~O.}\ \bibnamefont {Loutfy}}, \ and\
  \bibinfo {author} {\bibfnamefont {A.-M.}\ \bibnamefont {Hor}},\ }\href
  {\doibase 10.1016/0584-8539(87)80212-X} {\bibfield  {journal} {\bibinfo
  {journal} {Spectrochimica Acta Part A: Molecular Spectroscopy}\ }\textbf
  {\bibinfo {volume} {43}},\ \bibinfo {pages} {725} (\bibinfo {year}
  {1987})}\BibitemShut {NoStop}%
\bibitem [{\citenamefont {Wieting}\ \emph {et~al.}(1980)\citenamefont
  {Wieting}, \citenamefont {Grisel},\ and\ \citenamefont
  {L\'evy}}]{Wieting1980}%
  \BibitemOpen
  \bibfield  {author} {\bibinfo {author} {\bibfnamefont {T.~J.}\ \bibnamefont
  {Wieting}}, \bibinfo {author} {\bibfnamefont {A.}~\bibnamefont {Grisel}}, \
  and\ \bibinfo {author} {\bibfnamefont {F.}~\bibnamefont {L\'evy}},\ }\href
  {\doibase 10.1016/0378-4363(80)90256-9} {\bibfield  {journal} {\bibinfo
  {journal} {Physica B+C}\ }\textbf {\bibinfo {volume} {99}},\ \bibinfo {pages}
  {337} (\bibinfo {year} {1980})}\BibitemShut {NoStop}%
\bibitem [{\citenamefont {Ghosh}\ and\ \citenamefont
  {Maiti}(1983)}]{Ghosh1983}%
  \BibitemOpen
  \bibfield  {author} {\bibinfo {author} {\bibfnamefont {P.~N.}\ \bibnamefont
  {Ghosh}}\ and\ \bibinfo {author} {\bibfnamefont {C.~R.}\ \bibnamefont
  {Maiti}},\ }\href {\doibase 10.1103/PhysRevB.28.2237} {\bibfield  {journal}
  {\bibinfo  {journal} {Phys. Rev. B}\ }\textbf {\bibinfo {volume} {28}},\
  \bibinfo {pages} {2237} (\bibinfo {year} {1983})}\BibitemShut {NoStop}%
\bibitem [{\citenamefont {Plechinger}\ \emph {et~al.}(2012)\citenamefont
  {Plechinger}, \citenamefont {Heydrich}, \citenamefont {Eroms}, \citenamefont
  {Weiss}, \citenamefont {Sch\"uller},\ and\ \citenamefont
  {Korn}}]{Plechinger2012}%
  \BibitemOpen
  \bibfield  {author} {\bibinfo {author} {\bibfnamefont {G.}~\bibnamefont
  {Plechinger}}, \bibinfo {author} {\bibfnamefont {S.}~\bibnamefont
  {Heydrich}}, \bibinfo {author} {\bibfnamefont {J.}~\bibnamefont {Eroms}},
  \bibinfo {author} {\bibfnamefont {D.}~\bibnamefont {Weiss}}, \bibinfo
  {author} {\bibfnamefont {C.}~\bibnamefont {Sch\"uller}}, \ and\ \bibinfo
  {author} {\bibfnamefont {T.}~\bibnamefont {Korn}},\ }\href {\doibase
  10.1063/1.4751266} {\bibfield  {journal} {\bibinfo  {journal} {Applied
  Physics Letters}\ }\textbf {\bibinfo {volume} {101}},\ \bibinfo {pages}
  {101906} (\bibinfo {year} {2012})}\BibitemShut {NoStop}%
\bibitem [{\citenamefont {Zeng}\ \emph {et~al.}(2012)\citenamefont {Zeng},
  \citenamefont {Zhu}, \citenamefont {Liu}, \citenamefont {Fan}, \citenamefont
  {Cui},\ and\ \citenamefont {Zhang}}]{Zeng2012}%
  \BibitemOpen
  \bibfield  {author} {\bibinfo {author} {\bibfnamefont {H.}~\bibnamefont
  {Zeng}}, \bibinfo {author} {\bibfnamefont {B.}~\bibnamefont {Zhu}}, \bibinfo
  {author} {\bibfnamefont {K.}~\bibnamefont {Liu}}, \bibinfo {author}
  {\bibfnamefont {J.}~\bibnamefont {Fan}}, \bibinfo {author} {\bibfnamefont
  {X.}~\bibnamefont {Cui}}, \ and\ \bibinfo {author} {\bibfnamefont {Q.~M.}\
  \bibnamefont {Zhang}},\ }\href {\doibase 10.1103/PhysRevB.86.241301}
  {\bibfield  {journal} {\bibinfo  {journal} {Physical Review B}\ }\textbf
  {\bibinfo {volume} {86}},\ \bibinfo {pages} {241301} (\bibinfo {year}
  {2012})}\BibitemShut {NoStop}%
\bibitem [{\citenamefont {Zhang}\ \emph {et~al.}(2013)\citenamefont {Zhang},
  \citenamefont {Han}, \citenamefont {Wu}, \citenamefont {Milana},
  \citenamefont {Lu}, \citenamefont {Li}, \citenamefont {Ferrari},\ and\
  \citenamefont {Tan}}]{Zhang2013}%
  \BibitemOpen
  \bibfield  {author} {\bibinfo {author} {\bibfnamefont {X.}~\bibnamefont
  {Zhang}}, \bibinfo {author} {\bibfnamefont {W.~P.}\ \bibnamefont {Han}},
  \bibinfo {author} {\bibfnamefont {J.~B.}\ \bibnamefont {Wu}}, \bibinfo
  {author} {\bibfnamefont {S.}~\bibnamefont {Milana}}, \bibinfo {author}
  {\bibfnamefont {Y.}~\bibnamefont {Lu}}, \bibinfo {author} {\bibfnamefont
  {Q.~Q.}\ \bibnamefont {Li}}, \bibinfo {author} {\bibfnamefont {A.~C.}\
  \bibnamefont {Ferrari}}, \ and\ \bibinfo {author} {\bibfnamefont {P.~H.}\
  \bibnamefont {Tan}},\ }\href {\doibase 10.1103/PhysRevB.87.115413} {\bibfield
   {journal} {\bibinfo  {journal} {Phys. Rev. B}\ }\textbf {\bibinfo {volume}
  {87}},\ \bibinfo {pages} {115413} (\bibinfo {year} {2013})}\BibitemShut
  {NoStop}%
\bibitem [{\citenamefont {Boukhicha}\ \emph {et~al.}(2013)\citenamefont
  {Boukhicha}, \citenamefont {Calandra}, \citenamefont {Measson}, \citenamefont
  {Lancry},\ and\ \citenamefont {Shukla}}]{Boukhicha2013}%
  \BibitemOpen
  \bibfield  {author} {\bibinfo {author} {\bibfnamefont {M.}~\bibnamefont
  {Boukhicha}}, \bibinfo {author} {\bibfnamefont {M.}~\bibnamefont {Calandra}},
  \bibinfo {author} {\bibfnamefont {M.-A.}\ \bibnamefont {Measson}}, \bibinfo
  {author} {\bibfnamefont {O.}~\bibnamefont {Lancry}}, \ and\ \bibinfo {author}
  {\bibfnamefont {A.}~\bibnamefont {Shukla}},\ }\href {\doibase
  10.1103/PhysRevB.87.195316} {\bibfield  {journal} {\bibinfo  {journal} {Phys.
  Rev. B}\ }\textbf {\bibinfo {volume} {87}},\ \bibinfo {pages} {195316}
  (\bibinfo {year} {2013})}\BibitemShut {NoStop}%
\bibitem [{\citenamefont {Tonndorf}\ \emph {et~al.}(2013)\citenamefont
  {Tonndorf}, \citenamefont {Schmidt}, \citenamefont {B\"ottger}, \citenamefont
  {Zhang}, \citenamefont {B\"orner}, \citenamefont {Liebig}, \citenamefont
  {Albrecht}, \citenamefont {Kloc}, \citenamefont {Gordan}, \citenamefont
  {Zahn}, \citenamefont {Michaelis~de Vasconcellos},\ and\ \citenamefont
  {Bratschitsch}}]{Tonndorf2013}%
  \BibitemOpen
  \bibfield  {author} {\bibinfo {author} {\bibfnamefont {P.}~\bibnamefont
  {Tonndorf}}, \bibinfo {author} {\bibfnamefont {R.}~\bibnamefont {Schmidt}},
  \bibinfo {author} {\bibfnamefont {P.}~\bibnamefont {B\"ottger}}, \bibinfo
  {author} {\bibfnamefont {X.}~\bibnamefont {Zhang}}, \bibinfo {author}
  {\bibfnamefont {J.}~\bibnamefont {B\"orner}}, \bibinfo {author}
  {\bibfnamefont {A.}~\bibnamefont {Liebig}}, \bibinfo {author} {\bibfnamefont
  {M.}~\bibnamefont {Albrecht}}, \bibinfo {author} {\bibfnamefont
  {C.}~\bibnamefont {Kloc}}, \bibinfo {author} {\bibfnamefont {O.}~\bibnamefont
  {Gordan}}, \bibinfo {author} {\bibfnamefont {D.~R.~T.}\ \bibnamefont {Zahn}},
  \bibinfo {author} {\bibfnamefont {S.}~\bibnamefont {Michaelis~de
  Vasconcellos}}, \ and\ \bibinfo {author} {\bibfnamefont {R.}~\bibnamefont
  {Bratschitsch}},\ }\href {\doibase 10.1364/OE.21.004908} {\bibfield
  {journal} {\bibinfo  {journal} {Opt. Express}\ }\textbf {\bibinfo {volume}
  {21}},\ \bibinfo {pages} {4908} (\bibinfo {year} {2013})}\BibitemShut
  {NoStop}%
\bibitem [{\citenamefont {Chen}\ \emph {et~al.}(2015)\citenamefont {Chen},
  \citenamefont {Zheng}, \citenamefont {Fuhrer},\ and\ \citenamefont
  {Yan}}]{Chen2015}%
  \BibitemOpen
  \bibfield  {author} {\bibinfo {author} {\bibfnamefont {S.-Y.}\ \bibnamefont
  {Chen}}, \bibinfo {author} {\bibfnamefont {C.}~\bibnamefont {Zheng}},
  \bibinfo {author} {\bibfnamefont {M.~S.}\ \bibnamefont {Fuhrer}}, \ and\
  \bibinfo {author} {\bibfnamefont {J.}~\bibnamefont {Yan}},\ }\href {\doibase
  10.1021/acs.nanolett.5b00092} {\bibfield  {journal} {\bibinfo  {journal}
  {Nano Letters}\ ,\ \bibinfo {pages} {2526}} (\bibinfo {year}
  {2015})}\BibitemShut {NoStop}%
\bibitem [{\citenamefont {Staiger}\ \emph {et~al.}(2015)\citenamefont
  {Staiger}, \citenamefont {Gillen}, \citenamefont {Scheuschner}, \citenamefont
  {Ochedowski}, \citenamefont {Kampmann}, \citenamefont {Schleberger},
  \citenamefont {Thomsen},\ and\ \citenamefont {Maultzsch}}]{Staiger2015}%
  \BibitemOpen
  \bibfield  {author} {\bibinfo {author} {\bibfnamefont {M.}~\bibnamefont
  {Staiger}}, \bibinfo {author} {\bibfnamefont {R.}~\bibnamefont {Gillen}},
  \bibinfo {author} {\bibfnamefont {N.}~\bibnamefont {Scheuschner}}, \bibinfo
  {author} {\bibfnamefont {O.}~\bibnamefont {Ochedowski}}, \bibinfo {author}
  {\bibfnamefont {F.}~\bibnamefont {Kampmann}}, \bibinfo {author}
  {\bibfnamefont {M.}~\bibnamefont {Schleberger}}, \bibinfo {author}
  {\bibfnamefont {C.}~\bibnamefont {Thomsen}}, \ and\ \bibinfo {author}
  {\bibfnamefont {J.}~\bibnamefont {Maultzsch}},\ }\href {\doibase
  10.1103/PhysRevB.91.195419} {\bibfield  {journal} {\bibinfo  {journal} {Phys.
  Rev. B}\ }\textbf {\bibinfo {volume} {91}},\ \bibinfo {pages} {195419}
  (\bibinfo {year} {2015})}\BibitemShut {NoStop}%
\bibitem [{\citenamefont {Lee}\ \emph {et~al.}(2010)\citenamefont {Lee},
  \citenamefont {Yan}, \citenamefont {Brus}, \citenamefont {Heinz},
  \citenamefont {Hone},\ and\ \citenamefont {Ryu}}]{Lee2010}%
  \BibitemOpen
  \bibfield  {author} {\bibinfo {author} {\bibfnamefont {C.}~\bibnamefont
  {Lee}}, \bibinfo {author} {\bibfnamefont {H.}~\bibnamefont {Yan}}, \bibinfo
  {author} {\bibfnamefont {L.~E.}\ \bibnamefont {Brus}}, \bibinfo {author}
  {\bibfnamefont {T.~F.}\ \bibnamefont {Heinz}}, \bibinfo {author}
  {\bibfnamefont {J.}~\bibnamefont {Hone}}, \ and\ \bibinfo {author}
  {\bibfnamefont {S.}~\bibnamefont {Ryu}},\ }\href {\doibase 10.1021/nn1003937}
  {\bibfield  {journal} {\bibinfo  {journal} {ACS Nano}\ }\textbf {\bibinfo
  {volume} {4}},\ \bibinfo {pages} {2695} (\bibinfo {year} {2010})}\BibitemShut
  {NoStop}%
\bibitem [{\citenamefont {Li}\ \emph {et~al.}(2012)\citenamefont {Li},
  \citenamefont {Zhang}, \citenamefont {Yap}, \citenamefont {Tay},
  \citenamefont {Edwin}, \citenamefont {Olivier},\ and\ \citenamefont
  {Baillargeat}}]{Li2012}%
  \BibitemOpen
  \bibfield  {author} {\bibinfo {author} {\bibfnamefont {H.}~\bibnamefont
  {Li}}, \bibinfo {author} {\bibfnamefont {Q.}~\bibnamefont {Zhang}}, \bibinfo
  {author} {\bibfnamefont {C.~C.~R.}\ \bibnamefont {Yap}}, \bibinfo {author}
  {\bibfnamefont {B.~K.}\ \bibnamefont {Tay}}, \bibinfo {author} {\bibfnamefont
  {T.~H.~T.}\ \bibnamefont {Edwin}}, \bibinfo {author} {\bibfnamefont
  {A.}~\bibnamefont {Olivier}}, \ and\ \bibinfo {author} {\bibfnamefont
  {D.}~\bibnamefont {Baillargeat}},\ }\href {\doibase 10.1002/adfm.201102111}
  {\bibfield  {journal} {\bibinfo  {journal} {Advanced Functional Materials}\
  }\textbf {\bibinfo {volume} {22}},\ \bibinfo {pages} {1385} (\bibinfo {year}
  {2012})}\BibitemShut {NoStop}%
\bibitem [{\citenamefont {Berkdemir}\ \emph {et~al.}(2013)\citenamefont
  {Berkdemir}, \citenamefont {Guti{\'e}rrez}, \citenamefont
  {Botello-M{\'e}ndez}, \citenamefont {Perea-L{\'o}pez}, \citenamefont
  {El{\'\i}as}, \citenamefont {Chia}, \citenamefont {Wang}, \citenamefont
  {Crespi}, \citenamefont {L{\'o}pez-Ur{\'\i}as}, \citenamefont {Charlier}
  \emph {et~al.}}]{Berkdemir2013}%
  \BibitemOpen
  \bibfield  {author} {\bibinfo {author} {\bibfnamefont {A.}~\bibnamefont
  {Berkdemir}}, \bibinfo {author} {\bibfnamefont {H.~R.}\ \bibnamefont
  {Guti{\'e}rrez}}, \bibinfo {author} {\bibfnamefont {A.~R.}\ \bibnamefont
  {Botello-M{\'e}ndez}}, \bibinfo {author} {\bibfnamefont {N.}~\bibnamefont
  {Perea-L{\'o}pez}}, \bibinfo {author} {\bibfnamefont {A.~L.}\ \bibnamefont
  {El{\'\i}as}}, \bibinfo {author} {\bibfnamefont {C.-I.}\ \bibnamefont
  {Chia}}, \bibinfo {author} {\bibfnamefont {B.}~\bibnamefont {Wang}}, \bibinfo
  {author} {\bibfnamefont {V.~H.}\ \bibnamefont {Crespi}}, \bibinfo {author}
  {\bibfnamefont {F.}~\bibnamefont {L{\'o}pez-Ur{\'\i}as}}, \bibinfo {author}
  {\bibfnamefont {J.-C.}\ \bibnamefont {Charlier}},  \emph {et~al.},\ }\href
  {\doibase 10.1038/srep01755} {\bibfield  {journal} {\bibinfo  {journal}
  {Scientific reports}\ }\textbf {\bibinfo {volume} {3}} (\bibinfo {year}
  {2013}),\ 10.1038/srep01755}\BibitemShut {NoStop}%
\bibitem [{\citenamefont {Loudon}(1964)}]{Loudon1964}%
  \BibitemOpen
  \bibfield  {author} {\bibinfo {author} {\bibfnamefont {R.}~\bibnamefont
  {Loudon}},\ }\href {\doibase 10.1080/00018736400101051} {\bibfield  {journal}
  {\bibinfo  {journal} {Advances in Physics}\ }\textbf {\bibinfo {volume}
  {13}},\ \bibinfo {pages} {423} (\bibinfo {year} {1964})}\BibitemShut
  {NoStop}%
\bibitem [{\citenamefont {Fathipour}\ \emph {et~al.}(2014)\citenamefont
  {Fathipour}, \citenamefont {Ma}, \citenamefont {Hwang}, \citenamefont
  {Protasenko}, \citenamefont {Vishwanath}, \citenamefont {Xing}, \citenamefont
  {Xu}, \citenamefont {Jena}, \citenamefont {Appenzeller},\ and\ \citenamefont
  {Seabaugh}}]{Fathipour2014}%
  \BibitemOpen
  \bibfield  {author} {\bibinfo {author} {\bibfnamefont {S.}~\bibnamefont
  {Fathipour}}, \bibinfo {author} {\bibfnamefont {N.}~\bibnamefont {Ma}},
  \bibinfo {author} {\bibfnamefont {W.~S.}\ \bibnamefont {Hwang}}, \bibinfo
  {author} {\bibfnamefont {V.}~\bibnamefont {Protasenko}}, \bibinfo {author}
  {\bibfnamefont {S.}~\bibnamefont {Vishwanath}}, \bibinfo {author}
  {\bibfnamefont {H.~G.}\ \bibnamefont {Xing}}, \bibinfo {author}
  {\bibfnamefont {H.}~\bibnamefont {Xu}}, \bibinfo {author} {\bibfnamefont
  {D.}~\bibnamefont {Jena}}, \bibinfo {author} {\bibfnamefont {J.}~\bibnamefont
  {Appenzeller}}, \ and\ \bibinfo {author} {\bibfnamefont {A.}~\bibnamefont
  {Seabaugh}},\ }\href {\doibase 10.1063/1.4901527} {\bibfield  {journal}
  {\bibinfo  {journal} {Applied Physics Letters}\ }\textbf {\bibinfo {volume}
  {105}},\ \bibinfo {pages} {192101} (\bibinfo {year} {2014})}\BibitemShut
  {NoStop}%
\bibitem [{\citenamefont {Lezama}\ \emph {et~al.}(2014)\citenamefont {Lezama},
  \citenamefont {Ubaldini}, \citenamefont {Longobardi}, \citenamefont
  {Giannini}, \citenamefont {Renner}, \citenamefont {Kuzmenko},\ and\
  \citenamefont {Morpurgo}}]{Lezama2014}%
  \BibitemOpen
  \bibfield  {author} {\bibinfo {author} {\bibfnamefont {I.~G.}\ \bibnamefont
  {Lezama}}, \bibinfo {author} {\bibfnamefont {A.}~\bibnamefont {Ubaldini}},
  \bibinfo {author} {\bibfnamefont {M.}~\bibnamefont {Longobardi}}, \bibinfo
  {author} {\bibfnamefont {E.}~\bibnamefont {Giannini}}, \bibinfo {author}
  {\bibfnamefont {C.}~\bibnamefont {Renner}}, \bibinfo {author} {\bibfnamefont
  {A.~B.}\ \bibnamefont {Kuzmenko}}, \ and\ \bibinfo {author} {\bibfnamefont
  {A.~F.}\ \bibnamefont {Morpurgo}},\ }\href {\doibase
  10.1088/2053-1583/1/2/021002} {\bibfield  {journal} {\bibinfo  {journal} {2D
  Materials}\ }\textbf {\bibinfo {volume} {1}},\ \bibinfo {pages} {021002}
  (\bibinfo {year} {2014})}\BibitemShut {NoStop}%
\bibitem [{\citenamefont {Lin}\ \emph {et~al.}(2014)\citenamefont {Lin},
  \citenamefont {Xu}, \citenamefont {Wang}, \citenamefont {Li}, \citenamefont
  {Yamamoto}, \citenamefont {Aparecido-Ferreira}, \citenamefont {Li},
  \citenamefont {Sun}, \citenamefont {Nakaharai}, \citenamefont {Jian},
  \citenamefont {Ueno},\ and\ \citenamefont {Tsukagoshi}}]{Lin2014}%
  \BibitemOpen
  \bibfield  {author} {\bibinfo {author} {\bibfnamefont {Y.-F.}\ \bibnamefont
  {Lin}}, \bibinfo {author} {\bibfnamefont {Y.}~\bibnamefont {Xu}}, \bibinfo
  {author} {\bibfnamefont {S.-T.}\ \bibnamefont {Wang}}, \bibinfo {author}
  {\bibfnamefont {S.-L.}\ \bibnamefont {Li}}, \bibinfo {author} {\bibfnamefont
  {M.}~\bibnamefont {Yamamoto}}, \bibinfo {author} {\bibfnamefont
  {A.}~\bibnamefont {Aparecido-Ferreira}}, \bibinfo {author} {\bibfnamefont
  {W.}~\bibnamefont {Li}}, \bibinfo {author} {\bibfnamefont {H.}~\bibnamefont
  {Sun}}, \bibinfo {author} {\bibfnamefont {S.}~\bibnamefont {Nakaharai}},
  \bibinfo {author} {\bibfnamefont {W.-B.}\ \bibnamefont {Jian}}, \bibinfo
  {author} {\bibfnamefont {K.}~\bibnamefont {Ueno}}, \ and\ \bibinfo {author}
  {\bibfnamefont {K.}~\bibnamefont {Tsukagoshi}},\ }\href {\doibase
  10.1002/adma.201305845} {\bibfield  {journal} {\bibinfo  {journal} {Advanced
  Materials}\ }\textbf {\bibinfo {volume} {26}},\ \bibinfo {pages} {3263}
  (\bibinfo {year} {2014})}\BibitemShut {NoStop}%
\bibitem [{\citenamefont {Pradhan}\ \emph {et~al.}(2014)\citenamefont
  {Pradhan}, \citenamefont {Rhodes}, \citenamefont {Feng}, \citenamefont {Xin},
  \citenamefont {Memaran}, \citenamefont {Moon}, \citenamefont {Terrones},
  \citenamefont {Terrones},\ and\ \citenamefont {Balicas}}]{Pradhan2014}%
  \BibitemOpen
  \bibfield  {author} {\bibinfo {author} {\bibfnamefont {N.~R.}\ \bibnamefont
  {Pradhan}}, \bibinfo {author} {\bibfnamefont {D.}~\bibnamefont {Rhodes}},
  \bibinfo {author} {\bibfnamefont {S.}~\bibnamefont {Feng}}, \bibinfo {author}
  {\bibfnamefont {Y.}~\bibnamefont {Xin}}, \bibinfo {author} {\bibfnamefont
  {S.}~\bibnamefont {Memaran}}, \bibinfo {author} {\bibfnamefont {B.-H.}\
  \bibnamefont {Moon}}, \bibinfo {author} {\bibfnamefont {H.}~\bibnamefont
  {Terrones}}, \bibinfo {author} {\bibfnamefont {M.}~\bibnamefont {Terrones}},
  \ and\ \bibinfo {author} {\bibfnamefont {L.}~\bibnamefont {Balicas}},\ }\href
  {\doibase 10.1021/nn501013c} {\bibfield  {journal} {\bibinfo  {journal} {ACS
  Nano}\ }\textbf {\bibinfo {volume} {8}},\ \bibinfo {pages} {5911} (\bibinfo
  {year} {2014})}\BibitemShut {NoStop}%
\bibitem [{\citenamefont {Ruppert}\ \emph {et~al.}(2014)\citenamefont
  {Ruppert}, \citenamefont {Aslan},\ and\ \citenamefont {Heinz}}]{Ruppert2014}%
  \BibitemOpen
  \bibfield  {author} {\bibinfo {author} {\bibfnamefont {C.}~\bibnamefont
  {Ruppert}}, \bibinfo {author} {\bibfnamefont {O.~B.}\ \bibnamefont {Aslan}},
  \ and\ \bibinfo {author} {\bibfnamefont {T.~F.}\ \bibnamefont {Heinz}},\
  }\href {\doibase 10.1021/nl502557g} {\bibfield  {journal} {\bibinfo
  {journal} {Nano Letters}\ }\textbf {\bibinfo {volume} {14}},\ \bibinfo
  {pages} {6231} (\bibinfo {year} {2014})}\BibitemShut {NoStop}%
\bibitem [{\citenamefont {Lezama}\ \emph {et~al.}(2015)\citenamefont {Lezama},
  \citenamefont {Arora}, \citenamefont {Ubaldini}, \citenamefont {Barreteau},
  \citenamefont {Giannini}, \citenamefont {Potemski},\ and\ \citenamefont
  {Morpurgo}}]{Lezama2015}%
  \BibitemOpen
  \bibfield  {author} {\bibinfo {author} {\bibfnamefont {I.~G.}\ \bibnamefont
  {Lezama}}, \bibinfo {author} {\bibfnamefont {A.}~\bibnamefont {Arora}},
  \bibinfo {author} {\bibfnamefont {A.}~\bibnamefont {Ubaldini}}, \bibinfo
  {author} {\bibfnamefont {C.}~\bibnamefont {Barreteau}}, \bibinfo {author}
  {\bibfnamefont {E.}~\bibnamefont {Giannini}}, \bibinfo {author}
  {\bibfnamefont {M.}~\bibnamefont {Potemski}}, \ and\ \bibinfo {author}
  {\bibfnamefont {A.~F.}\ \bibnamefont {Morpurgo}},\ }\href {\doibase
  10.1021/nl5045007} {\bibfield  {journal} {\bibinfo  {journal} {Nano Lett.}\
  }\textbf {\bibinfo {volume} {15}},\ \bibinfo {pages} {2336} (\bibinfo {year}
  {2015})}\BibitemShut {NoStop}%
\bibitem [{\citenamefont {Guo}\ \emph {et~al.}(2015)\citenamefont {Guo},
  \citenamefont {Yang}, \citenamefont {Yamamoto}, \citenamefont {Zhou},
  \citenamefont {Ishikawa}, \citenamefont {Ueno}, \citenamefont {Tsukagoshi},
  \citenamefont {Zhang}, \citenamefont {Dresselhaus},\ and\ \citenamefont
  {Saito}}]{Guo2015}%
  \BibitemOpen
  \bibfield  {author} {\bibinfo {author} {\bibfnamefont {H.}~\bibnamefont
  {Guo}}, \bibinfo {author} {\bibfnamefont {T.}~\bibnamefont {Yang}}, \bibinfo
  {author} {\bibfnamefont {M.}~\bibnamefont {Yamamoto}}, \bibinfo {author}
  {\bibfnamefont {L.}~\bibnamefont {Zhou}}, \bibinfo {author} {\bibfnamefont
  {R.}~\bibnamefont {Ishikawa}}, \bibinfo {author} {\bibfnamefont
  {K.}~\bibnamefont {Ueno}}, \bibinfo {author} {\bibfnamefont {K.}~\bibnamefont
  {Tsukagoshi}}, \bibinfo {author} {\bibfnamefont {Z.}~\bibnamefont {Zhang}},
  \bibinfo {author} {\bibfnamefont {M.~S.}\ \bibnamefont {Dresselhaus}}, \ and\
  \bibinfo {author} {\bibfnamefont {R.}~\bibnamefont {Saito}},\ }\href
  {\doibase 10.1103/PhysRevB.91.205415} {\bibfield  {journal} {\bibinfo
  {journal} {Phys. Rev. B}\ }\textbf {\bibinfo {volume} {91}},\ \bibinfo
  {pages} {205415} (\bibinfo {year} {2015})}\BibitemShut {NoStop}%
\bibitem [{\citenamefont {Lee}\ \emph {et~al.}(2015)\citenamefont {Lee},
  \citenamefont {Park}, \citenamefont {Son},\ and\ \citenamefont
  {Cheong}}]{Lee2015}%
  \BibitemOpen
  \bibfield  {author} {\bibinfo {author} {\bibfnamefont {J.-U.}\ \bibnamefont
  {Lee}}, \bibinfo {author} {\bibfnamefont {J.}~\bibnamefont {Park}}, \bibinfo
  {author} {\bibfnamefont {Y.-W.}\ \bibnamefont {Son}}, \ and\ \bibinfo
  {author} {\bibfnamefont {H.}~\bibnamefont {Cheong}},\ }\href {\doibase
  10.1039/C4NR05785F} {\bibfield  {journal} {\bibinfo  {journal} {Nanoscale}\
  }\textbf {\bibinfo {volume} {7}},\ \bibinfo {pages} {3229} (\bibinfo {year}
  {2015})}\BibitemShut {NoStop}%
\bibitem [{\citenamefont {Tan}\ \emph {et~al.}(2012)\citenamefont {Tan},
  \citenamefont {Han}, \citenamefont {Zhao}, \citenamefont {Wu}, \citenamefont
  {Chang}, \citenamefont {Wang}, \citenamefont {Wang}, \citenamefont {Bonini},
  \citenamefont {Marzari},\ and\ \citenamefont {Pugno}}]{Tan2012}%
  \BibitemOpen
  \bibfield  {author} {\bibinfo {author} {\bibfnamefont {P.~H.}\ \bibnamefont
  {Tan}}, \bibinfo {author} {\bibfnamefont {W.~P.}\ \bibnamefont {Han}},
  \bibinfo {author} {\bibfnamefont {W.~J.}\ \bibnamefont {Zhao}}, \bibinfo
  {author} {\bibfnamefont {Z.~H.}\ \bibnamefont {Wu}}, \bibinfo {author}
  {\bibfnamefont {K.}~\bibnamefont {Chang}}, \bibinfo {author} {\bibfnamefont
  {H.}~\bibnamefont {Wang}}, \bibinfo {author} {\bibfnamefont {Y.~F.}\
  \bibnamefont {Wang}}, \bibinfo {author} {\bibfnamefont {N.}~\bibnamefont
  {Bonini}}, \bibinfo {author} {\bibfnamefont {N.}~\bibnamefont {Marzari}}, \
  and\ \bibinfo {author} {\bibfnamefont {N.}~\bibnamefont {Pugno}},\ }\href
  {http://www.nature.com/nmat/journal/v11/n4/abs/nmat3245.html} {\bibfield
  {journal} {\bibinfo  {journal} {Nature materials}\ }\textbf {\bibinfo
  {volume} {11}},\ \bibinfo {pages} {294} (\bibinfo {year} {2012})}\BibitemShut
  {NoStop}%
\bibitem [{\citenamefont {Lui}\ \emph {et~al.}(2014)\citenamefont {Lui},
  \citenamefont {Ye}, \citenamefont {Keiser}, \citenamefont {Xiao},\ and\
  \citenamefont {He}}]{Lui2014}%
  \BibitemOpen
  \bibfield  {author} {\bibinfo {author} {\bibfnamefont {C.~H.}\ \bibnamefont
  {Lui}}, \bibinfo {author} {\bibfnamefont {Z.}~\bibnamefont {Ye}}, \bibinfo
  {author} {\bibfnamefont {C.}~\bibnamefont {Keiser}}, \bibinfo {author}
  {\bibfnamefont {X.}~\bibnamefont {Xiao}}, \ and\ \bibinfo {author}
  {\bibfnamefont {R.}~\bibnamefont {He}},\ }\href {\doibase 10.1021/nl501678j}
  {\bibfield  {journal} {\bibinfo  {journal} {Nano Letters}\ }\textbf {\bibinfo
  {volume} {14}},\ \bibinfo {pages} {4615} (\bibinfo {year}
  {2014})}\BibitemShut {NoStop}%
\bibitem [{\citenamefont {Michel}\ and\ \citenamefont
  {Verberck}(2012)}]{Michel2012}%
  \BibitemOpen
  \bibfield  {author} {\bibinfo {author} {\bibfnamefont {K.~H.}\ \bibnamefont
  {Michel}}\ and\ \bibinfo {author} {\bibfnamefont {B.}~\bibnamefont
  {Verberck}},\ }\href {\doibase 10.1103/PhysRevB.85.094303} {\bibfield
  {journal} {\bibinfo  {journal} {Physical Review B}\ }\textbf {\bibinfo
  {volume} {85}},\ \bibinfo {pages} {094303} (\bibinfo {year}
  {2012})}\BibitemShut {NoStop}%
\bibitem [{\citenamefont {Umari}\ \emph {et~al.}(2001)\citenamefont {Umari},
  \citenamefont {Pasquarello},\ and\ \citenamefont {Dal~Corso}}]{Umari2001}%
  \BibitemOpen
  \bibfield  {author} {\bibinfo {author} {\bibfnamefont {P.}~\bibnamefont
  {Umari}}, \bibinfo {author} {\bibfnamefont {A.}~\bibnamefont {Pasquarello}},
  \ and\ \bibinfo {author} {\bibfnamefont {A.}~\bibnamefont {Dal~Corso}},\
  }\href {\doibase 10.1103/PhysRevB.63.094305} {\bibfield  {journal} {\bibinfo
  {journal} {Phys. Rev. B}\ }\textbf {\bibinfo {volume} {63}},\ \bibinfo
  {pages} {094305} (\bibinfo {year} {2001})}\BibitemShut {NoStop}%
\bibitem [{\citenamefont {Wirtz}\ \emph {et~al.}(2005)\citenamefont {Wirtz},
  \citenamefont {Lazzeri}, \citenamefont {Mauri},\ and\ \citenamefont
  {Rubio}}]{Wirtz2005}%
  \BibitemOpen
  \bibfield  {author} {\bibinfo {author} {\bibfnamefont {L.}~\bibnamefont
  {Wirtz}}, \bibinfo {author} {\bibfnamefont {M.}~\bibnamefont {Lazzeri}},
  \bibinfo {author} {\bibfnamefont {F.}~\bibnamefont {Mauri}}, \ and\ \bibinfo
  {author} {\bibfnamefont {A.}~\bibnamefont {Rubio}},\ }\href {\doibase
  10.1103/PhysRevB.71.241402} {\bibfield  {journal} {\bibinfo  {journal} {Phys.
  Rev. B}\ }\textbf {\bibinfo {volume} {71}},\ \bibinfo {pages} {241402}
  (\bibinfo {year} {2005})}\BibitemShut {NoStop}%
\bibitem [{\citenamefont {Chakraborty}\ \emph {et~al.}(2013)\citenamefont
  {Chakraborty}, \citenamefont {Matte}, \citenamefont {Sood},\ and\
  \citenamefont {Rao}}]{Chakraborty2013}%
  \BibitemOpen
  \bibfield  {author} {\bibinfo {author} {\bibfnamefont {B.}~\bibnamefont
  {Chakraborty}}, \bibinfo {author} {\bibfnamefont {H.~S. S.~R.}\ \bibnamefont
  {Matte}}, \bibinfo {author} {\bibfnamefont {A.~K.}\ \bibnamefont {Sood}}, \
  and\ \bibinfo {author} {\bibfnamefont {C.~N.~R.}\ \bibnamefont {Rao}},\
  }\href {\doibase 10.1002/jrs.4147} {\bibfield  {journal} {\bibinfo  {journal}
  {Journal of Raman Spectroscopy}\ }\textbf {\bibinfo {volume} {44}},\ \bibinfo
  {pages} {92} (\bibinfo {year} {2013})}\BibitemShut {NoStop}%
\bibitem [{\citenamefont {Carvalho}\ \emph {et~al.}(2015)\citenamefont
  {Carvalho}, \citenamefont {Malard}, \citenamefont {Alves}, \citenamefont
  {Fantini},\ and\ \citenamefont {Pimenta}}]{Carvalho2015}%
  \BibitemOpen
  \bibfield  {author} {\bibinfo {author} {\bibfnamefont {B.~R.}\ \bibnamefont
  {Carvalho}}, \bibinfo {author} {\bibfnamefont {L.~M.}\ \bibnamefont
  {Malard}}, \bibinfo {author} {\bibfnamefont {J.~M.}\ \bibnamefont {Alves}},
  \bibinfo {author} {\bibfnamefont {C.}~\bibnamefont {Fantini}}, \ and\
  \bibinfo {author} {\bibfnamefont {M.~A.}\ \bibnamefont {Pimenta}},\ }\href
  {\doibase 10.1103/PhysRevLett.114.136403} {\bibfield  {journal} {\bibinfo
  {journal} {Phys. Rev. Lett.}\ }\textbf {\bibinfo {volume} {114}},\ \bibinfo
  {pages} {136403} (\bibinfo {year} {2015})}\BibitemShut {NoStop}%
\bibitem [{\citenamefont {Favron}\ \emph {et~al.}(2015)\citenamefont {Favron},
  \citenamefont {Gaufr{\`e}s}, \citenamefont {Fossard}, \citenamefont
  {Phaneuf-L{'}Heureux}, \citenamefont {Tang}, \citenamefont {L{\'e}vesque},
  \citenamefont {Loiseau}, \citenamefont {Leonelli}, \citenamefont
  {Francoeur},\ and\ \citenamefont {Martel}}]{Favron2015}%
  \BibitemOpen
  \bibfield  {author} {\bibinfo {author} {\bibfnamefont {A.}~\bibnamefont
  {Favron}}, \bibinfo {author} {\bibfnamefont {E.}~\bibnamefont {Gaufr{\`e}s}},
  \bibinfo {author} {\bibfnamefont {F.}~\bibnamefont {Fossard}}, \bibinfo
  {author} {\bibfnamefont {A.-L.}\ \bibnamefont {Phaneuf-L{'}Heureux}},
  \bibinfo {author} {\bibfnamefont {N.~Y.}\ \bibnamefont {Tang}}, \bibinfo
  {author} {\bibfnamefont {P.~L.}\ \bibnamefont {L{\'e}vesque}}, \bibinfo
  {author} {\bibfnamefont {A.}~\bibnamefont {Loiseau}}, \bibinfo {author}
  {\bibfnamefont {R.}~\bibnamefont {Leonelli}}, \bibinfo {author}
  {\bibfnamefont {S.}~\bibnamefont {Francoeur}}, \ and\ \bibinfo {author}
  {\bibfnamefont {R.}~\bibnamefont {Martel}},\ }\href
  {http://dx.doi.org/10.1038/nmat4299} {\bibfield  {journal} {\bibinfo
  {journal} {Nature materials}\ } (\bibinfo {year} {2015})},\ \bibinfo {note}
  {doi:10.1038/nmat4299}\BibitemShut {NoStop}%
\bibitem [{\citenamefont {Zhao}\ \emph {et~al.}(2014)\citenamefont {Zhao},
  \citenamefont {Luo}, \citenamefont {Zhang}, \citenamefont {Wu}, \citenamefont
  {Bai}, \citenamefont {Wang}, \citenamefont {Jia}, \citenamefont {Peng},
  \citenamefont {Liu}, \citenamefont {Quek},\ and\ \citenamefont
  {Xiong}}]{Zhao2014}%
  \BibitemOpen
  \bibfield  {author} {\bibinfo {author} {\bibfnamefont {Y.}~\bibnamefont
  {Zhao}}, \bibinfo {author} {\bibfnamefont {X.}~\bibnamefont {Luo}}, \bibinfo
  {author} {\bibfnamefont {J.}~\bibnamefont {Zhang}}, \bibinfo {author}
  {\bibfnamefont {J.}~\bibnamefont {Wu}}, \bibinfo {author} {\bibfnamefont
  {X.}~\bibnamefont {Bai}}, \bibinfo {author} {\bibfnamefont {M.}~\bibnamefont
  {Wang}}, \bibinfo {author} {\bibfnamefont {J.}~\bibnamefont {Jia}}, \bibinfo
  {author} {\bibfnamefont {H.}~\bibnamefont {Peng}}, \bibinfo {author}
  {\bibfnamefont {Z.}~\bibnamefont {Liu}}, \bibinfo {author} {\bibfnamefont
  {S.~Y.}\ \bibnamefont {Quek}}, \ and\ \bibinfo {author} {\bibfnamefont
  {Q.}~\bibnamefont {Xiong}},\ }\href {\doibase 10.1103/PhysRevB.90.245428}
  {\bibfield  {journal} {\bibinfo  {journal} {Phys. Rev. B}\ }\textbf {\bibinfo
  {volume} {90}},\ \bibinfo {pages} {245428} (\bibinfo {year}
  {2014})}\BibitemShut {NoStop}%
\bibitem [{\citenamefont {Baroni}\ \emph {et~al.}(2001)\citenamefont {Baroni},
  \citenamefont {de~Gironcoli}, \citenamefont {Dal~Corso},\ and\ \citenamefont
  {Giannozzi}}]{Baroni2001}%
  \BibitemOpen
  \bibfield  {author} {\bibinfo {author} {\bibfnamefont {S.}~\bibnamefont
  {Baroni}}, \bibinfo {author} {\bibfnamefont {S.}~\bibnamefont
  {de~Gironcoli}}, \bibinfo {author} {\bibfnamefont {A.}~\bibnamefont
  {Dal~Corso}}, \ and\ \bibinfo {author} {\bibfnamefont {P.}~\bibnamefont
  {Giannozzi}},\ }\href {\doibase 10.1103/RevModPhys.73.515} {\bibfield
  {journal} {\bibinfo  {journal} {Rev. Mod. Phys.}\ }\textbf {\bibinfo {volume}
  {73}},\ \bibinfo {pages} {515} (\bibinfo {year} {2001})}\BibitemShut
  {NoStop}%
\bibitem [{\citenamefont {Giannozzi}\ \emph {et~al.}(2009)\citenamefont
  {Giannozzi}, \citenamefont {Baroni}, \citenamefont {Bonini}, \citenamefont
  {Calandra}, \citenamefont {Car}, \citenamefont {Cavazzoni}, \citenamefont
  {Ceresoli}, \citenamefont {Chiarotti}, \citenamefont {Cococcioni},
  \citenamefont {Dabo}, \citenamefont {Corso}, \citenamefont {de~Gironcoli},
  \citenamefont {Fabris}, \citenamefont {Fratesi}, \citenamefont {Gebauer},
  \citenamefont {Gerstmann}, \citenamefont {Gougoussis}, \citenamefont
  {Kokalj}, \citenamefont {Lazzeri}, \citenamefont {Martin-Samos},
  \citenamefont {Marzari}, \citenamefont {Mauri}, \citenamefont {Mazzarello},
  \citenamefont {Paolini}, \citenamefont {Pasquarello}, \citenamefont
  {Paulatto}, \citenamefont {Sbraccia}, \citenamefont {Scandolo}, \citenamefont
  {Sclauzero}, \citenamefont {Seitsonen}, \citenamefont {Smogunov},
  \citenamefont {Umari},\ and\ \citenamefont {Wentzcovitch}}]{Giannozzi2009}%
  \BibitemOpen
  \bibfield  {author} {\bibinfo {author} {\bibfnamefont {P.}~\bibnamefont
  {Giannozzi}}, \bibinfo {author} {\bibfnamefont {S.}~\bibnamefont {Baroni}},
  \bibinfo {author} {\bibfnamefont {N.}~\bibnamefont {Bonini}}, \bibinfo
  {author} {\bibfnamefont {M.}~\bibnamefont {Calandra}}, \bibinfo {author}
  {\bibfnamefont {R.}~\bibnamefont {Car}}, \bibinfo {author} {\bibfnamefont
  {C.}~\bibnamefont {Cavazzoni}}, \bibinfo {author} {\bibfnamefont
  {D.}~\bibnamefont {Ceresoli}}, \bibinfo {author} {\bibfnamefont {G.~L.}\
  \bibnamefont {Chiarotti}}, \bibinfo {author} {\bibfnamefont {M.}~\bibnamefont
  {Cococcioni}}, \bibinfo {author} {\bibfnamefont {I.}~\bibnamefont {Dabo}},
  \bibinfo {author} {\bibfnamefont {A.~D.}\ \bibnamefont {Corso}}, \bibinfo
  {author} {\bibfnamefont {S.}~\bibnamefont {de~Gironcoli}}, \bibinfo {author}
  {\bibfnamefont {S.}~\bibnamefont {Fabris}}, \bibinfo {author} {\bibfnamefont
  {G.}~\bibnamefont {Fratesi}}, \bibinfo {author} {\bibfnamefont
  {R.}~\bibnamefont {Gebauer}}, \bibinfo {author} {\bibfnamefont
  {U.}~\bibnamefont {Gerstmann}}, \bibinfo {author} {\bibfnamefont
  {C.}~\bibnamefont {Gougoussis}}, \bibinfo {author} {\bibfnamefont
  {A.}~\bibnamefont {Kokalj}}, \bibinfo {author} {\bibfnamefont
  {M.}~\bibnamefont {Lazzeri}}, \bibinfo {author} {\bibfnamefont
  {L.}~\bibnamefont {Martin-Samos}}, \bibinfo {author} {\bibfnamefont
  {N.}~\bibnamefont {Marzari}}, \bibinfo {author} {\bibfnamefont
  {F.}~\bibnamefont {Mauri}}, \bibinfo {author} {\bibfnamefont
  {R.}~\bibnamefont {Mazzarello}}, \bibinfo {author} {\bibfnamefont
  {S.}~\bibnamefont {Paolini}}, \bibinfo {author} {\bibfnamefont
  {A.}~\bibnamefont {Pasquarello}}, \bibinfo {author} {\bibfnamefont
  {L.}~\bibnamefont {Paulatto}}, \bibinfo {author} {\bibfnamefont
  {C.}~\bibnamefont {Sbraccia}}, \bibinfo {author} {\bibfnamefont
  {S.}~\bibnamefont {Scandolo}}, \bibinfo {author} {\bibfnamefont
  {G.}~\bibnamefont {Sclauzero}}, \bibinfo {author} {\bibfnamefont {A.~P.}\
  \bibnamefont {Seitsonen}}, \bibinfo {author} {\bibfnamefont {A.}~\bibnamefont
  {Smogunov}}, \bibinfo {author} {\bibfnamefont {P.}~\bibnamefont {Umari}}, \
  and\ \bibinfo {author} {\bibfnamefont {R.~M.}\ \bibnamefont {Wentzcovitch}},\
  }\href {http://stacks.iop.org/0953-8984/21/i=39/a=395502} {\bibfield
  {journal} {\bibinfo  {journal} {Journal of Physics: Condensed Matter}\
  }\textbf {\bibinfo {volume} {21}},\ \bibinfo {pages} {395502} (\bibinfo
  {year} {2009})}\BibitemShut {NoStop}%
\end{thebibliography}

%%%%%%%%%%%%%%%%%%%%%%%%%%%%%%%%%%%%%%%%%%%%%%%%%%%%%%%%%%%%%%%%%%%%%%%%%%%%%%


%%%%%%%%%%%%%%%%%%SUPPORTING INFO%%%%%%%%%%%%%%%%%%%%%%%%%%%%%%%%%%%%%%%%%%%%%%

\onecolumngrid
\newpage
\begin{center}
{\Large\textbf{Supplementary Information}}
\end{center}

\setcounter{equation}{0}%reset counter
\setcounter{figure}{0}
\setcounter{section}{0}
\renewcommand{\theequation}{S\arabic{equation}}
\renewcommand{\thefigure}{S\arabic{figure}}
\renewcommand{\thesection}{S\arabic{section}}
\linespread{1.4}

This document contains the following sections:
\begin{itemize}
\item Force constant model  (Section \ref{MDLMQT})
\item Bulk phonon frequencies  (Section \ref{BPF})
\item Normal mode displacements (Section \ref{NMD})
\item Empirical bond polarizability model (Section \ref{BPM})
\item Raman Spectra from the bond polarizability model (Section \ref{RBPM})
\item Ab-initio bulk phonon frequencies (Section \ref{AI})
\item Additional Raman measurements (Section \ref{ARM})
\end{itemize}

%\clearpage

%\newpage

\section{Force constant model}
\label{MDLMQT}

\begin{figure}[!htb]
\begin{center}
\includegraphics[width=1\linewidth]{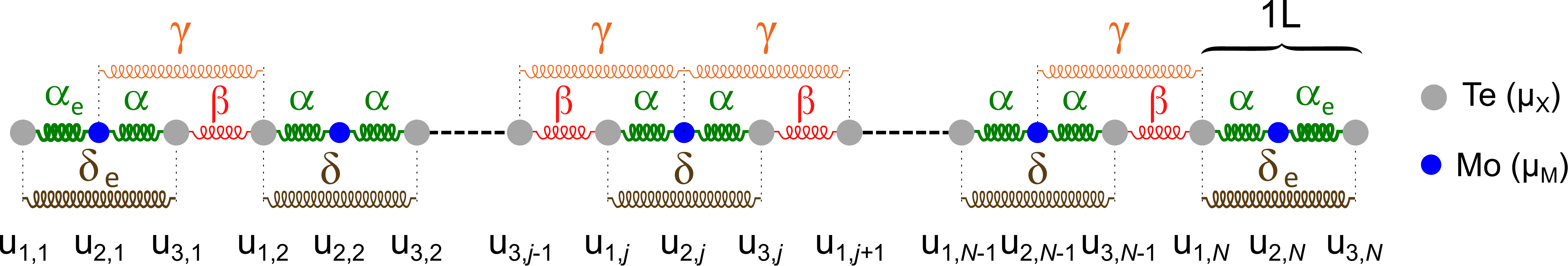}
\caption{Schematic of the finite linear chain model. $\mu_M$ ($\mu_X$) is the mass per unit area of the Mo (Te) atom. $\alpha$ and $\beta$ are the (intra-plane) force constants that connect the first nearest neighbor atoms. $\gamma$ and $\delta$ are the (inter-plane) force constants that connect the second nearest neighbor atoms. $u_{i,j}$ is the displacement, with respect to the equilibrium position, of the $i^{\rm th}$ atom ($i=1,3$ for Te and $i=2$ for Mo) in the $j^{\rm th}$ MoTe$_2$ layer ($j\in \llbracket 1,N \rrbracket$).}
\label{figSI_model}
\end{center}
\end{figure}

As explained in the main text, $N$-layer MoTe$_2$ is modeled as a one-dimensional finite linear chain composed of $2N$ Te atoms of mass per unit area $\mu_X$ and $N$ Mo atoms of mass per unit area $\mu_M$ (see Fig. \ref{figSI_model}) \cite{Luo2013}. Within one MoTe$_2$ layer, nearest neighbor Mo and Te atoms and the pair of second nearest neighbor Te atoms are connected by springs with force constants per unit area $\alpha$ and $\delta$ respectively. Interlayer interactions are described by two force constants per unit area $\beta$ and $\gamma$ between nearest neighbor Te atoms belonging to adjacent layers and between second nearest neighbor Mo and Te atoms, respectively. To account for surface effects, we consider effective force constants $\alpha_e$ and $\delta_e$ for the first and $N^{\rm th}$ layer. Since substrate effects have been shown to have a negligible influence on the Raman modes of MX$_2$\cite{Lee2010,Yamamoto2014,Luo2013,Luo2013b,Zhao2013}, we assume that the two extreme layers are only connected to one layer, \textit{i.e.,} we do not include an additional spring constant that would account for coupling of one of the outer layers to a substrate. 

We note $u_{i,j}$ the displacement, with respect to the equilibrium position, of the $i^{\rm th}$ atom ($i=1,3$ for Te and $i=2$ for Mo) in the $j^{\rm th}$ MoTe$_2$ layer ($j\in \llbracket 1,N \rrbracket$). We can then write the equations of motion using Newton's law. These equations form a system of $3N$ coupled differential equations that can be written as
\begin{equation}
\frac{\textrm{d}^2\mathcal{U}}{\textrm{d}t^2}=-\mathcal{D}\:\mathcal{U},
\label{eq_matrice}
\end{equation}
with the displacement vector $\mathcal{U} =\left(\begin{mmatrix} u_{1,1}, & u_{2,1}, & u_{3,1}, & \hdots, & u_{i,j}, & \hdots,  & u_{1,N}, & u_{2,N}, & u_{3,N} \end{mmatrix}\right)$ and the $3N\times 3N$ dynamical matrix 

\setcounter{MaxMatrixCols}{13}
\arraycolsep=1pt
$\mathcal{D}=\left(\begin{mmatrix}
\frac{\alpha_{\textrm{e}}+\delta{\textrm{e}}}{\mu_X}  & -\frac{\alpha_{\textrm{e}}}{\mu_X} & -\frac{\delta_{\textrm{e}}}{\mu_X} & \cdots & 0 & 0 & 0 & 0 & 0 & \cdots & 0 & 0 & 0 \\

-\frac{\alpha_{\textrm{e}}}{\mu_M}  & \frac{\alpha_{\textrm{e}}+\alpha+\gamma}{\mu_M} & -\frac{\alpha}{\mu_M} & \ddots & \vdots & \vdots & \vdots & \vdots & \vdots & \ddots & \vdots & \vdots & \vdots \\

-\frac{\delta_{\textrm{e}}}{\mu_X}  & -\frac{\alpha}{\mu_X} & \frac{\alpha+\beta+\gamma+\delta_{\textrm{e}}}{\mu_X} & \ddots & \vdots & \vdots & \vdots & \vdots & \vdots & \ddots & \vdots & \vdots & \vdots 
\\

0  & -\frac{\gamma}{\mu_X} & -\frac{\beta}{\mu_X} & \ddots & \vdots & \vdots & \vdots & \vdots & \vdots & \ddots & \vdots & \vdots & \vdots \\

\vdots  & 0 & -\frac{\gamma}{\mu_M} & \ddots & \vdots & \vdots & \vdots & \vdots & \vdots & \ddots & \vdots & \vdots & \vdots \\

\vdots  & \vdots & 0 & \ddots & \vdots & \vdots & \vdots & \vdots & \vdots & \ddots & \vdots & \vdots & \vdots \\

\vdots  & \vdots & \vdots & \ddots & 0 & \vdots & \vdots & \vdots & \vdots & \ddots & \vdots & \vdots & \vdots \\ 

\vdots  & \vdots & \vdots & \ddots & -\frac{\delta}{\mu_X} & 0 & \vdots & \vdots & \vdots & \ddots & \vdots & \vdots & \vdots \\ 

\vdots  & \vdots & \vdots & \ddots & -\frac{\alpha}{\mu_M} & -\frac{\gamma}{\mu_M} & 0 & \vdots & \vdots & \ddots & \vdots & \vdots & \vdots \\

\vdots  & \vdots & \vdots & \ddots & \frac{\alpha+\beta+\gamma+\delta}{\mu_X} & -\frac{\beta}{\mu_X} & -\frac{\gamma}{\mu_X} & 0 & \vdots & \ddots & \vdots & \vdots & \vdots \\

\vdots  & \vdots & \vdots & \ddots & -\frac{\beta}{\mu_X}& \frac{\alpha+\beta+\gamma+\delta}{\mu_X} & -\frac{\alpha}{\mu_X} & -\frac{\delta}{\mu_X} & 0 & \ddots & \vdots & \vdots & \vdots \\

\vdots  & \vdots & \vdots & \ddots & -\frac{\gamma}{\mu_M} & -\frac{\alpha}{\mu_M} & 2\frac{\alpha+\gamma}{\mu_M} & -\frac{\alpha}{\mu_M} & -\frac{\gamma}{\mu_M} & \ddots & \vdots & \vdots & \vdots \\

\vdots  & \vdots & \vdots & \ddots & 0 & -\frac{\delta}{\mu_X} & -\frac{\alpha}{\mu_X} & \frac{\alpha+\beta+\gamma+\delta}{\mu_X} & -\frac{\beta}{\mu_X} & \ddots & \vdots & \vdots & \vdots \\

\vdots  & \vdots & \vdots & \ddots & \vdots & 0 & -\frac{\gamma}{\mu_X} & -\frac{\beta}{\mu_X} & \frac{\alpha+\beta+\gamma+\delta}{\mu_X} & \ddots & \vdots & \vdots & \vdots \\

\vdots  & \vdots & \vdots & \ddots & \vdots & \vdots & 0 & -\frac{\gamma}{\mu_M} & -\frac{\alpha}{\mu_M} & \ddots & \vdots & \vdots & \vdots \\

\vdots  & \vdots & \vdots & \ddots & \vdots & \vdots & \vdots & 0 & -\frac{\delta}{\mu_X} & \ddots & \vdots & \vdots & \vdots \\

\vdots  & \vdots & \vdots & \ddots & \vdots & \vdots & \vdots & \vdots & 0 & \ddots & \vdots & \vdots & \vdots \\

\vdots  & \vdots & \vdots & \ddots & \vdots & \vdots & \vdots & \vdots & \vdots & \ddots & \vdots & \vdots & \vdots \\

\vdots  & \vdots & \vdots & \ddots & \vdots & \vdots & \vdots & \vdots & \vdots & \ddots & 0 & \vdots & \vdots \\

\vdots  & \vdots & \vdots & \ddots & \vdots & \vdots & \vdots & \vdots & \vdots & \ddots &  -\frac{\delta}{\mu_M} & 0 & \vdots \\

\vdots  & \vdots & \vdots & \ddots & \vdots & \vdots & \vdots & \vdots & \vdots & \ddots &  -\frac{\beta}{\mu_X} & -\frac{\gamma}{\mu_X} & 0\\
 
\vdots  & \vdots & \vdots & \ddots & \vdots & \vdots & \vdots & \vdots & \vdots & \ddots &  \frac{\alpha+\beta+\gamma+\delta}{\mu_X} & -\frac{\alpha}{\mu_X} & -\frac{\delta_{\textrm{e}}}{\mu_X}\\

\vdots  & \vdots & \vdots & \ddots & \vdots & \vdots & \vdots & \vdots & \vdots & \ddots &  -\frac{\alpha}{\mu_M} & \frac{\alpha_{\textrm{e}}+\alpha+\gamma}{\mu_M} & -\frac{\alpha_{\textrm{e}}}{\mu_M}\\

0  & 0 & 0 & \cdots & 0 & 0 & 0 & 0 & 0 & \cdots &  -\frac{\delta_{\textrm{e}}}{\mu_X} & -\frac{\alpha_{\textrm{e}}}{\mu_X} & \frac{\alpha_{\textrm{e}}+\delta_{\textrm{e}}}{\mu_X}
\end{mmatrix}\right)$.\\

To find the normal modes, one has to seek for sinusoidal solutions. For this kind of solutions, Eq. \eqref{eq_matrice} becomes
\begin{equation}
\mathcal{D}\:\mathcal{U}=\omega^2 \:\mathcal{U}.
\end{equation}
Therefore, the $3N$ normal modes, with eigenfrequencies $\omega_k$ and normal displacements $\mathcal{U}^k$ ($k\in\llbracket 1,3N \rrbracket$), are obtained by diagonalizing the dynamical matrix $\mathcal{D}$.

\section{Bulk phonon frequencies}
\label{BPF}
To obtain the frequencies of the six bulk normal modes, we use the same model as in the previous section \ref{MDLMQT} except that we apply the Born von Karman periodic boundary conditions to take into account the infinite size of the crystal. In this case, the unit cell of this one-dimensional Bravais lattice contains the three atoms of one layer. For the $n^{\rm th}$ layer, we suppose that the equilibrium positions are $na$ for the Mo atom and $na-d$ and $na+d$ for the two Te atoms. Thus, Mo atoms belonging to adjacent layers are separated by $a$. With the same notation as in section \ref{MDLMQT}, we seek for solutions in the form of a plane wave with frequency $\omega$ and wave vector $k$ : $u_{j,n} = A_j e^{-i(\omega t - kna)}$ where $j=1,3$ for Te and $j=2$ for Mo, and $A_j$ are constants to be determined,  whose ratio specify the relative amplitude and phase of vibration of the atoms within each layer. By substituting $u_{j,n}$ into the equations of motion, we obtain three homogeneous equations in terms of $A_j$. These equations will have a non-zero solution provided that the determinant of the coefficients vanishes. This yields
\begin{multline}
\left[ \mu_X\omega^2 -(\alpha+\beta+\gamma+\delta)\right]^2\left[\mu_M\omega^2-2(\alpha+\gamma)\right]\\
+(\alpha+\gamma e^{ika})^2(\delta+\beta e^{-ika})+(\alpha+\gamma e^{-ika})^2(\delta+\beta e^{ika})\\
-\left[\mu_M\omega^2-2(\alpha+\gamma)\right](\delta+\beta e^{ika})(\delta+\beta e^{-ika})\\
-2\left[ \mu_X\omega^2 -(\alpha+\beta+\gamma+\delta)\right](\alpha+\gamma e^{ika})(\alpha+\gamma e^{-ika}) = 0.
\label{eq_det}
\end{multline} 
The Born von Karman boundary condition leads to $N$ nonequivalent values of $k$ given by $k=\frac{2\pi}{a}\frac{p}{N}$ with $p$ an integer. Eq. \eqref{eq_det} does not need be solved for every $k$. In fact, for the six bulk normal modes, the displacements of the three atoms within one layer are either in-phase or out-of-phase with the displacements of the atoms of adjacent layers. Therefore, $k=0$ or $k=\frac{\pi}{a}$ respectively. By Solving Eq. \eqref{eq_det} with $k=0$ and $k=\frac{\pi}{a}$ and using the symmetry of the atomic displacements, we can get the expression of the six bulk frequencies associated with the low- (LSM, LBM), mid- (iX and oX) and high-frequency (iMX, oMX) modes \cite{Luo2013}.
\begin{align}
\omega_\tr{low}^-  &= 0,\label{eq_low-}\\ 
\omega_\tr{low}^+  &= \frac{\alpha+\gamma+2\beta}{2\mu_X}+ \frac{\alpha+\gamma}{\mu_M}-\sqrt{\left(\frac{\alpha+\gamma+2\beta}{2\mu_X}- \frac{\alpha+\gamma}{\mu_M}\right)^2+2\frac{(\alpha-\gamma)^2}{\mu_X\mu_M}}, \label{eq_low+}\\
\omega_\tr{mid}^- &= \frac{\alpha+\gamma+2\delta}{\mu_X}, \label{eq_mid-} \\ 
\omega_\tr{mid}^+ &= \frac{\alpha+\gamma+2\delta+2\beta}{\mu_X}, \label{eq_mid+} \\ 
\omega_\tr{high}^- &= \frac{(2\mu_X+\mu_M)(\alpha+\gamma)}{\mu_X\mu_M}, \label{eq_high-}\\  
\omega_\tr{high}^+  &= \frac{\alpha+\gamma+2\beta}{2\mu_X}+ \frac{\alpha+\gamma}{\mu_M}+\sqrt{\left(\frac{\alpha+\gamma+2\beta}{2\mu_X}- \frac{\alpha+\gamma}{\mu_M}\right)^2+2\frac{(\alpha-\gamma)^2}{\mu_X\mu_M}}. \label{eq_high+}
\end{align}

From the value of the force constants extracted from the fit of our experimental data (see Table II of the main text), we notice that $\left|\alpha\right|\gg\left|\beta\right|,\left|\gamma\right|,\left|\delta\right|$. Thus, we can perform Taylor developments of Eqs. \eqref{eq_low+} and \eqref{eq_high+} to get more convenient expressions
\begin{align}
\omega_\tr{low}^+  &\approx 4\frac{\beta+2\gamma}{\mu}, \label{eq_dl_low+}\\
\omega_\tr{high}^+  &\approx \frac{\alpha\mu^2+2\beta\mu_M^2+\gamma(2\mu_X-\mu_M)^2}{\mu \mu_X\mu_M}, \label{eq_dl_high+}
\end{align}
where $\mu=2\mu_X+\mu_M$ is the mass per unit area of the unit cell. The relative difference between the results of Eqs. \eqref{eq_dl_low+}/\eqref{eq_dl_high+}  and the exact values obtained using Eqs. \eqref{eq_low+}/\eqref{eq_high+}, respectively, is lower than 1\textperthousand.

An interesting quantity than can be deduced from the expressions of the bulk frequencies, for low-, mid- and high-frequency modes, is the bulk Davydov splitting $\Delta\omega=\omega^+ - \omega^-$. Again by performing Taylor expansions, we get the following expressions for the Davydov splitting
\begin{align}
\Delta\omega_\tr{low}  &\approx 2\sqrt{\frac{\beta+2\gamma}{\mu}},\label{eq_dav_low} \\
\Delta\omega_\tr{mid}  &\approx \frac{\beta}{\alpha}\left(1-\frac{\gamma+2\delta}{2\alpha}\right)\sqrt{\frac{\alpha}{\mu_X}}, \label{eq_dav_mid} \\
\Delta\omega_\tr{high} &\approx \left(\frac{\mu_M^2}{\mu^2}\frac{\beta}{\alpha}-\frac{4\mu_X\mu_M}{\mu^2}\frac{\gamma}{\alpha}\right) \sqrt{\frac{\alpha\mu}{\mu_X\mu_M}} \label{eq_dav_high}.
\end{align}
The deviation of the results of Eqs. \eqref{eq_dav_low} and \eqref{eq_dav_mid}  from the exact values deduced from Eqs. \eqref{eq_low-}-\eqref{eq_mid+} is lower than $1\%$, and the deviation of the results of Eqs. \eqref{eq_dav_high} from the exact values deduced from Eqs. \eqref{eq_high-} and \eqref{eq_high+} is lower than $10\%$.

Interestingly, the high-frequency Davydov splitting (Eq. \eqref{eq_dav_high}) is the only one that can be negative since $\alpha\gg\beta,\gamma,\delta$ . If $\frac{\mu_M}{4\mu_X}\frac{\beta}{\gamma}\geq1$ the splitting is normal and the bulk high-frequency in-phase mode has a lower frequency than the bulk high-frequency out of phase mode. Otherwise, the splitting is anomalous, as it has been reported for the iMX mode in bulk transition metal dichalcogenides~\cite{Wieting1980}.

%\clearpage

\section{Normal mode displacements}
\label{NMD}
\begin{figure}[!htb]
\begin{center}
\includegraphics[scale=0.65]{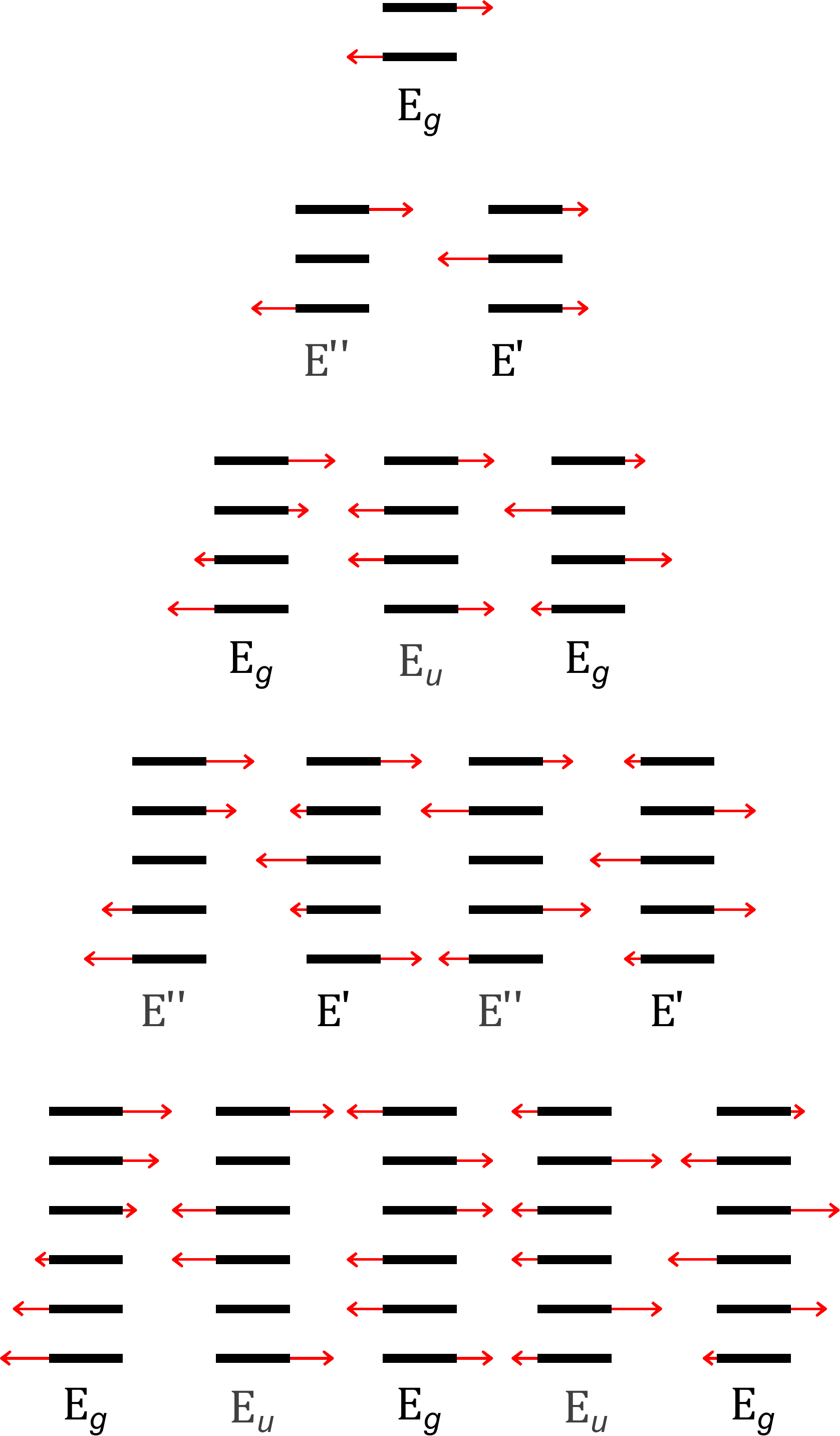}
\caption{Calculated normal displacements associated with the LSM in $N=1$ to $N=6$ layers MoTe$_2$. The size of the arrows is proportional to the amplitude of $u_{i,j}^k$ of the normal displacement obtained from the solution of Eq. \eqref{eq_matrice}. The frequencies of the modes increase from left to right. The irreductible representation of each normal mode is indicated. The modes that are Raman-active in our geometry appear in black. The other modes appear in grey.}
\label{figSI_mvt_LSM}
\end{center}
\end{figure}

\begin{figure}[!htb]
\begin{center}
\includegraphics[scale=0.42]{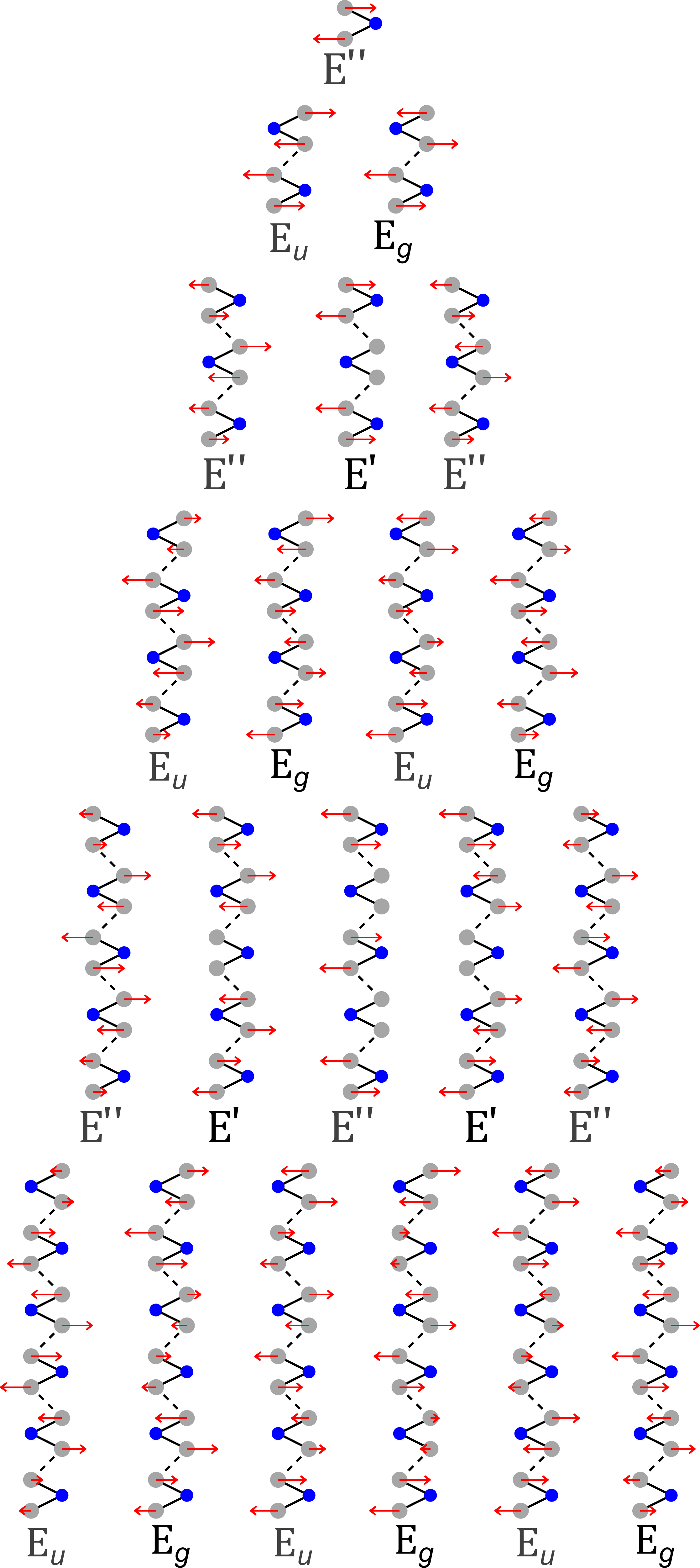}
\caption{Same as Fig.~\ref{figSI_mvt_LSM} for the iX modes in $N=1$ to $N=6$ layers MoTe$_2$. }
\label{figSI_mvt_iX}
\end{center}
\end{figure}

\begin{figure}[!htb]
\begin{center}
\includegraphics[scale=0.42]{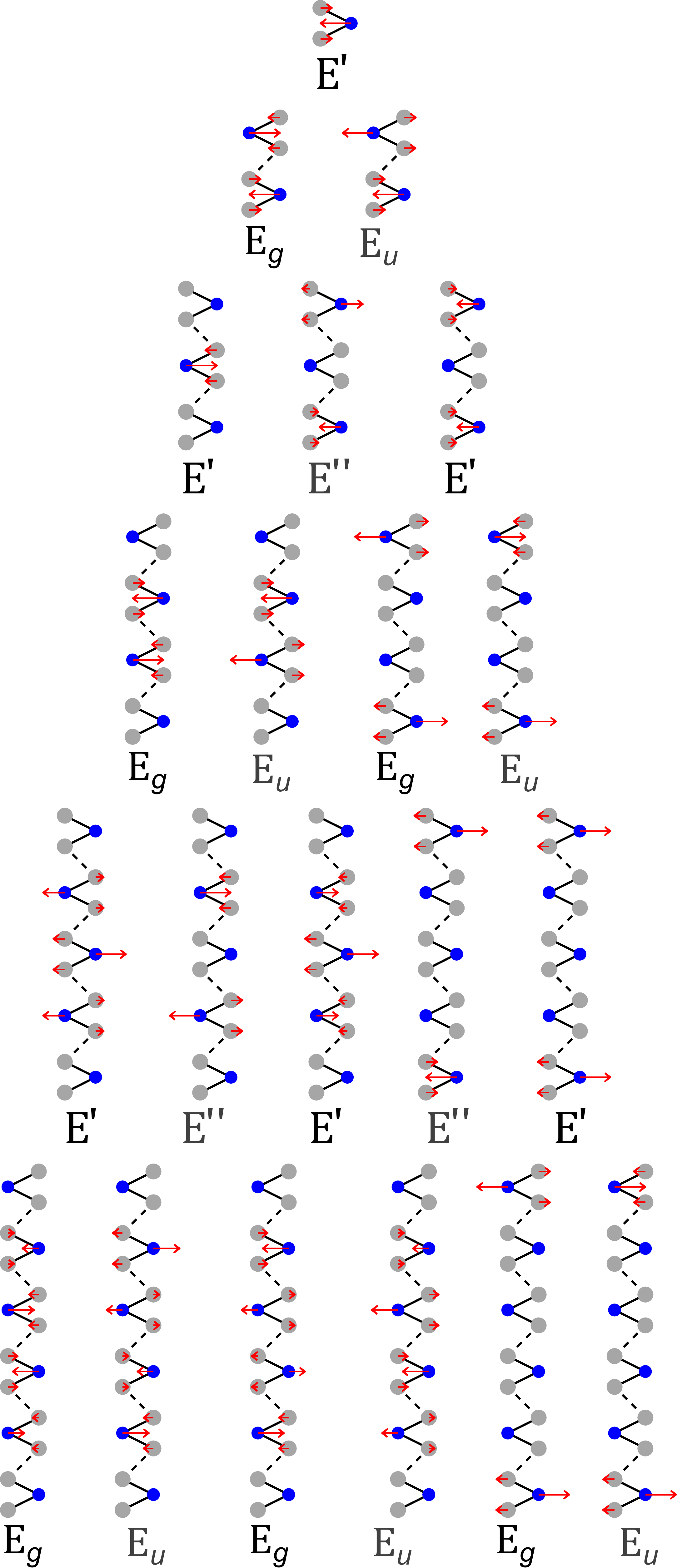}
\caption{Same as Fig.~\ref{figSI_mvt_LSM} for the iMX modes in $N=1$ to $N=6$ layers MoTe$_2$. }
\label{figSI_mvt_iMX}
\end{center}
\end{figure}

\begin{figure}[!htb]
\begin{center}
\includegraphics[scale=0.65]{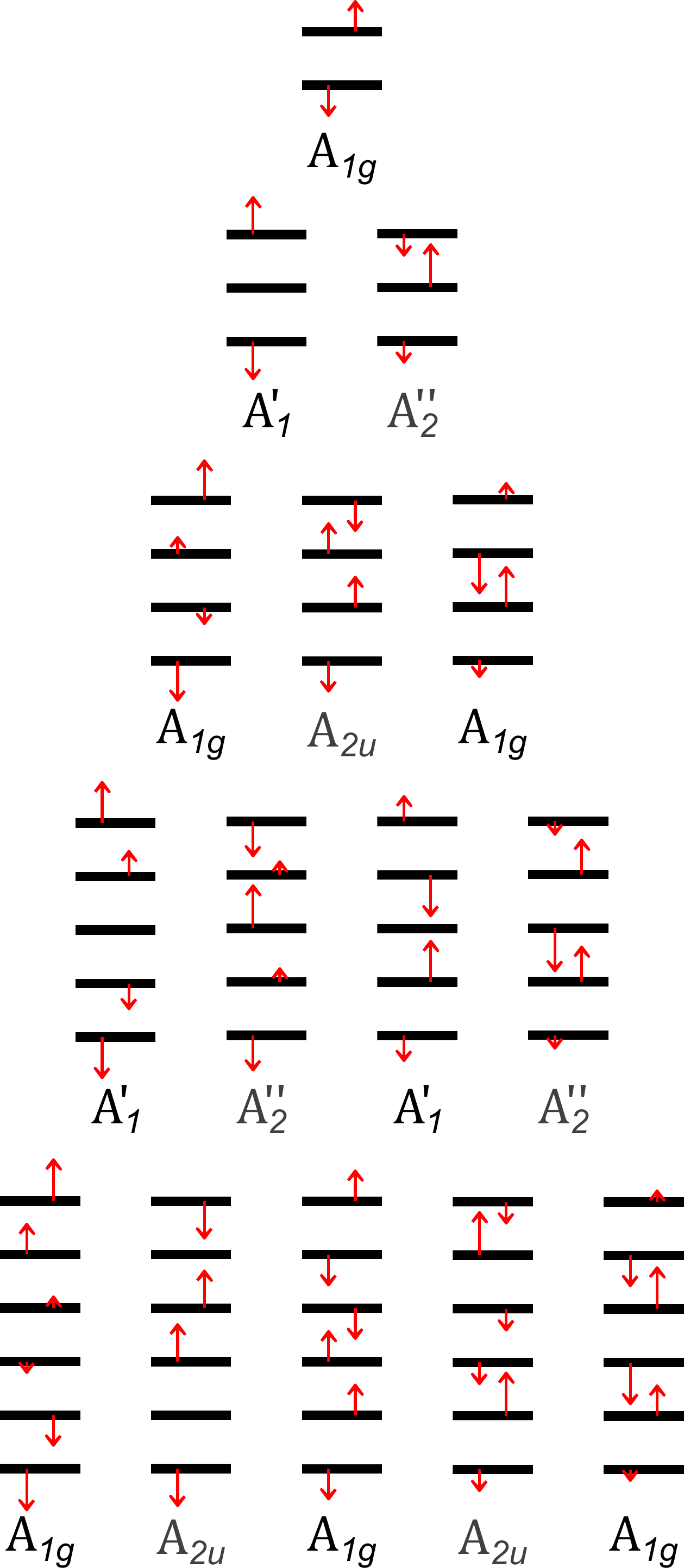}
\caption{Same as Fig.~\ref{figSI_mvt_LSM} for the LBM in $N=1$ to $N=6$ layers MoTe$_2$. }
\label{figSI_mvt_LBM}
\end{center}
\end{figure}

\begin{figure}[!htb]
\begin{center}
\includegraphics[scale=0.42]{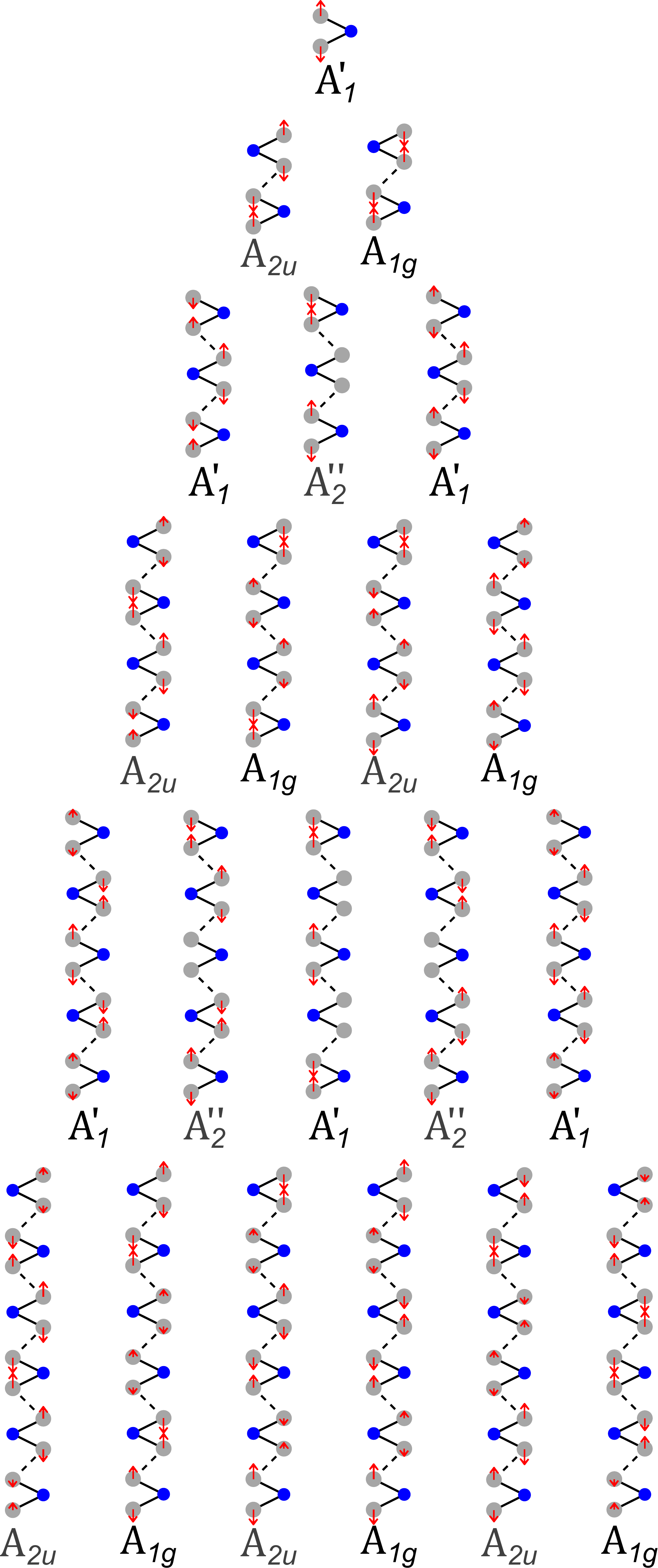}
\caption{Same as Fig.~\ref{figSI_mvt_LSM} for the oX modes in $N=1$ to $N=6$ layers MoTe$_2$. }
\label{figSI_mvt_oX}
\end{center}
\end{figure}

\begin{figure}[!htb]
\begin{center}
\includegraphics[scale=0.42]{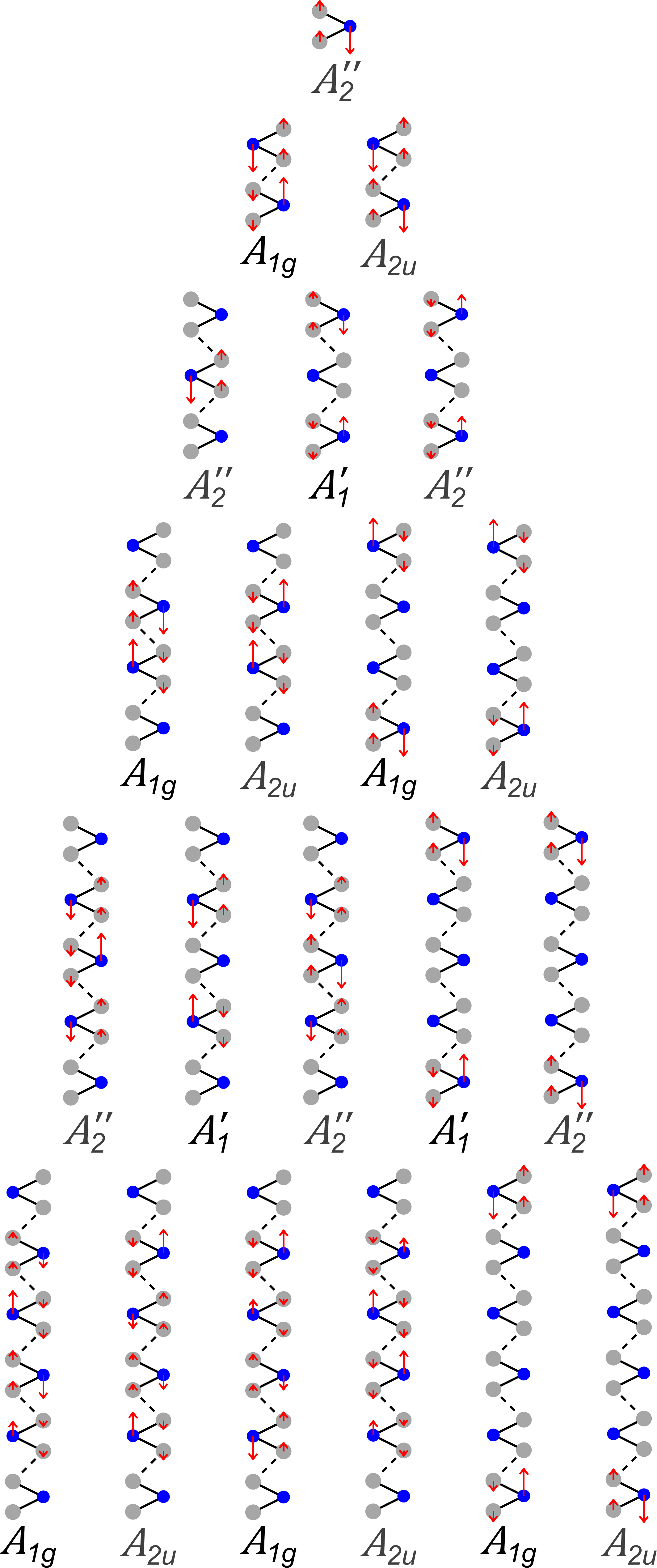}
\caption{Same as Fig.~\ref{figSI_mvt_LSM} for the oMX modes in $N=1$ to $N=6$ layers MoTe$_2$.}
\label{figSI_mvt_oMX}
\end{center}
\end{figure}

Figures \ref{figSI_mvt_LSM} to \ref{figSI_mvt_oMX} show the normal mode displacements associated with the LSM, iX, iMX, LBM, oX, and oMX  modes in $N$-layer MoTe$_2$.

\clearpage 

\section{The empirical bond polarizability model}
\label{BPM}
In the Placzek approximation, the Raman intensity for non resonant first order scattering is given by
\begin{equation}
I^{\nu} \propto |\mathbf{e}_i \cdot \mathbf{A}^{\nu} \cdot \mathbf{e}_s|^2 \frac{1}{\omega_{\nu}}(n_{\nu}+1),
\end{equation}
where $\mathbf{e}_i$ and $\mathbf{e}_s$ are the polarizations of the incident and the scattered light respectively, $\omega_{\nu}$ is the Raman frequency, and $n_{\nu} = [\exp(\hbar \omega_{\nu}/k_B T)-1]^{-1}$ with $T$ being the temperature. The Raman tensor $\mathbf{A}^{\nu}$ is related to the change of the polarizability
$\alpha$ under atomic displacement. It can be calculated as
\begin{align}
  A^{\nu}_{ij} = \sum_{k \gamma} B_{ij}^{k \gamma}  \frac{\xi^{\nu}_{k
  \gamma}}{\sqrt{M_{\gamma}}}
\end{align}
where $\xi^{\nu}_{k \gamma}$ is the $k^{\rm th}$ Cartesian component of atom $\gamma$ of the $\nu^{\rm th}$ orthonormal vibrational eigenvector $M_{\gamma}$ is the atomic mass, and
\begin{equation}
\begin{aligned}
  B_{ij}^{k \gamma} &= \frac{\partial^3 \mathcal{E}}{\partial E_i \partial E_j
  \partial u_{k \gamma}} 
  = \frac{\partial \alpha_{ij}}{\partial u_{k \gamma}}.
\end{aligned}
\end{equation}
$\mathcal{E}$ is the total energy in the unit cell, $E$ is a uniform electric field, $u_{k \gamma}$ is the $k^{\rm th}$ component of the atomic displacement $\bm{u}$ of the atom $\gamma$ and $\alpha_{i j}$ is the electronic polarizability tensor.\\
The bond polarizability model approximates the total
polarizability of the atom as the sum of the individual
bond polarizabilities. Under the assumption that the bonds have cylindrical
symmetry, the polarizability tensor for a single bond can be written as
\begin{equation}
\begin{aligned}
\alpha^b_{ij} &= \frac{1}{3}  ( 2 \alpha_p + \alpha_l) \delta_{ij} + (
   \alpha_l - \alpha_p)  \left( \hat{R}_i  \hat{R}_j - \frac{1}{3} \delta_{ij}
   \right) \\
 & = \alpha_p \delta_{ij} + ( \alpha_l - \alpha_p)
   \hat{R}_i \hat{R}_j
\end{aligned}
\end{equation}
where $\alpha_l$ and $\alpha_p$ are longitudinal and perpendicular
polarizabilities of the bond and $\hat{R}_i$ are the components of the unit
vector along the bond. Moreover, in the model, the polarizabilities depend only on
the length of the bond ($R = \sqrt{\bm{R}\cdot \bm{R}}$).
The bond vector $\bm{R}$ joining atoms $\gamma$ to atom $\gamma'$ is given by
\begin{equation}
 \bm{R} = \bm{R}_\gamma - \bm{R}_{\gamma'} + \bm{u}_\gamma - \bm{u}_{\gamma'}
\end{equation}
and
\begin{equation}
\begin{aligned}
 \frac{\partial \alpha ( R)}{\partial u_{k \gamma}} &= \frac{\partial
   \alpha ( R)}{\partial R} \frac{\partial R}{\partial u_{k \gamma}} \\
 & = \frac{\partial
   \alpha ( R)}{\partial R}  \frac{1}{2 \sqrt{\mathbf{R}\cdot\mathbf{R}}}  \left(
   \mathbf{R} \cdot \frac{\partial \mathbf{R}}{\partial u_{k \gamma}} +
   \frac{\partial \mathbf{R}}{\partial u_{k \gamma}} \cdot \mathbf{R}
   \right) \\
   & = \frac{\partial \alpha ( R)}{\partial R}  \frac{1}{2R}
   ( - 2 R_k )
\end{aligned}
\end{equation}
Hence,
\begin{equation}
  \frac{\partial \alpha ( R)}{\partial u_{k \gamma} ( l)} = \alpha' \hat{R}_k
\end{equation}
where $\alpha' =$-$\frac{\partial \alpha ( R)}{\partial R}$.
The contribution of a particular bond $b$ to the $B$ tensor is therefore
\begin{equation}
\frac{\partial \alpha^b_{ij}}{\partial u_{k \gamma}} = \alpha_p'  \hat{R}_k
   \delta_{ij} + ( \alpha_l' - \alpha_p') \hat{R}_i \hat{R}_j \hat{R}_k + (
   \alpha_l - \alpha_p) ( ( \partial_k \hat{R}_i) \hat{R}_j + ( \partial_k
   \hat{R}_j) \hat{R}_i)
\end{equation}
with
\begin{equation}
\partial_k\hat{R}_i=-\frac{1}{R}\left(\delta_{ik}-\hat{R}_i\hat{R}_k\right)
\end{equation}
In the $2H$ structure of MoTe$_2$, the Molybdenum atom is bonded to six Tellurium atoms and the Tellurium atom is bonded to three Molybdenum atoms. In our calculations, we neglect the weak inter-layer bonds. That is the reason why we don't get a finite peak for the $E^2_{2g}$ (shear) mode with this model.\\
The calculated $B$ tensors are as follows:
\begin{center}
\begin{tabular}{ l c r }
$ B_x (\text{Te}_1) = \left(\begin{array}{ccc}
     0 & \frac{-p}{2} & q\\
     \frac{-p}{2} & 0 & 0\\
     q & 0 & 0
   \end{array}\right) $ &
$ B_y ( \text{Te}_1) = \left(\begin{array}{ccc}
     \frac{-p}{2} & 0 & 0\\
     0 & \frac{p}{2} & q\\
     0 & q & 0
   \end{array}\right) $ &
 $ B_z ( \text{Te}_1) = \left(\begin{array}{ccc}
     a & 0 & 0\\
     0 & a & 0\\
     0 & 0 & b
   \end{array}\right) $ \\ \\
$ B_x ( \text{Mo}_1) = \left(\begin{array}{ccc}
     0 & p & 0\\
     p & 0 & 0\\
     0 & 0 & 0
   \end{array}\right) $ &
$ B_y ( \text{Mo}_1) = \left(\begin{array}{ccc}
     p & 0 & 0\\
     0 & -p & 0\\
     0 & 0 & 0
   \end{array}\right) $ &
 $ B_z ( \text{Mo}_1) = \left(\begin{array}{ccc}
     0 & 0 & 0\\
     0 & 0 & 0\\
     0 & 0 & 0
   \end{array}\right) $ \\ \\
   $ B_x ( \text{Te}_2) = \left(\begin{array}{ccc}
     0 & \frac{-p}{2} & -q\\
     \frac{-p}{2} & 0 & 0\\
     -q & 0 & 0
   \end{array}\right) $ &
$ B_y ( \text{Te}_2) = \left(\begin{array}{ccc}
     \frac{-p}{2} & 0 & 0\\
     0 & \frac{p}{2} & -q\\
     0 & -q & 0
   \end{array}\right) $ &
 $ B_z ( \text{Te}_2) = \left(\begin{array}{ccc}
     -a & 0 & 0\\
     0 & -a & 0\\
     0 & 0 & -b
   \end{array}\right) $ \\ \\ 
$ B_x ( \text{Te}_3) = \left(\begin{array}{ccc}
     0 & \frac{p}{2} & q\\
     \frac{p}{2} & 0 & 0\\
     q & 0 & 0
   \end{array}\right) $ &
$ B_y ( \text{Te}_3) = \left(\begin{array}{ccc}
     \frac{p}{2} & 0 & 0\\
     0 & \frac{-p}{2} & q\\
     0 & q & 0
   \end{array}\right) $ &
 $ B_z ( \text{Te}_3) = \left(\begin{array}{ccc}
     a & 0 & 0\\
     0 & a & 0\\
     0 & 0 & b
   \end{array}\right) $ \\ \\
 $ B_x ( \text{Mo}_2) = \left(\begin{array}{ccc}
     0 & -p & 0\\
     -p & 0 & 0\\
     0 & 0 & 0
   \end{array}\right) $ &
$ B_y ( \text{Mo}_2) = \left(\begin{array}{ccc}
     -p & 0 & 0\\
     0 & p & 0\\
     0 & 0 & 0
   \end{array}\right) $ &
 $ B_z ( \text{Mo}_2) = \left(\begin{array}{ccc}
     0 & 0 & 0\\
     0 & 0 & 0\\
     0 & 0 & 0
   \end{array}\right) $ \\ \\
$ B_x ( \text{Te}_4) = \left(\begin{array}{ccc}
     0 & \frac{p}{2} & -q\\
     \frac{p}{2} & 0 & 0\\
     -q & 0 & 0
   \end{array}\right) $ &
$ B_y ( \text{Te}_4) = \left(\begin{array}{ccc}
     \frac{p}{2} & 0 & 0\\
     0 & \frac{-p}{2} & -q\\
     0 & -q & 0
   \end{array}\right) $ &
 $ B_z ( \text{Te}_4) = \left(\begin{array}{ccc}
     -a & 0 & 0\\
     0 & -a & 0\\
     0 & 0 & -b
   \end{array}\right) $ \\ \\ 
\end{tabular}
\end{center}
The values of the constants $a$, $b$, $p$ and $q$, in terms of the polarizabilities and their derivatives, obtained after substituting the MoTe$_2$ lattice parameters are:
\begin{equation}
\begin{aligned}
a &= -0.24\alpha_l - 0.56\alpha'_l + 0.24\alpha_t - 1.40\alpha'_t \\
b & = 0.48\alpha_l - 0.84\alpha'_l - 0.48\alpha_t - 1.12\alpha'_t \\
p & = 0.28\alpha_l + 0.64\alpha'_l - 0.28\alpha_t - 0.64\alpha'_t \\
q & = 0.18\alpha_l - 0.56\alpha'_l - 0.18\alpha_t + 0.56\alpha'_t
\end{aligned}
\end{equation}

\section{Raman spectra from the bond polarizability model}
\label{RBPM}

\begin{figure}[h]
%\subfloat[iX Model]{\includegraphics[scale=0.6]{Raman-iX.png}} 
%\subfloat[oX Mode]{\includegraphics[scale=0.6]{Raman-oX.png}}
%\subfloat[iMX Mode]{\includegraphics[scale=0.6]{Raman-iMX.png}}
%\subfloat[oMX Mode]{\includegraphics[scale=0.6]{Raman-oMX.png}} 
\includegraphics[width=0.8\linewidth]{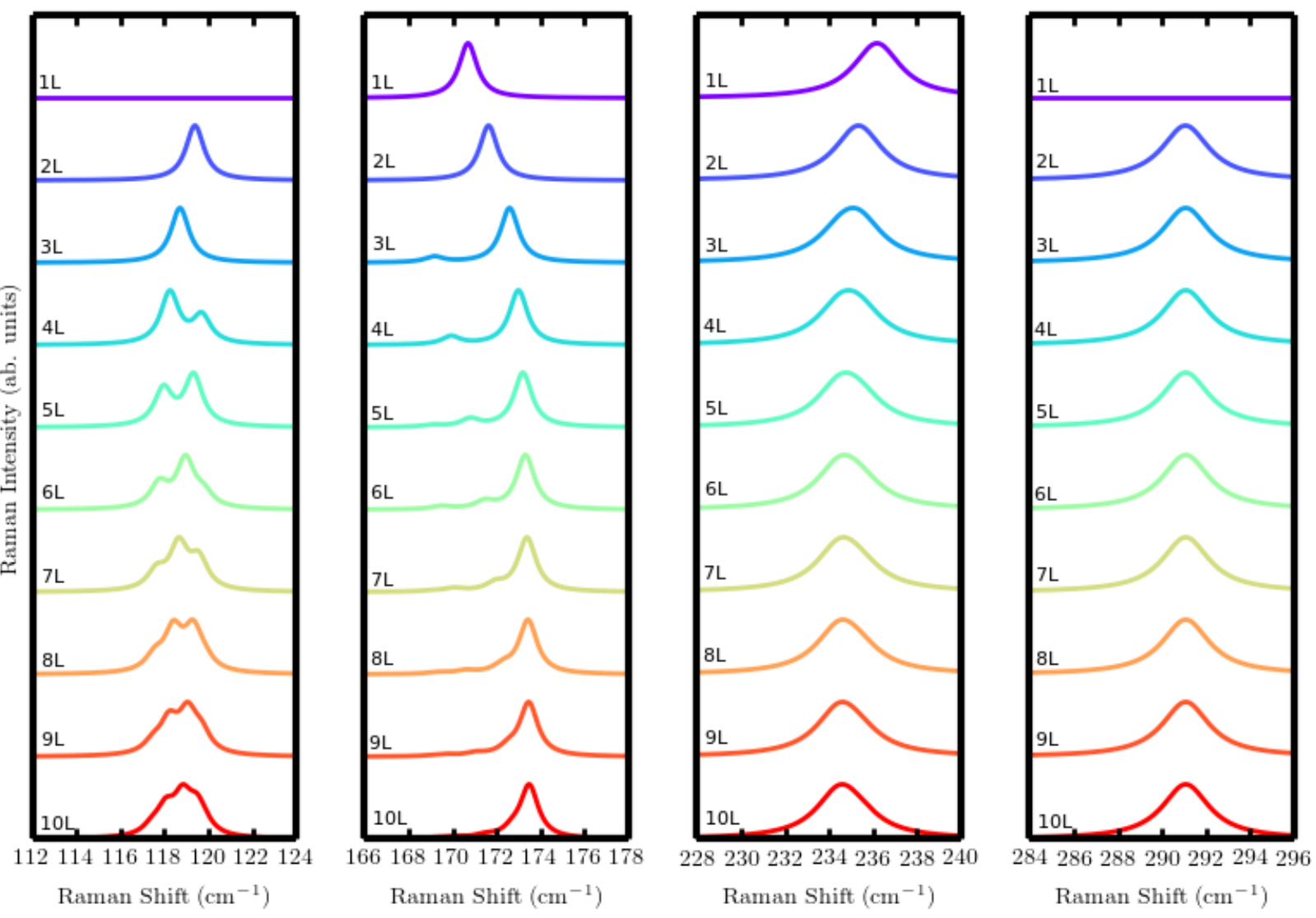}
\caption{Raman Spectra obtained with the bond polarizability model.}
\label{raman-theory}
\end{figure}

With the bond polarizability model, we can assign a Raman intensity to each Raman frequency obtained with the
force constant model. While active Raman modes agree with group theory and experimental data, the model does not include Raman resonance effects, and some discrepancies arise when we compare with experimental spectra. Figure \ref{raman-theory} shows the theoretical Raman spectra for the modes iX, oX, iMX, and oMX. The spectra for different $n$ values are offset for clarity. The iX mode has $\lfloor N/2 \rfloor$ active subfeatures in $N-$layer MoTe$_2$ and is in nice
agreement with the spectra of Fig. 3.\\
For the iMX mode, there are two close peaks for $N\geq 3$; one being an inner mode (with lower frequency) and the other being a surface mode (with higher frequency). The difference in their frequencies is around 0.5 cm$^{-1}$.The absolute intensity of the inner mode increases almost linearly with number of layers, as there are more layers vibrating, whereas the absolute intensity of the surface mode is independent of number of layers, as only the outer molecules are vibrating. As a result, the relative intensity of the surface mode drops as $N$ increases, and thus the maximum of the combined peak shifts to smaller
frequencies as $N$ increases.\\
In the case of the oX mode our model reproduces the $\lceil N/2 \rceil$ active subfeatures well, but it fails in
describing the observed relative intensity between peaks. The highest frequency peak in the oX-mode feature has much larger relative intensity than the rest of the peaks in the model, which is in contrast to our measurements, where all the phonons 
have comparable intensities. We can understand this result from the shape of the $B_z$ 
tensors shown above. The largest contribution occurs when all the Te layers are out-of-phase with the neighboring Te layers. In case two layers vibrate 
in-phase, they cancel each other, and therefore increasing the number of layers in-phase reduces
the intensity drastically. The similar intensity of the $A'_1$ peaks suggest the participation of Raman resonance effects.\\
For the oMX mode, the surface phonons, which are split from the inner modes, have much larger intensity than the inner modes in the model. From the Raman tensors, if tellurium atoms in the same layer vibrate in-phase and with 
identical amplitude, the intensity is identically zero. Only differences in the amplitudes can generate Raman
signal, which will be very small in comparison with the intensity of the oX modes. The largest amplitude differences
are due to surface phonons, which carry most of the contribution to the Raman intensity and it results in the
sole peak observed in the Raman spectra. A feature assigned to the inner modes is slightly visible in the measured spectra at $E_L = 1.96$~eV (see Fig. 3(c) in the main text) but not at $E_L = 2.33$ eV, which, again, suggests a resonance effect at $E_L = 1.96$ eV, similar to the oX mode.

\clearpage

\section{Ab-initio bulk phonon modes}
\label{AI}

\begin{figure}[!htb]
\begin{center}
\includegraphics[scale=0.25]{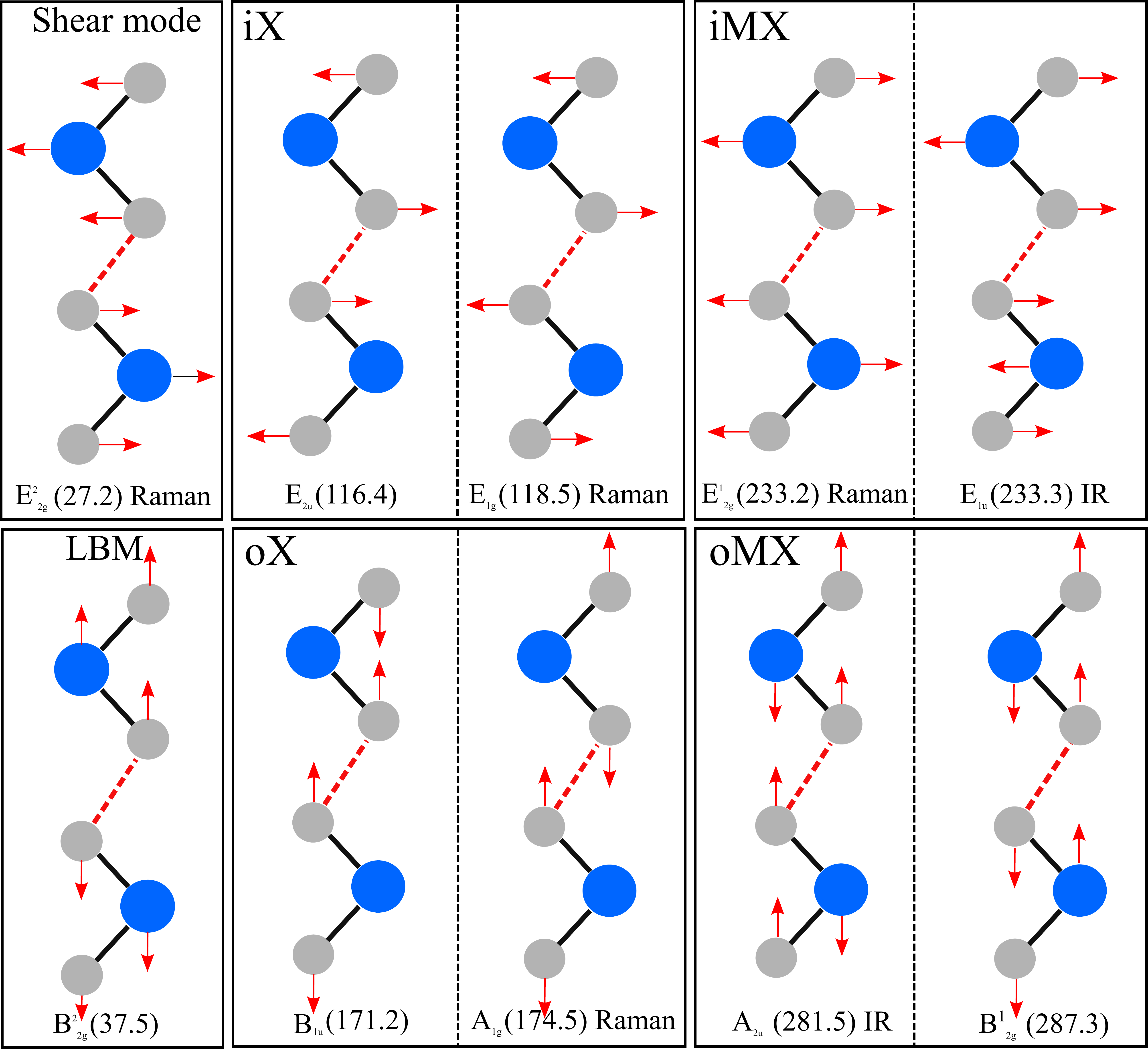}
\caption{Optical phonon modes of bulk MoTe$_2$. In the first row, modes with in-plane polarization in ascending order
of frequency. In the second row, the out-of-plane modes are shown. Davydov pairs of phonon modes are plotted in one
box.}
\label{bulk-modes-ab}
\end{center}
\end{figure}

We have complemented the empirical model with \textit{ab-initio} calculations of the 
phonon modes for bulk MoTe$_2$. We have used density functional perturbation theory (DFPT)~\cite{Baroni2001} as implemented in the Quantum Espresso code~\cite{Giannozzi2009}. We used the local-density approximation (LDA) which does not properly take
into account van der Waals interaction between the layers but nevertheless
gives decent result for the phonons of many layered systems because it overestimates the weak covalent part of the inter-layer bonding.
The energy cutoff is 80 Ry, and the Monkhorst-Pack sampling of the $\bf{k}-$grid is $12\times 12\times 4$. The optimized lattice vectors are $a=3.499$ \AA and $c=13.829$ \AA. Figure \ref{bulk-modes-ab} shows the ab-initio frequencies together with the corresponging phonon eigenvectors. We have grouped the phonon modes in Davydov pairs, except the shear and layer-breathing modes (whose ``Davydov partner'' would be a zero frequency acoustic mode).
We observe a positive Davydov splitting for the iX, oX and oMX modes and a negative Davydov splitting for the mode iMX mode, 
as reported for other transition metal dichalcogenides~\cite{Wieting1980}. The calculated Davydov-splitting of the iX mode is 2.1 cm$^{-1}$ (compared
to the 2.7 cm$^{-1}$ extrapolated from the measurements). For the oX mode, we obtain a splitting of 3.2 cm$^{-1}$ (compared to 4.7 cm$^{-1}$ from the 
experiments). The agreement between calculations and experiment is not perfect because of the lack of a proper treatment of van der Waals interaction with
local exchange-correlation functionals. The predicted Davydov-Splitting for the oMX mode is 5.8 cm$^{-1}$. For the calculation of the frequency of the $A_{2u}$ mode, we have not taken into account the coupling to an external-electric field (Lydanne-Sachs-Teller interaction) that would lead to an up-shift of this mode. However, in layers of finite width, this interaction is absent.

\clearpage

\section{Additional Raman measurements}
\label{ARM}

\begin{figure}[!htb]
\begin{center}
\includegraphics[width=1\linewidth]{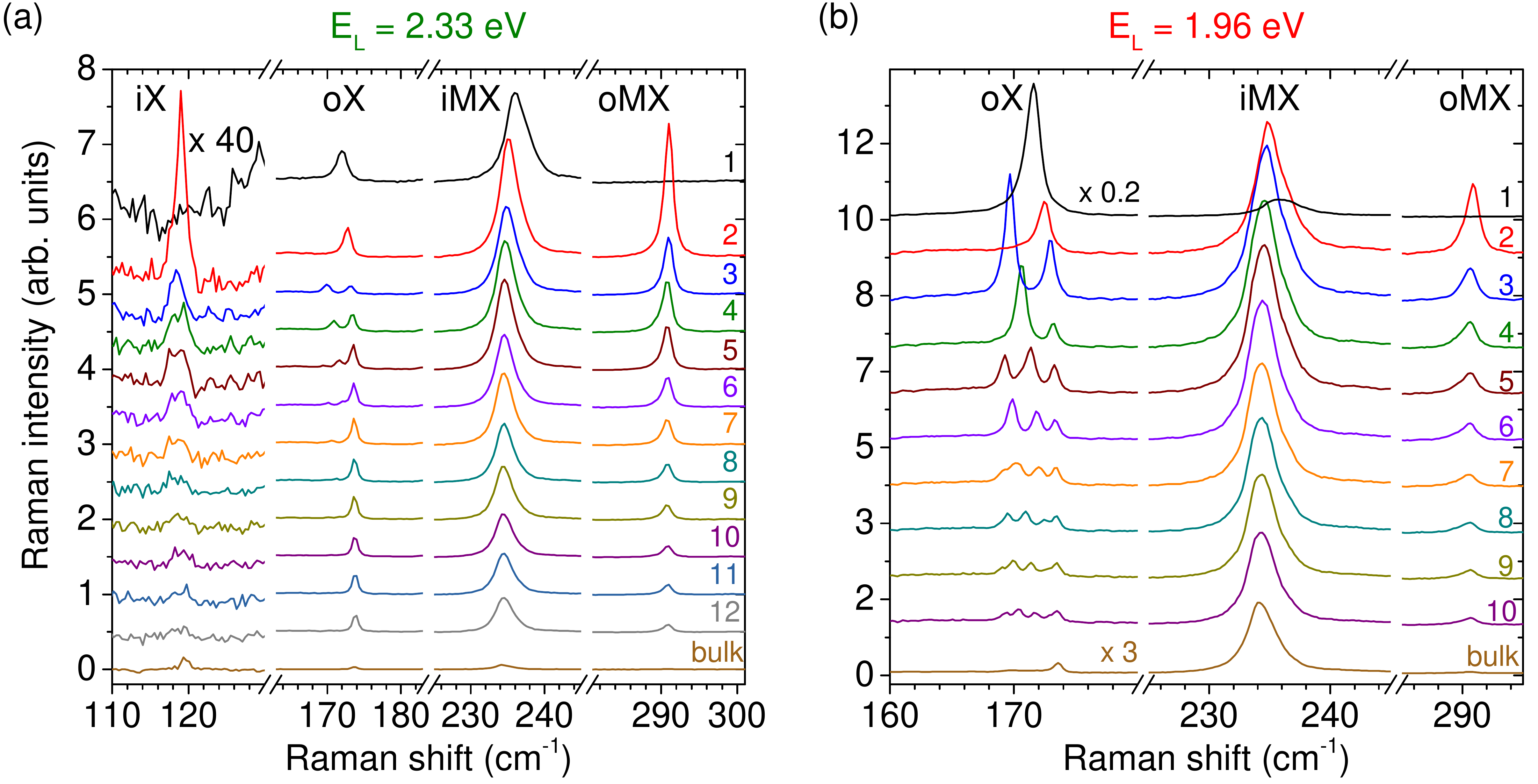}
\caption{Micro-Raman spectra of $N-$layer MoTe$_2$ recorded under the same conditions at a photon energy of (a) $2.33~\tr{eV}$ and (b) $1.96~\tr{eV}$. The spectra are vertically offset for clarity.}
\label{figSI_spectres}
\end{center}
\end{figure}

Figure~\ref{figSI_spectres} shows the raw Raman spectra of $N$-layer MoTe$_2$ recorded at $E_{\rm L}=2.33~\tr{eV}$ and $E_{\rm L}=1.96~\tr{eV}$. Note that the iX mode has not been studied at $E_{\rm L}=1.96~\tr{eV}$ due to the relatively large bandwidth of our Notch filter at $E_{\rm L}=1.96~\tr{eV}$.

\clearpage

\begin{figure}[!ht]
\begin{center}
\includegraphics[width=1\linewidth]{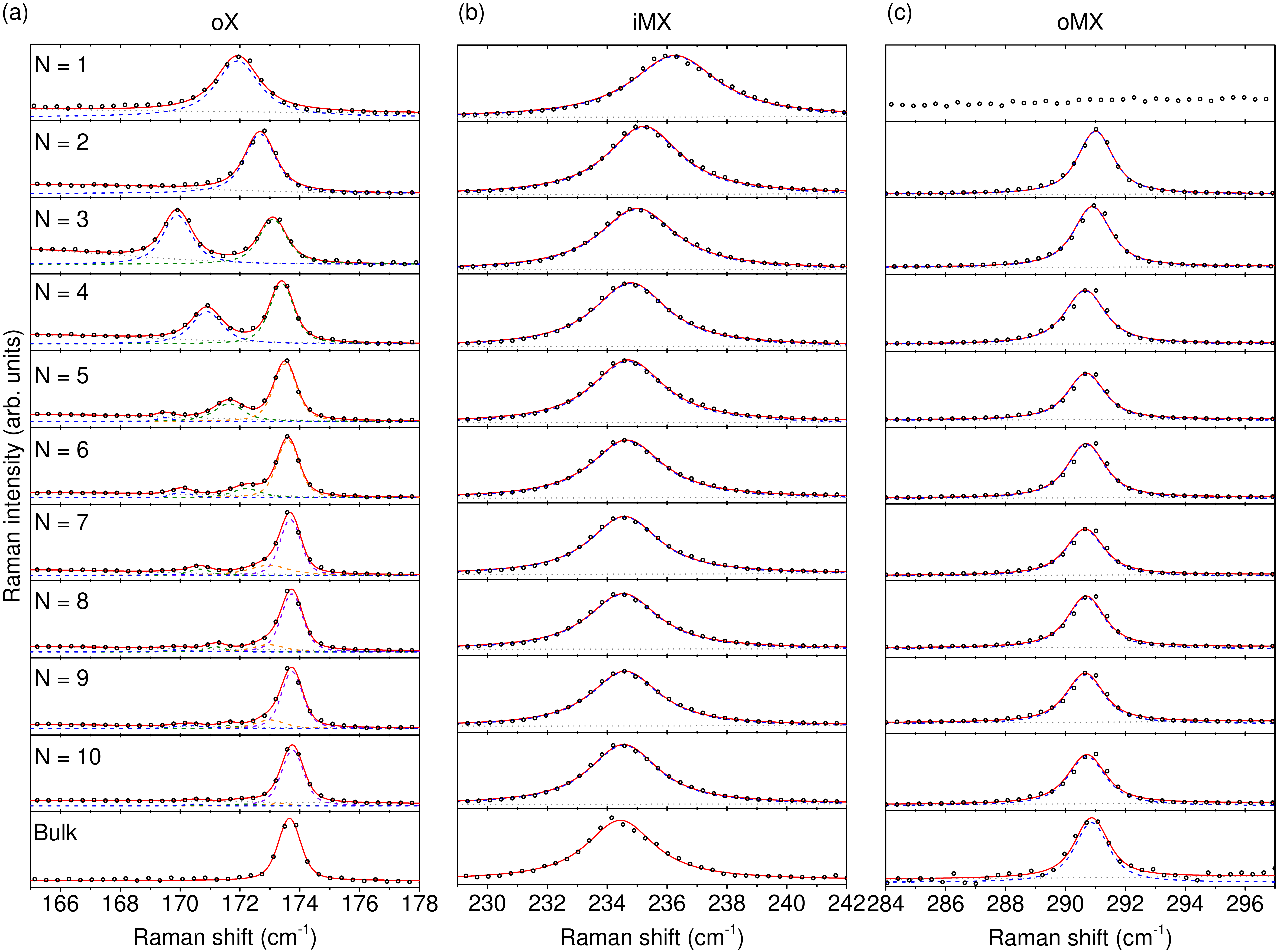}
\caption{Normalized Micro-Raman spectra of the (a) oX, (b) iMX, and (c) oMX mode-features in $N-$layer MoTe$_2$ recorded at $E_{\rm L}=2.33~\tr{eV}$. The measured Raman features (symbols) are fit to Voigt profiles (solid lines). For the modes that show a Davydov splitting, each subpeak is represented with a colored dashed line. A featureless background (grey dashed line) has been considered when necessary.}
\label{figSI_spectres_532}
\end{center}
\end{figure}

Figure~\ref{figSI_spectres_532} shows Raman spectra of the oX, iMX and oMX modes in $N$-layer MoTe$_2$ recorded at $E_{\rm L}=2.33~\tr{eV}$. The results recorded at $E_{\rm L}=1.96~\tr{eV}$ are discussed in the main manuscript. At $E_{\rm L}=2.33~\tr{eV}$, the Davydov splitting also appears clearly for the oX feature, although the highest energy subfeature contains most of the oscillator strength for $N\geq6$. The iMX feature also downshifts as $N$ increases and no appreciable splitting can be resolved. However, in contrast with our results at $E_{\rm L}=1.96~\tr{eV}$, the oMX feature does not exhibit any measurable splitting at $E_{\rm L}=2.33~\tr{eV}$. 

%\clearpage

\clearpage 

\begin{figure}[!ht]
\begin{center}
\includegraphics[width=1\linewidth]{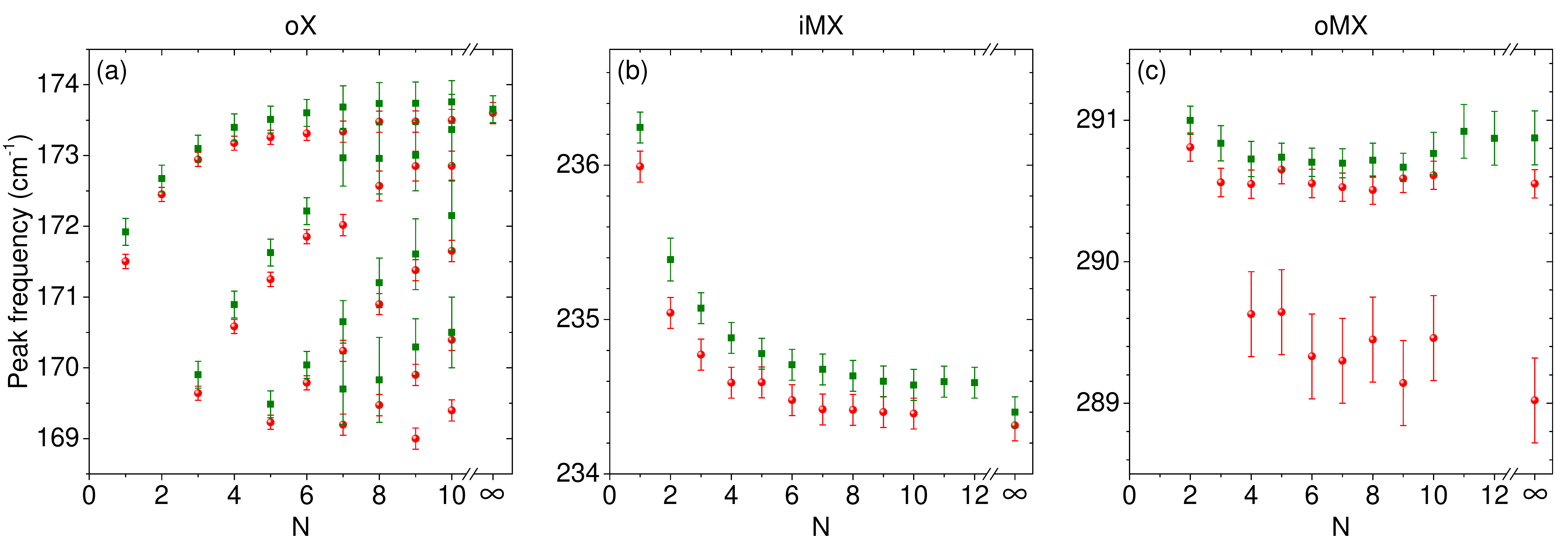}
\caption{Frequencies of the (a) oX, (b) iMX and (c) oMX modes extracted from Voigt fits as a function of the number of layers $N$. Green squares (red circles) correspond to data recorded at $E_{\rm L}=2.33~\tr{eV}$ ($E_{\rm L}=1.96~\tr{eV}$).}
\label{figSI_fan_diagrams}
\end{center}
\end{figure}

The corresponding fan diagrams associated with oX-, iMX and oMX-mode frequencies recorded at $E_{\rm L}=2.33~\tr{eV}$ are shown in Fig.~\ref{figSI_fan_diagrams}, together with the fan diagrams for oX and oMX modes extracted from the data recorded at $E_{\rm L}=1.96~\tr{eV}$ and discussed in details in the main text. These two sets of data are very consistent with each other. Still, we can notice a small rigid shift of approximately $0.2~\tr{cm}^{-1}$ which is smaller than the resolution of our experimental setup. This shift presumably arises from uncertainties (below our spectral resolution) in the calibration of our spectrometer. 
Importantly such a small shift has a negligible influence on the determination of the force constants. Indeed, the latter vary by less  $1.5~\%$ if one uses the oX, oMX and iMX  frequencies recorded at $E_{\rm L}=2.33~\tr{eV}$ instead of their values recorded at $E_{\rm L}=1.96~\tr{eV}$ in the global fitting procedure described in the main manuscript.

\bigskip
\begin{center}
\rule [0pt]{6cm}{0.5mm}
\end{center}

%\clearpage

\end{document}